\documentclass[manuscript,screen]{acmart}
\usepackage{url} 
\usepackage{natbib}
\usepackage{hyperref}
\usepackage{hyperref}

\AtBeginDocument{%
  }

\setcopyright{acmlicensed}
\copyrightyear{2025}
\acmYear{2025}
\acmDOI{X.X}

\acmISBN{978-1-4503-XXXX-X/2018/06}

\usepackage{tikz}
\usetikzlibrary{positioning,shapes.geometric}

\usepackage{url}

\PassOptionsToPackage{final}{graphicx}
\usepackage{graphicx}
\setkeys{Gin}{draft=false}

\usepackage{caption}
\usepackage{subcaption} 

\usepackage{booktabs,array,longtable,ragged2e}
\usepackage{makecell}

\usepackage{enumitem}
\usepackage{pdflscape}

\usepackage{pgfplots}
\pgfplotsset{compat=1.18}
\newcommand{\sparkbar}[2]{%
  \begin{tikzpicture}[baseline=(b.base)]
    \path (0,0) node[inner sep=0pt] (b) {};
    \draw[draw=black!25,fill=black!10] (0,0) rectangle (2.8,0.14);
    \pgfmathsetmacro{\w}{(max(#1,0)/#2)*2.8}
    \fill[black!55] (0,0) rectangle (\w,0.14);
  \end{tikzpicture}%
}

\usepackage{graphicx}
\usepackage{caption}
\usepackage{subcaption}                 
\usepackage{placeins}

\usepackage[export]{adjustbox}   
\usepackage{caption}
\usepackage{subcaption}

\usepackage{caption}
\usepackage{enumitem,tikz,xcolor}
\newcommand{\badge}[2]{%
  \tikz[baseline]\node[
    anchor=base, rounded corners=2pt,
    inner xsep=6pt, inner ysep=1.5pt,
    draw=black!25, fill=black!5,
    font=\footnotesize
  ]{#1\textsuperscript{#2}};%
}
\setlist[itemize]{leftmargin=1.2em,itemsep=2pt,topsep=2pt}

\begin{document}

\title{From Superficial Outputs to Superficial Learning: Risks of Large Language Models in Education}

\author{Iris Delikoura}
\email{idelikoura@connect.ust.hk}
\affiliation{%
  \institution{The Hong Kong University of Science and Technology}
  \city{Hong Kong}
  \country{China}
}

\author{Yi R. (May) Fung}
\email{yrfung@cse.ust.hk}
\affiliation{%
  \institution{The Hong Kong University of Science and Technology}
  \city{Hong Kong}
  \country{China}}

\author{Pan Hui}
\email{panhui@ust.hk}
\affiliation{%
  \institution{The Hong Kong University of Science and Technology, Guangzhou Campus}
  \city{Guangzhou}
  \country{China}
}

\renewcommand{\shortauthors}{}

\begin{abstract}

Large Language Models (LLMs) are transforming education by enabling personalization, feedback, and knowledge access, while also raising concerns about risks to students and learning systems. Yet empirical evidence on these risks remains fragmented. This paper presents a systematic review of 70 empirical studies across computer science, education, and psychology. Guided by four research questions, we examine: (i) which applications of LLMs in education have been most frequently explored; (ii) how researchers have measured their impact; (iii) which risks stem from such applications; and (iv) what mitigation strategies have been proposed. We find that research on LLMs clusters around three domains: operational effectiveness, personalized applications, and interactive learning tools. Across these, model-level risks include superficial understanding, bias, limited robustness, anthropomorphism, hallucinations, privacy concerns, and knowledge constraints. When learners interact with LLMs, these risks extend to cognitive and behavioural outcomes, including reduced neural activity, over-reliance, diminished independent learning skills, and a loss of student agency. To capture this progression, we propose an LLM-Risk Adapted Learning Model that illustrates how technical risks cascade through interaction and interpretation to shape educational outcomes. As the first synthesis of empirically assessed risks, this review provides a foundation for responsible, human-centred integration of LLMs in education.

\end{abstract}





\maketitle

\section{Introduction}
The integration of generative Artificial Intelligence (AI) into education is consistently gaining momentum, grounded in decades of research within the broader AI domain \cite{Kurt}. As a transformative force, AI offers innovative solutions to both teaching and learning, from personalising educational pathways to optimising instructional processes \cite{jose_cognitive_2025, srinivasa}. AI-powered tools create more interactive and adaptable learning environments by offering continuous responses, real-time feedback and enabling opportunities for self-directed learning \cite{Tapalova}. The core technologies driving this transformation include intelligent dialogue systems powered by LLMs, which adapt instruction to individual learners, and predictive analytics, which monitor and enhance performance. These adaptive learning systems possess the ability to analyse learner behaviour and progress in order to customise educational content, ensuring that students receive instruction tailored to their unique needs \cite{zhai}. This techno-optimistic narrative has become ubiquitous in academic and policy discourse, reflecting a broader consensus that emphasizes the revolutionary and positive potential of AI in education \cite{Nguyen, Klayklung}.

\par While such discourse often emphasizes potential benefits, this review adopts a more critical perspective. Rather than reiterating narratives that position AI as a universal solution to educational challenges, we investigate the specific technical and pedagogical risks introduced by LLM integration in educational settings. Accordingly, we emphasize the need for interdisciplinary collaboration among educators, developers, and policymakers to promote human-centred and contextually sensitive applications of LLMs within broader socio-technical learning ecosystems. Although situated within the wider landscape of AI in education, our analysis specifically focuses on LLM-based technologies—a rapidly evolving domain with far-reaching implications. By foregrounding empirical evidence, this review contributes a grounded understanding of the challenges posed by LLMs and establishes a foundation for informed, interdisciplinary dialogue on their responsible implementation.

\subsection{Research Questions}
We address the following questions:
\begin{itemize}
    \item RQ1: Which educational applications of LLMs have been most frequently empirically examined in the literature?
    \item RQ2: What are the predominant methods used to assess these applications?
    \item RQ3: Which risks stem from such applications?
    \item RQ4: Which mitigation strategies have been proposed?
\end{itemize}

\subsection{Contributions}

We examine the literature on LLMs in education, focusing on both the applications that have been studied and the methodologies employed to evaluate their impact. Building on this foundation, we synthesize the risks reported across technical and pedagogical domains. In the computer science literature, risks are often framed at the system level, such as superficial understanding, limited robustness, algorithmic bias, data privacy, and hallucinations. In contrast, educational and psychological studies emphasize how these systems affect learners’ experiences and cognitive processes, with concerns including diminished critical thinking, metacognitive disengagement, over-reliance, and reduced student agency. To capture a comprehensive picture, this survey organizes evidence in a way that systematically connects system-level risks with their pedagogical consequences, providing an integrated framework for interpreting fragmented findings across disciplines.

This survey makes the following contributions:
\begin{itemize}

    \item A novel taxonomy of LLM applications in education, organizing existing studies into coherent categories.
    \item A synthesis of the methodologies used to evaluate the impact of LLMs in educational contexts.  
    \item An identification of the most frequently reported risks within each application category.
    \item A summary of mitigation strategies mapped to key stakeholders—educators, developers, students, and policymakers—across LLM applications.  
    \item A conceptual LLM-Risk Adapted Learning Model, derived from the literature, that maps how risks emerge across the learning cycle.
\end{itemize}

\par Our systematic literature review differentiates itself from prior reviews on LLMs in education, many of which have examined their impact within specific domains such as computer science education \cite{Raihan2025}, computer programming instruction \cite{Pereira2025}, or have provided broad surveys of trends, challenges, and applications in education \cite{pelaez_2024,Dong_2024,guizani_systematic_2025}. In contrast, we focus exclusively on empirically documented applications of LLMs in education and the risks that follow. To the best of our knowledge, this is the first review to analyse empirical studies of LLMs in education with an explicit focus on risks, synthesizing insights from traditionally siloed domains within a unified analytical framework.

\par This paper is organized as follows. Section 2 provides background on the risks posed by LLMs across three domains: technical, pedagogical, and societal. Section 3 outlines the review methodology, detailing the Preferred Reporting Items for Systematic Reviews and Meta-Analyses (PRISMA)–based search and selection process and the development of categories to address RQ1. Sections 4–6 present the findings, focusing on the identified applications, the methods used to evaluate their impact (RQ2), the associated risks (RQ3), and proposed mitigation strategies (RQ4). Section 7 discusses these findings and introduces the LLM-Risk Adapted Learning Model. Finally, Section 8 summarizes the main conclusions of our systematic review and their implications.

\section{Background \& Related Work}

This section provides the essential background for understanding the risks posed by LLMs. First, we begin by outlining the core principles and operational mechanisms of LLMs, followed by an introduction to the key risk categories relevant to their educational use. These risks are organized into three domains: (i) inherent technical, (ii) pedagogical, and (iii) societal. Together, these areas of concern establish the conceptual foundation for the systematic review that follows.

\subsection{Large-Language Models}

LLMs are developed through training on massive corpora that include textbooks, essays, online forums, and academic websites, enabling them to mimic expert-like responses across a wide range of domains \cite{harvey}. LLMs are based on the transformer model which allows them to assign dynamic weights to different parts of the input sequence. This enables them to selectively focus on the most contextually relevant information at each step of processing \cite{xu}. Built on this foundation, the core objective of an LLM is to predict the most probable next token based on the preceding sequence. This auto-regressive process allows the model to generate fluent and contextually coherent language, making it highly effective across a wide range of natural language processing tasks. In educational settings, these capabilities have made LLMs increasingly popular among learners, who use them as accessible, on-demand tools for academic support. Such tools include ChatGPT, Bard, DeepSeek, Claude, Gemini, and Copilot. Students use them to clarify complex concepts, paraphrase challenging texts, brainstorm essay ideas, solve problems, and even generate full written assignments. 

\subsection{Inherent Technical Risks of LLMs}

\subsubsection{Toxic Content}
The large-scale corpora used for LLM training contain toxic, biased and noisy data which may lead to the generation of harmful content by the model and increase the risk of exploitation \cite{hernandez2022}. Even after pre-processing, studies show that LLMs may still generate pornographic, vulgar, violent, or biased content. This is largely due to the challenges of fully auditing massive training datasets \cite{hartmann2023, mcgee2023}. Recent research has demonstrated that both proprietary and open-source LLMs remain susceptible to adversarial prompting techniques that exploit these models. Experiments conducted across a range of models—including GPT-4, GPT-4o-mini, LLaMA3-405B-Instruct, Cohere, and Gemini—reveal that they can be manipulated to produce fabricated explanations that imitate scientific reasoning, thereby framing biased or discriminatory views as if supported by empirical evidence \cite{yubin2025}. Such vulnerabilities also expose models to jail-breaking attacks, wherein safety and alignment constraints are bypassed through subtle, deceptive input phrasing. This highlights a critical structural weakness in even the most advanced LLMs: their inability to discern adversarial intent when it is cloaked in authoritative or scholarly language. These models lack the capacity for holistic understanding or meaningful inference, and are unable to interpret psychological or behavioural nuances \cite{borji2023s}. As a result, they cannot recognize the harmful content or false narratives embedded in their training data, and thus rely entirely on human oversight for regulation. An additional concern stems from the black box nature of transformer based models making it challenging to trace how specific inputs lead to particular outputs \cite{bommasani2022}. This opacity, combined with their lack of robustness (where even minor changes in input phrasing can yield unreliable or misleading responses) is especially problematic in educational settings, where transparency, consistency, and accountability are essential.

\subsubsection{Misinformation} This arises as a significant risk stemming from the structural design of LLMs. When these models are trained on datasets that contain a blend of accurate and misleading information, they are prone to reproducing and amplifying falsehoods. Students may rely on model outputs that are factually incorrect, fabricated, or outdated—mistaking linguistic fluency for epistemic reliability. Such confusion can erode trust in both the learning process and knowledge systems, especially when learners lack the critical skills to distinguish between authoritative information and plausible-sounding but incorrect responses. The circulation of misinformation not only undermines learners' academic development but may also shape distorted world-views, perpetuate harmful stereotypes, or propagate pseudoscience. Moreover, the anthropomorphic characteristics of LLMs, displaying apparent fluency, coherence and confidence in their responses, may lead users  to relate to the LLMs as sentient or authoritative entities. Despite their surface level plausibility, the outputs are ultimately probabilistic constructs, lacking genuine semantic intent or contextual awareness. In this sense, LLMs can be considered a "stochastic parrot", repeating learned linguistic forms without grasping their meaning, or purpose, and lacking in nuance and authentic engagement \cite{Bender}. When students rely on these outputs as a substitute for their own reasoning or critical engagement, the learning process is fundamentally compromised. Genuine learning requires the active construction of meaning, integration of knowledge, and reflective engagement with content. These processes cannot occur through passive consumption of syntactically correct but semantically hollow responses. Without this deeper cognitive work, learners risk mistaking linguistic fluency for understanding, thereby undermining the very goals of education \cite{Stuchlikova}.  

\subsubsection{Amplification of Bias}
Research has demonstrated that LLMs encode both explicit and implicit associations, often linking certain names, dialects, or identity markers with specific traits, occupations, or behaviours, frequently reflecting dominant cultural norms. For instance, prompts involving male-associated names have been found to yield more confident or positive assessments than equivalent prompts using female-associated names \cite{salinas}. When integrated into educational tools, such biases can subtly shape students’ experiences, undermining their self-perception, sense of fairness, and motivation to learn. In addition to content-level bias, emerging research highlights the influence of persona conditioning on the model (i.e. the roles assigned to the model during interaction). When prompted to act as a “teacher” or “student,” an LLM’s tone, reasoning, and formality may shift, potentially amplifying prescriptive or authoritative behaviour \cite{weissburgg}. In some cases, this role-dependent behaviour has also been associated with increased toxicity or stereotyping \cite{cheng,deshpande}. Similarly, the narrative framing of prompts, whether in first-person or third-person, can significantly affect the objectivity and coherence of generated responses. While third-person prompts tend to produce more detached and accurate outputs, first-person formulations often elicit more emotionally charged or biased content \cite{suzgun2024}. These framing effects remain under-explored in education, where even subtle shifts in interaction style could meaningfully influence how students interpret and respond to LLM-generated material. Importantly, bias is not confined to the internal workings of the models themselves, but extends to their broader deployment within educational systems. From maths performance predictors to automated essay scoring and learning analytics dashboards, bias can emerge through input design, user interaction, and the interpretation of outputs \cite{chai, akgun}. These effects are particularly pronounced among younger students and learners from historically marginalized groups, where biased model behaviour can reinforce existing inequalities in educational opportunities and outcomes \cite{stureborg}.

\subsubsection{Hallucinations}
These occur when LLMs generate outputs that deviate from factual information or misinterpret the context in which a prompt is given. Such errors can be broadly categorized into four types: intrinsic, extrinsic, amalgamated, and non-factual hallucinations. Intrinsic hallucinations stem from limitations in the model’s internal knowledge or representation \cite{liu2024}, while extrinsic hallucinations arise from misinterpretations of external prompts or multi-modal inputs \cite{bai2025}. Amalgamated hallucinations involve the inaccurate merging of multiple facts, leading to distorted or hybridized claims \cite{zhang2024}. Non-factual hallucinations occur when the model outputs statements that directly contradict established facts \cite{yu2024}. These phenomena are driven by underlying mechanisms such as knowledge overshadowing, insufficient internal representation, context misalignment, and failures in information extraction. Given the reliance of learners on perceived authoritative output, such hallucinations present significant epistemic and pedagogical challenges in educational settings. 

\subsubsection{Privacy Violations}
Concerns about structural vulnerabilities in LLMs, particularly with regard to privacy breaches further complicate their responsible deployment in educational contexts. Although training datasets are expected to exclude personal or sensitive information \cite{Abadi}, redaction practices are often inconsistent or insufficient. As a result, models may inadvertently memorize and later reproduce identifiable data when triggered by specific prompts \cite{Ramaswamy}. This poses significant risks in educational settings, where LLMs may be used to process student work, communications, or assessment data. The inability of many LLMs to reliably safeguard sensitive information raises serious ethical and legal concerns, especially in jurisdictions with stringent data protection regulations pertaining to minors and educational records.

\subsection{Pedagogical Risks of LLMs}
At its foundation, education serves as a structured process for cultivating higher-order cognitive abilities alongside essential social and practical skills. This is achieved through active participation in the learning process, ranging from basic information retrieval to more advanced practices such as analysing, evaluating, and creating \cite{bloom}. By engaging in these processes, students build the cognitive flexibility and \textit{metacognitive awareness} necessary to adapt to complex real-world challenges. Two theoretical perspectives are particularly useful for analysing the pedagogical risks posed by LLMs: Cognitive Load Theory (CLT) \cite{sweller1988} and Metacognitive Theory \cite{Flavell}.

CLT examines the mental effort used when learning and problem-solving. It distinguishes between:
\begin{itemize}
    \item Intrinsic Load: Inherent complexity of the material.
    \item Extraneous Load: Stems from the way information is presented.
    \item Germane Load: Involves the mental effort required to integrate new knowledge in long-term memory.
\end{itemize}

Metacognitive Theory complements this perspective by examining how one manages learning (e.g. the ability to regulate one’s own thinking through planning, reflection and evaluation). More specifically, this encompasses:
\begin{itemize}
    \item Metacognitive Knowledge: Awareness of cognitive strategies and task demands (Planning).
    \item Metacognitive Experiences: The subjective judgments and feelings that arise during learning (Awareness \& Reflection).
    \item Metacognitive Monitoring: Ongoing evaluation of progress and strategy effectiveness (Evaluation \& Adjustment).
\end{itemize}

While CLT explains \textit{how} the distribution of cognitive effort influences opportunities for mental regulation, metacognitive theory explains \textit{what} specific processes learners engage in to regulate their thinking. Metacognition requires significant mental effort; processes such as planning, reflection, and evaluation can only occur when sufficient cognitive capacity remains after processing the content. The integration of LLM tools, such as ChatGPT, poses risks to these processes \cite{jose_cognitive_2025}. By reducing extraneous load through simplified presentation, LLMs can make learning tasks more efficient. Yet, this convenience that may also diminish germane load (the productive effort that builds understanding) so that metacognitive processes are not triggered at all, simply because learners do not need to stop and think.

\subsubsection{Cognitive Risks}

LLMs have been shown to reduce cognitive load by streamlining information retrieval and simplifying comprehension, often outperforming traditional methods such as web searches in terms of efficiency and accessibility \cite{stadler2024}. By synthesizing and structuring information in a readily digestible form, LLMs lower the need for learners to actively seek out and integrate content. However, while this reduction in extraneous cognitive load may enhance productivity and ease of task completion, it does not necessarily translate into improved learning outcomes. Research suggests that users often engage less deeply with content when using LLMs, potentially undermining the germane cognitive load \cite{stadler2024}. This decline in cognitive engagement raises significant concerns about the over-reliance on LLM tools in educational settings, where sustained mental effort is essential for developing transferable understanding and long-term retention. Dependence on such tools not only risks homogenizing students’ reasoning patterns but may also inhibit creative thinking and critical evaluation, thereby diminishing their ability to question, interpret, and build upon information \cite{Kosmyna_2025}. As students increasingly rely on LLM-generated content, their active participation in the learning process may lead to the reduction of \textit{student agency} (i.e. students’ active involvement in shaping their learning through choice, responsibility, and reflection). This risks depriving learners of the opportunity to acquire the intellectual and ethical skills necessary for constructing reasoned arguments and cultivating an authentic academic voice \cite{bouchard}. The disconnect becomes especially problematic when LLMs are positioned to perform roles traditionally reserved for educators such as teaching complex material, offering formative feedback, or grading written assignments. The dual dynamic where LLMs are used to simulate educator roles, while students also rely on the same systems for generating content or solving problems, creates a feedback loop. In such cases, both instruction and response generation are outsourced to the same probabilistic system, reducing the opportunity for genuine intellectual exchange and critical development.

\par Research indicates that students are increasingly inclined to accept LLM-generated content without critical scrutiny, engaging with it passively rather than as active constructors of knowledge \cite{jose_cognitive_2025}. This shift is concerning in light of epistemic cognition theory \cite{perry}, which highlights how learners’ beliefs about knowledge shape the ways they evaluate and justify information. As students come to view LLM systems as epistemically authoritative, they may adopt a deferential stance, bypassing essential processes of reasoning, questioning, and justification, and instead become reliant on these systems \cite{lee}. Individual differences in the extent of the impact of LLMs have further been identified. Specifically, more competent learners have been found to engage with LLMs strategically, using them to revisit, integrate, and refine information in ways that supports the construction of meaningful knowledge structures, reducing cognitive strain while maintaining high levels of mental engagement \cite{yang2024}. In contrast, less competent learners are more likely to rely on the immediacy and convenience of LLM outputs, bypassing essential learning processes such as paraphrasing, reflection, or synthesis.

\subsubsection {Memory Erosion}

The increasing reliance on LLMs for cognitive offloading may contribute to the erosion of memory retention. Emerging research shows that prolonged exposure to LLM-generated content is associated with measurable declines in memory performance \cite{akgun2024}. One key indicator of this effect is the ability to accurately quote one’s own work—a task that reflects not only surface-level recall but also deeper semantic integration. Studies show that participants who use LLMs have significantly more difficulty recalling quotes from their own essays compared to those who do not rely on such tools. The complete absence of accurate quoting in LLM group sessions, suggests that these learners engage in shallow memory encoding and exhibit limited semantic processing of the material \cite{Kosmyna_2025}. These findings raise significant concerns about the long-term cognitive implications of LLM-assisted writing, particularly for educational tasks that require sustained attention, deep comprehension, and the internalization of knowledge for future retrieval.

\subsubsection {Academic Integrity}
The use of LLMs in academic settings has been linked to an increase in cases of misconduct \cite{thorp2023}, which can compromise the credibility and reliability of scholarly work. As students and researchers increasingly rely on LLM-generated content, detecting plagiarism and academic dishonesty has become more challenging, undermining efforts to maintain academic integrity \cite{wu2023}. Beyond formal infractions, integrity is also compromised at the psychological level. Students who use LLMs for academic support often report a conflicted sense of authorship and ownership over their work \cite{Kosmyna_2025}. This erosion of academic integrity is closely tied to the psychological dimensions of learning, as described by Self-Determination Theory (SDT) \cite{ryan}. According to SDT, there are three basic psychological needs that are essential for intrinsic motivation and  meaningful learning: autonomy, competence, and relatedness. When learners outsource intellectual effort to LLM systems, their sense of autonomy can diminish, as they become less self-directed and more dependent on external tools. Similarly, the perception of competence may be undermined when students rely on LLM-generated answers rather than engaging in problem-solving or critical reasoning themselves. Finally, the need for relatedness—a sense of connection with others—may disappear in LLM-dominated learning environments, where collaborative dialogue and human feedback are replaced by solitary interactions with machines.

\subsection{Societal Risks of LLMs}
These harms concern the systemic and societal effects of widespread LLM deployment, including the emergence of an AI divide, data colonialism, policy and regulatory gaps, and environmental impacts. 

\subsubsection{AI Divide}
These risks relate to the broader societal implications of large-scale LLM development and deployment. Carter et al. \cite{Carter} offer a comprehensive framework that positions the AI divide as a distinct sub-dimension of the broader digital divide. They conceptualize the AI divide as encompassing three levels of inequality: access to AI technologies (first-level), the ability to effectively use them (second-level), and the outcomes resulting from AI engagement (third-level). These disparities can manifest at the individual, institutional, or national level. LLM training and inference is heavily concentrated in a small number of economically advantaged countries. Due to the high financial costs associated with developing these models, only the wealthiest companies, institutions, and individuals have the capacity to deploy LLM technologies \cite{sathish2024}. This uneven distribution of technological power presents a significant educational risk: learners and institutions in low and middle-income countries may be excluded from access to cutting-edge educational tools, deepening existing global inequalities in digital literacy, academic opportunity, and participation in AI-driven learning innovation. Even within high-income countries, as reported in the Netherlands \cite{wang_2024}, groups such as unskilled users remain at particular risk. These individuals tend to be older, possess lower levels of formal education, and show limited confidence in navigating digital technologies effectively. This disparity can be understood as an algorithmic knowledge gap or algorithmic awareness gap, where certain groups are systematically excluded not due to lack of access alone, but due to insufficient skills and understanding required to navigate AI tools effectively. It is clear that most commercially available models are not designed with developmental considerations or pedagogical objectives in mind, leading to a misalignment between system behaviour and the goals of an inclusive educational and societal environment. Without careful domain-specific tuning or context-aware guardrails, LLMs risk perpetuating norms and assumptions that conflict with core educational values such as fairness and accessibility.

\subsubsection{Data Colonialism}

The concept of data colonialism, first introduced by Couldry and Mejias \cite{couldry2019}, describes how powerful nations in the \textit{Global North} (i.e. economically developed countries, primarily in North America and Western Europe) exert control over data flows, often to the detriment of countries in the \textit{Global South} (i.e. less wealthy or developing countries). Unlike traditional colonialism, which relied on the physical extraction of natural resources, data colonialism operates through the digitization and commodification of human life. It brings a new power imbalance to the global data landscape, where the Global North can disproportionately control data flows and, in turn, influence the digital economy and knowledge infrastructures \cite{kwet}. A key dynamic driving this imbalance is the skewed value exchange: while the Global South often supplies raw data and human labour, it receives little in the way of compensation, influence, or long-term benefit. This asymmetry has serious implications for education. Learners and schools in low- and middle-income countries often contribute to the development of AI systems, either directly through participation in pilot studies or educational platforms, or indirectly through outsourced data annotation and content moderation. A well-documented example is the case of Kenyan workers hired to label harmful content for OpenAI’s language models. While their labour was essential in refining safety layers for systems like ChatGPT, they were paid minimal wages and exposed to traumatic material, all while remaining invisible in the global AI supply chain \cite{perrigo2023}. Similarly, students' behavioural and performance data, collected through their interaction with educational apps, can be used to improve LLMs without any guarantee that those systems will reflect their educational contexts or be made available to them equitably. For example, a student’s writing may help train an automated feedback system, but such models are typically aligned with dominant linguistic norms (e.g. standard British or American English) and evaluative frameworks rooted in Global North educational standards. Cultural expressions, local storytelling styles, or indigenous knowledge practices may be misinterpreted as errors or flagged as deviations from the norm. Without meaningful governance frameworks and participatory design processes, the promise of LLMs in education risks deepening inequalities.

\subsubsection{Environmental Harm}
The environmental impact of LLMs spans their entire life cycle, from resource extraction and hardware manufacturing to deployment, operation and eventual disposal. LLMs rely heavily on graphics processing units (GPUs), whose production begins with the extraction of rare metals such as tantalum, cobalt, lithium, and copper. These processes are environmentally destructive—contributing to deforestation, soil and water pollution, and loss of biodiversity. Critically, mining operations often take place in low-income regions, where local communities bear the brunt of ecological degradation without reaping the technological benefits \cite{hosseini}. In areas such as Kolwezi in the Democratic Republic of Congo, cobalt mining has been linked to hazardous working conditions and the documented use of child labour. The manufacturing stage involves plastics and silicon in an energy-intensive process that emits significant amounts of carbon dioxide and industrial waste. In deployment and operation, environmental costs rise further. LLM systems rely on large-scale data centres, which consume vast amount of energy not only for computation during training and inference, but also for the extensive cooling systems needed to maintain performance. Training a single large model like GPT-3 (with 175 billion parameters) consumes approximately 1287 megawatt-hours (MWh) of electricity and emits over 550 tons of CO$_2$ \cite{alzoubi}. By 2030, data centre energy use is projected to exceed 8\% of electricity consumption in the United States and 5\% in Europe \cite{sachs2023}. This phase compounds environmental strain, especially when powered by non-renewable energy sources. Finally, the disposal stage presents further challenges. GPUs and associated hardware eventually enter the global stream of electronic waste. Many of these components are non-recyclable, and even recyclable parts often require specialized processing that is inaccessible in many regions. This contributes to mounting waste management problems, including toxic landfill accumulation and environmental degradation in communities near e-waste processing sites.

\subsubsection{Policy and Regulatory Gaps}

The absence of sector-specific regulation has facilitated the rapid integration of commercial LLM tools into educational contexts, often without sufficient assessment of their safety, developmental appropriateness, or pedagogical soundness. While over 300 national and international AI policy initiatives are currently in place or under development, the majority focuses on capacity-building in higher education and workforce retraining to meet the demands of technological transformation \cite{funa}. However, the lack of a unified policy framework for guiding LLM use by educators and students has led to inconsistent practices and widespread uncertainty. In this regulatory vacuum, concerns have emerged around academic integrity and misuse, with Oxbridge banning the use of ChatGPT in 2023. On the other hand, some view tools such as ChatGPT as holding transformative potential for learning and assessment \cite{kasneci2023}. In the absence of robust governance mechanisms, educational institutions are left to interpret and implement LLM technologies independently, increasing the likelihood of ethical oversights, inequitable access, and pedagogical misuse.

\subsection{Intersection of Risks}
\par Figure 1 presents a Venn diagram categorizing the multifaceted risks posed by LLMs in educational contexts into three domains: inherent technical risks, pedagogical risks, and societal risks. Each circle highlights domain-specific concerns: technical risks include issues such as limited robustness and the black-box nature of LLMs; pedagogical risks emphasize the impact on cognitive processes, lack of critical thinking, memory erosion and compromised academic integrity, while societal risks capture the structural concerns such as the AI-driven digital divide and environmental harm. The convergence points in the Venn diagram represent risks that span across multiple domains. For example, toxic content, bias amplification, hallucinations, and data privacy violations arise from the intersection of technical, pedagogical, and societal concerns, reflecting how these risks are both system-level and learner-facing. Similarly, high resource requirements (technical and societal) and data colonialism (pedagogical and societal) illustrate how infrastructure and power imbalances affect both access and learning. These overlaps highlight that many harms are not isolated, but emerge from the interaction between technology, pedagogy, and society.

\begin{figure}[htbp]
  \centering
  \includegraphics[width=0.8\textwidth,trim=8pt 0pt 8pt 6pt,clip]{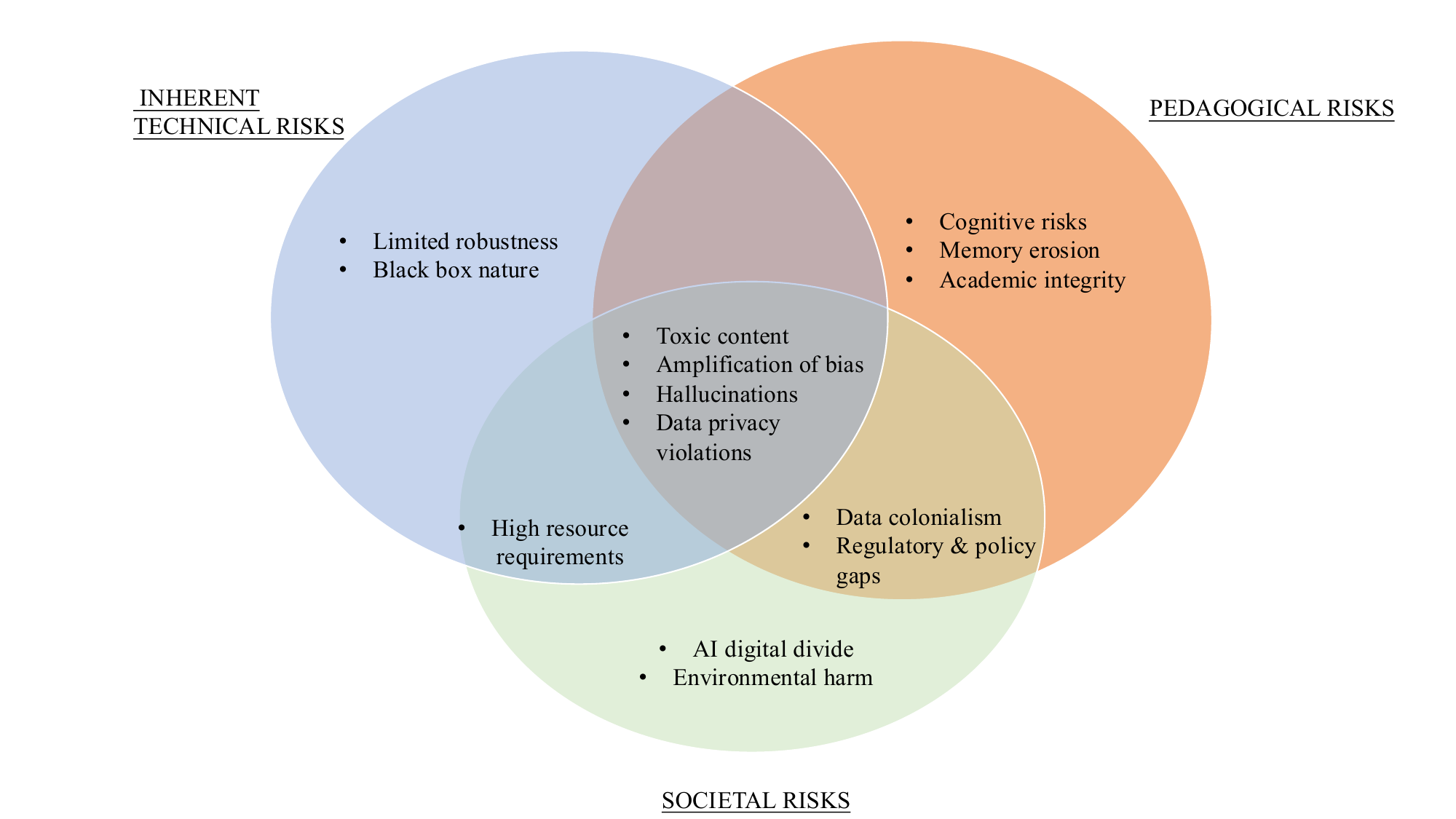}
  \Description{Venn diagram showing overlap of LLM risks.}\caption{Venn Diagram of LLM Risks}
  \label{fig:llm-risks}
\end{figure}

\section{Methodology}

We conducted a PRISMA-guided systematic review \cite{page_prisma_2021} to ensure transparency and reproducibility. The protocol covered (i) search and screening, (ii) eligibility, and (iii) data extraction and synthesis into an application–risk taxonomy.

\subsection{Search and Screening}
We searched five databases spanning computing and education: ACM DL, IEEE Xplore, Web of Science, ScienceDirect, and ERIC. Queries targeted three dimensions—(1) LLM/GenAI technologies, (2) educational contexts, and (3) risk constructs—adapted to each database’s syntax (see Table 1). The time window was 2015–2025 (search date: 20 May 2025). 

\begin{table}[t]
  \caption{Keyword dimensions used to build database-specific queries.}
  \label{tab:keywords-llm-risks}
  \footnotesize
  \begin{tabular}{@{}p{3.2cm}p{9.8cm}@{}}
    \toprule
    \textbf{Dimension} & \textbf{Examples} \\ \midrule
    LLM / GenAI & LLM(s), Large language model(s), AI, ChatGPT, GPT-4/3.5, Gemini, Claude, Copilot \\
    Education   & education, pedagogy, higher/secondary education, university, student \\
    Risks       & risk(s), bias, privacy, hallucinations, robustness, misinformation, trust, over-reliance, critical thinking \\ 
    \bottomrule
  \end{tabular}
\end{table}

\subsection{Eligibility}
\textbf{Included:} peer-reviewed journal or top-tier conference papers; empirical studies of LLM-based tools in education that report risks or outcomes from which risks can be inferred; English, full-text accessible.  
\textbf{Excluded:} perception-only studies without empirical risk evidence; model shoot-outs without educational risk analysis; secondary/theoretical work (SLRs, meta-analyses, opinion pieces).

\begin{figure}[t]
  \centering
  \includegraphics[width=\columnwidth,trim=6pt 0pt 7pt 6pt,clip]{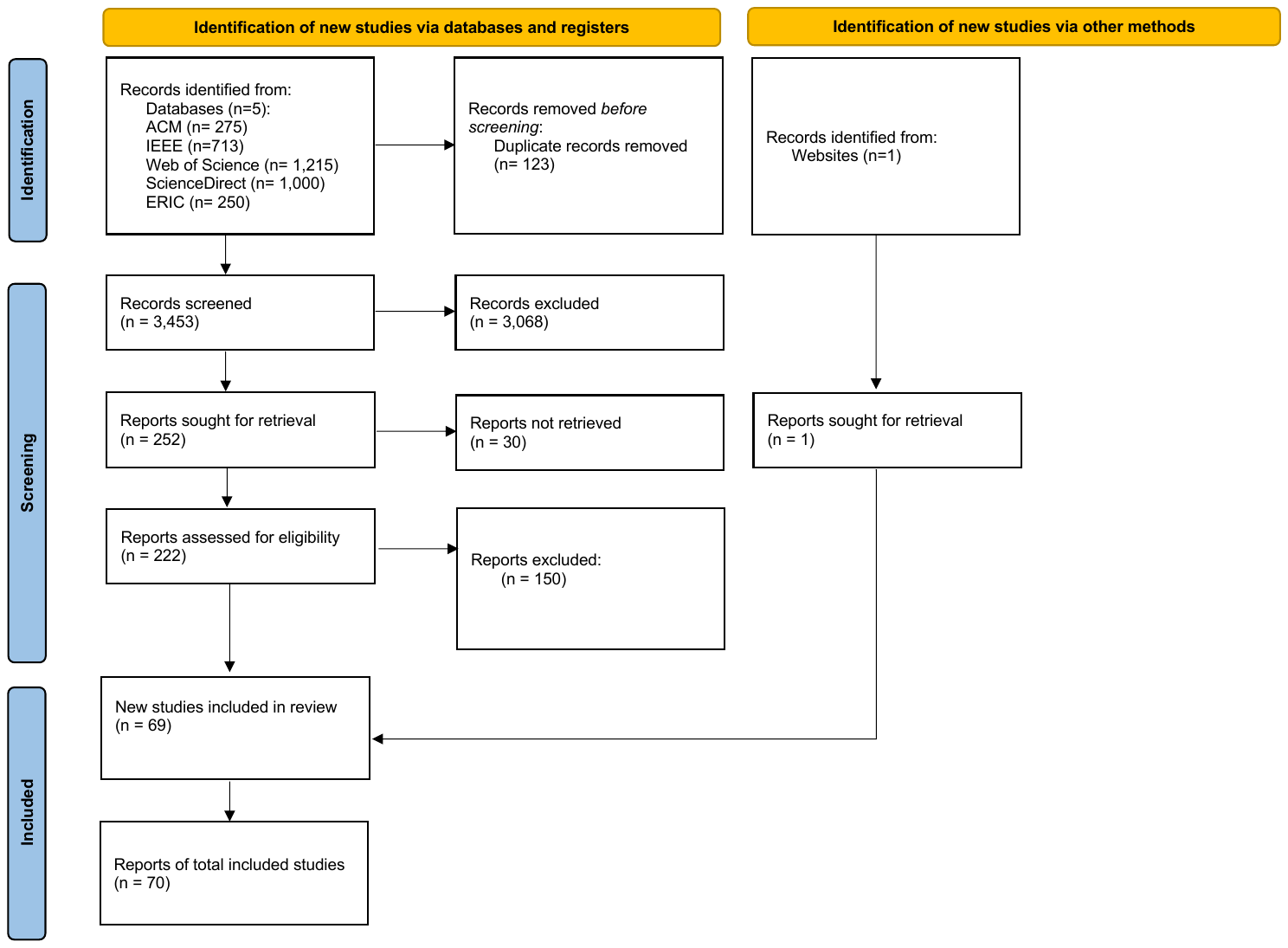}
  \Description{PRISMA flow diagram showing identification, screening, eligibility, and inclusion steps.}
  \caption{PRISMA flow of identification, screening, eligibility, and inclusion.}
  \label{fig:prisma}
\end{figure}

\subsection{Study Selection and Coding}
The search returned 3,453 records (See Figure 2). We used Rayyan \cite{rayaan} for deduplication and screening. Three reviewers independently screened titles/abstracts; conflicts were resolved by discussion to consensus. After screening and full-text review, 69 studies were included, all between 2023 and 2025. Three \cite{Idowu2024,rzepka_fairness_2022,song2025} out of the 69 papers examined AI systems  (algorithmic decision making or bias perception), not directly related to LLMs. One additional study \cite{Kosmyna_2025} widely cited in the community was included via forward references, for a total of \textbf{70} papers. For each study we extracted: authorship, method, LLM model, educational context, risk types, and outcome measures. 

\subsection{Descriptive Synthesis of Reviewed Studies}
We grouped studies into three application areas—\textit{operational effectiveness}, \textit{personalized applications}, and \textit{interactive learning tools}—and mapped reported risks to these areas to derive a structured taxonomy. Figure 3 shows the application distribution. Figure 4 breaks down subject areas by application. Operational effectiveness (a) is split between the humanities (43\%), with examples such as feedback generation in English and social science essays, and STEM (53\%), including engineering and medical contexts. A small “Other” category (4\%) captures studies examining whether LLMs can mimic human writing—for example, simulating personal experiences—and whether such outputs can be distinguished from human-authored text. Personalized applications (b) are dominated by “Other” (50\%), focusing on how LLMs respond to student demographics. The remaining studies fall into the humanities (20\%), such as content generation for English or law courses, and STEM (30\%), including medical imagery for student training. Interactive learning tools (c) are distributed across STEM (35\%), with studies centred on engineering and programming simulations, the humanities (42\%), including religious education and writing performance, and “Other” (23\%), reflecting research on student revision behaviour spanning multiple disciplines.

\begin{figure}[t]
  \centering
  \includegraphics[
    width=0.60\textwidth,
    height=0.28\textheight,
    keepaspectratio,
    trim=1.2cm 1.1cm 1.2cm 1.1cm,clip 
  ]{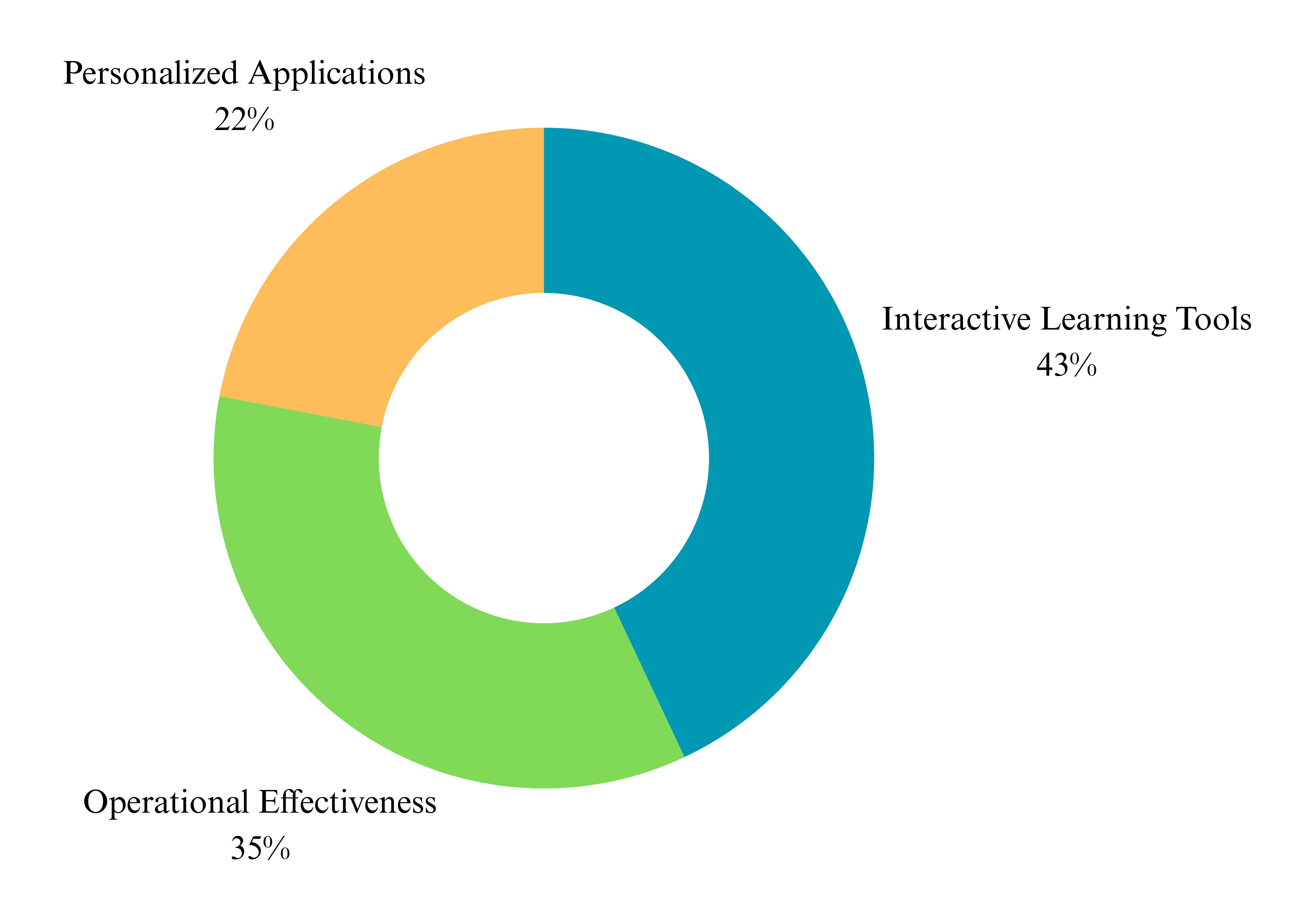}
  \caption{Distribution of reviewed papers across LLM application categories.}
  \Description{Pie chart showing the distribution of reviewed papers across LLM application categories.}
  \label{fig:dist-apps}
  \vspace{-6pt} 
\end{figure}

\begin{figure}[t]
  \centering
  \includegraphics[
    width=0.95\linewidth,
  ]{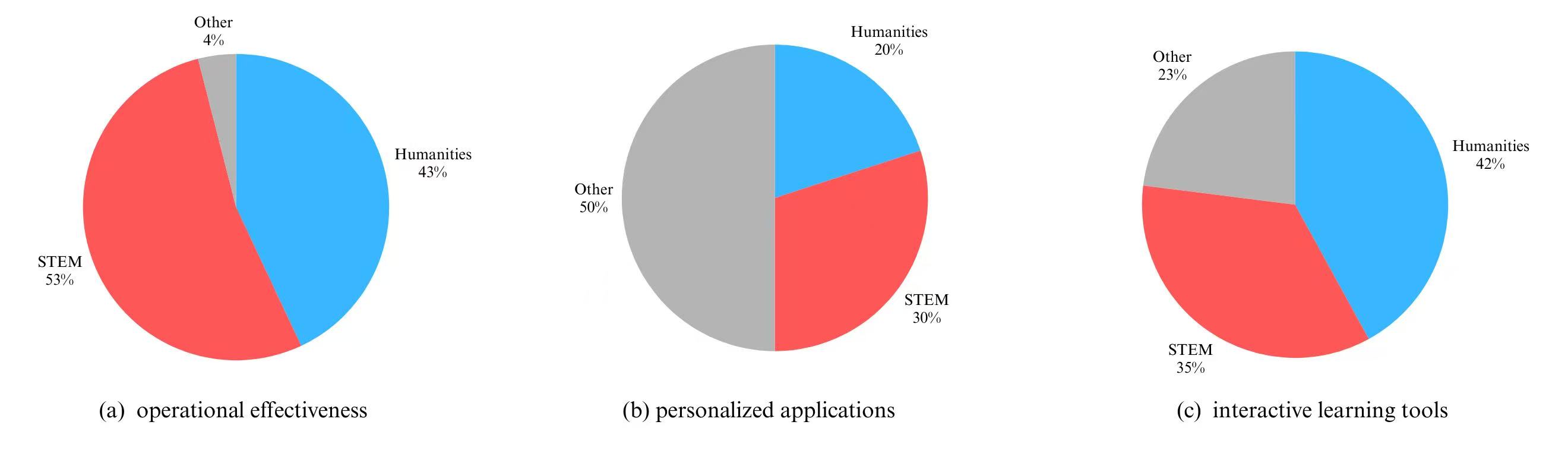}
  \caption{Subject-area distributions by application: (a) operational effectiveness, (b) personalized applications, (c) interactive learning tools.}
  \Description{Pie chart showing subject-area distributions by application: operational effectiveness, personalized applications, and interactive learning tools.}
  \label{fig:subject-breakdown}
  \vspace{-8pt} 
\end{figure}
\par For each application area, we note the number of times each LLM system was used. Most papers employ GPT models, mainly GPT-3.5, and to a lesser degree GPT-4. Other systems—Llama 2, Gemini, Bard, PaLM, BERT—were uncommon. 
Several studies did not report the exact GPT version (shown as “GPT (unspecified)"), or used specialized tools (e.g., WisdomBot, GitHub Copilot).

\begin{itemize}
  \item \textit{Operational effectiveness — LLM technology:}\\[2pt]
  \badge{GPT-3.5}{11}\;
  \badge{GPT-4}{9}\;
  \badge{LLaMA2}{1}\;
  \badge{Gemini}{1}\;
  \badge{BingAI}{1}\;
  \badge{Bard}{2}\;
  \badge{GPT (unspecified)}{9}

  \item \textit{Personalized applications — LLM technology:}\\[2pt]
  \badge{GPT-3.5}{24}\;
  \badge{GPT-4}{2}\;
  \badge{Bard}{1}\;
  \badge{Claude}{1}\;
  \badge{PaLM}{1}\;
  \badge{BERT}{1}\;
  \badge{GPT (unspecified)}{4}

  \item \textit{Interactive learning tools — LLM technology:}\\[2pt]
  \badge{GPT-3.5}{4}\;
  \badge{GPT-4}{2}\;
  \badge{WisdomBot}{1}\;
  \badge{LLaMA}{1}\;
  \badge{GitHub Copilot}{1}\;
  \badge{GPT (unspecified)}{12}
\end{itemize}

\begin{figure}
  \centering
  \includegraphics[width=0.8\linewidth]{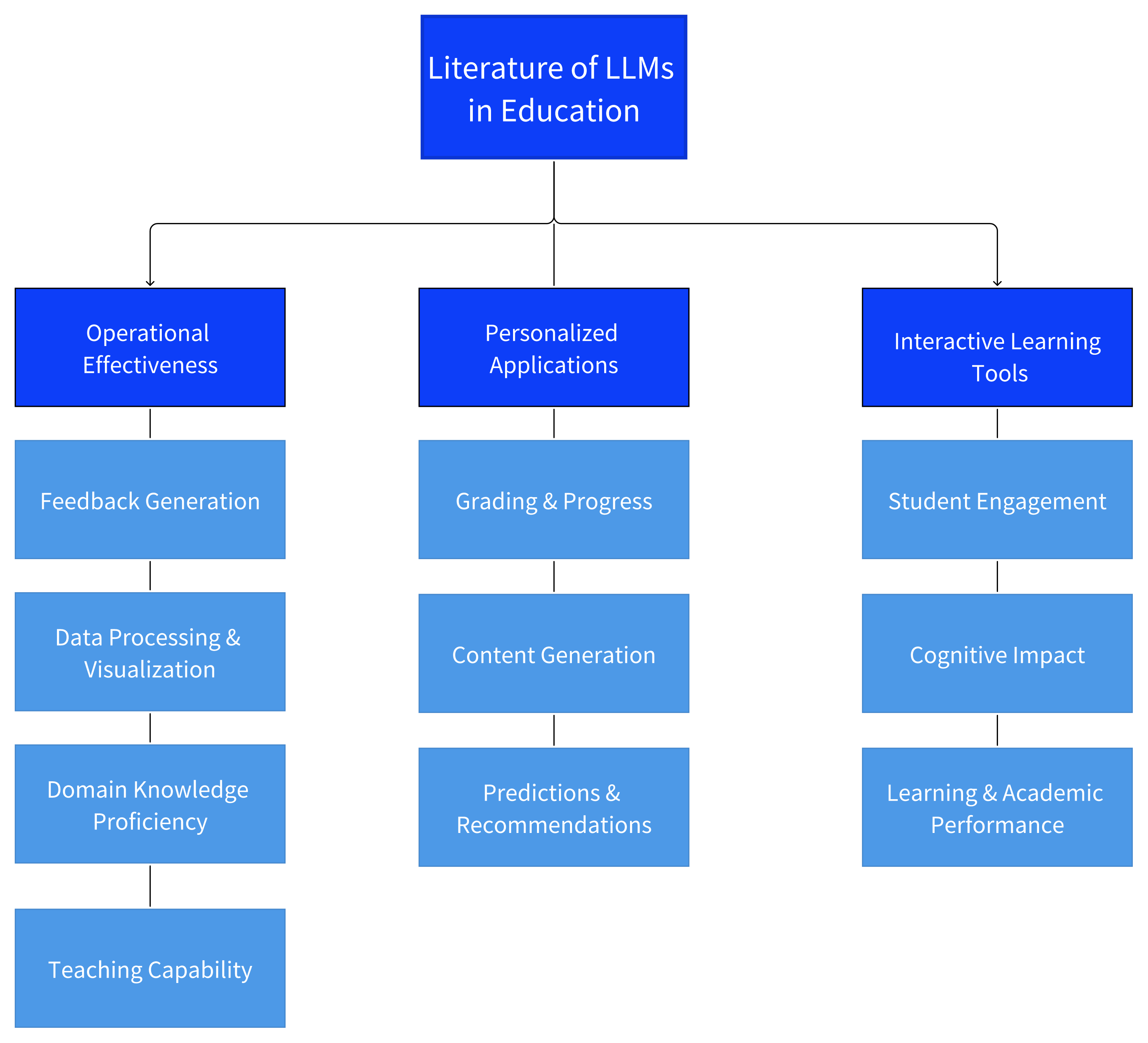} 
  \caption{Descriptive Synthesis of Reviewed Studies}
  \Description{A diagram showing the descriptive synthesis of reviewed studies.}
  \label{}
\end{figure}

\par In the remainder of the paper, we present a descriptive synthesis of the reviewed studies, organized into the three applications areas/sections and their relevant subsections, as shown in Figure 5. Each section highlights the methodological approaches adopted in the studies, the risks associated with LLM use in educational contexts, and the mitigation strategies reported in the literature.

\section{\textbf{Operational Effectiveness}}
Operational effectiveness refers to the extent to which LLMs can successfully perform core educational applications, which are used by both educators and students. Common applications for educators include feedback generation \cite{awidi2024,lin2024,trikoili2025,kooli} and data management tasks \cite{jiang2024, kwak2024}. Students, on the other hand, often use LLM tools to address subject-specific questions, making it crucial to evaluate the model's capacity to generate accurate and domain-relevant responses \cite{danesh2024,davies2024,borges2024,huang23,cooper2023,watts3}. In addition to factual correctness, several studies assess the teaching capabilities of LLMs, paying particular attention to how the content is delivered, including aspects such as tone, clarity and overall alignment with pedagogical principles \cite{Gong, Poucke}. 

\subsection{Feedback Generation} Feedback is a cornerstone of effective learning, providing learners with targeted, constructive insights into their current understanding and guiding them towards improvement. It is essential to differentiate between feedback and grading, with feedback aimed at supporting learning through guidance, prompting reflection, and fostering development, whereas grading typically serves as a summative evaluation of performance, assigning scores or marks to certify achievement. Generating high-quality written feedback is both cognitively demanding and time-intensive, especially for educators managing large cohorts. As such, there is increasing interest in leveraging LLMs to support the feedback process, making empirical investigations into the reliability and quality of LLM-generated feedback especially pertinent. 
\par The studies investigating feedback generation span a variety of domains, including engineering \cite{awidi2024}, critical thinking assessments \cite{trikoili2025}, learner essays \cite{lin2024}, social science essays \cite{kooli}, and STEM subjects such as computing, mathematics, and science \cite{Li}. All these papers employ comparative evaluation methodologies, systematically assessing the quality of LLM-generated feedback in relation to that offered by human instructors. In each case, LLMs and human educators are provided with the same grading rubric and student assignment, and the feedback is evaluated, primarily through quantitative methods. Importantly, all LLM evaluations are conducted in isolated sessions without prior context or memory, ensuring that feedback is produced independently for each instance. This approach allows for a fair and controlled comparison between the LLM and human-generated responses.
\par To analyse the consistency and quality of the feedback, researchers employ a range of statistical techniques. GPT and human assessors are compared using Pearson correlation coefficient, Mann–Whitney U tests, t-tests, Wilcoxon signed-rank tests, and error/sensitivity analysis. Such tools enable detailed evaluation of the alignment between LLMs and human markers in terms of scoring accuracy, reliability, and depth of feedback across diverse educational contexts and tasks \cite{awidi2024, trikoili2025, banihashem2024, lin2024}. Further granularity is provided in \cite{lin2024}, where feedback from humans and LLMs is categorized and coded by form (direct, indirect, reformulation, metalinguistic), scope (focused/unfocused), and level (local/global), along with measures of accuracy and redundancy. This mixed-methods approach allows for a detailed comparison between corrective feedback provided by GPT and that written by human instructors. Building on these foundational comparisons, Trikoli et al. \cite{trikoili2025} go beyond simple score alignment, and also measure the justification of LLM-generated feedback through a deductive thematic approach, analysing the reasons that were given behind each assigned score. In doing so, they examine whether ChatGPT generates meaningful and justified explanations, and the extent to which the reasoning matches that of the human grader. 

\par The general conclusion of the above studies is that teachers’ written corrective feedback typically includes logical rigor with a balance of both direct descriptive corrections and indirect cues \cite{trikoili2025}, which are essential for encouraging learners to engage in self-reflection and develop independent problem-solving skills. This feedback is also characterized by contextual sensitivity and pedagogical intentionality, reflecting teachers’ expertise and their broader aim of fostering deep learning rather than surface-level correction. In contrast, ChatGPT’s feedback often lacks these qualities, rarely employs indirect strategies, and tends to produce repetitive corrections. The model also lacks robustness, exhibiting system fatigue and inconsistencies, including occasional overcorrections or unnecessary changes to student writing, which may impede rather than support learning outcomes \cite{lin2024}. Additionally, ChatGPT demonstrates a superficial understanding of student work, struggling to interpret subjective or creative responses \cite{Li}. Koowli \cite{kooli} reports a moderate positive correlation between human grading and ChatGPT scores (Pearson’s r = 0.46), suggesting partial alignment but not a strong agreement. This correlation suggests that although there is some overlap between human assessors and ChatGPT in interpreting and scoring responses, LLMs do not consistently replicate human evaluative judgments. Notably, ChatGPT tends to be more conservative in its evaluations and shows greater variability in scoring compared to human assessors \cite{trikoili2025}.

\subsection{Data Processing \& Visualization}
Power BI is a tool widely used in universities to visualize and analyse student data. There is growing interest in integrating it with LLMs such as ChatGPT to simplify data exploration, reduce the need for technical expertise, and enhance data-driven decision-making in academic settings. Jiang et al. \cite{jiang2024} uploaded higher education Diversity, Equity, and Inclusion (DEI) data into Power BI, compiled from multiple Excel sheets (e.g. demographics, majors, enrolment). ChatGPT was then instructed to merge and analyse these datasets, generate visualizations, and produce summary reports. The study examined the feasibility of this integration, evaluated its performance, considered data security implications, and identified technical limitations. Although synthetic data was used to mitigate risk, a real-world implementation would involve sensitive student information, including IDs and enrolment records, raising serious concerns about data privacy. Several challenges were noted including the outdated training data of ChatGPT, limitations in API responsiveness and inconsistencies in the quality of AI-generated outputs. At present, platforms like ChatGPT do not meet the data protection standards required for secure API-based interactions, rendering them unsuitable for managing confidential educational datasets. Further research examined ChatGPT and LLaMA2s ability to  perform curriculum tagging tasks across different educational contexts \cite{kwak2024}. The study evaluated whether LLMs can classify student-generated text responses by tagging them with the correct learning objectives or curricular standards, a task that is typically complex due to variation in standards, language, and context. The findings reveal that GPT-3.5 displays a high level of alignment with the US Common Core State Standards (CCSS), limited alignment with the curriculum in England, and no measurable alignment with educational standards from Ireland, South Korea, or Maharashtra. These results reinforce earlier research, highlighting the US-centric nature of LLMs. By systematically comparing GPT-3.5’s performance across multiple international curricula, this study underscores the need for greater geographical and cultural inclusivity in the training data and evaluation benchmarks of LLMs.

\par LLMs were also tested on their capacity to evaluate engineering drawings and distinguish between correct and incorrect designs \cite{Abdul}. The findings show that while LLMs can recognize basic symbols, they struggle with more complex visual interpretation tasks. Specifically, ChatGPT-4 displays difficulty in accurately reading detailed features and attempts to generate 3D models based on schematics, leading to frequent errors or irrelevant outputs. These challenges highlight the technical barriers of using LLMs for accurate visual-data processing in specialized academic domains, such as engineering.

\subsection{Domain Knowledge Proficiency} The knowledge capabilities of LLMs have been widely explored, particularly in relation to their performance within domain-specific academic contexts. Several studies evaluate their ability to answer standardized examination questions, especially in medical and health-related disciplines such as periodontology \cite{danesh2024}, public health, and nursing. These investigations typically involve having ChatGPT take professional examinations—such as the DFPH exam \cite{davies2024}, a periodontics in-service assessment \cite{danesh2024}, and a nursing license test \cite{huang23}—to benchmark its subject-matter proficiency. Additional studies have explored ChatGPT’s ability to solve complex STEM and science education problems \cite{borges2024, cooper2023}, conduct research \cite{Mormul2024}, as well as its reliability in evaluating bilingual translation quality in real-world contexts \cite{qin2024}. Watts \cite{watts3} investigates the model’s ability to demonstrate mechanistic reasoning in organic chemistry assignments, comparing its responses to those of students. The key findings of the above papers highlight how ChatGPT can successfully complete a wide range of university-level tasks—including exams, term papers, research, and programming assignments, with performance comparable to or exceeding average student results.

\par While LLMs successfully pass the evaluated exams, several limitations are consistently observed. The model frequently hallucinates information, particularly when responding to open-ended or context-specific questions \cite{davies2024} and it encounters language-related challenges \cite{huang23}. A key technical constraint is the recency of ChatGPT’s training data, which can yield outdated or inaccurate responses, including fabricated events, hyperlinks, and academic citations \cite{malinka2023}. Notably, multiple studies emphasize that ChatGPT’s outputs, whether accurate or flawed, are often indistinguishable from human generated answers \cite{malinka2023,davies2024,borges2024}, highlighting its anthropomorphic characteristics, raising significant concerns for academic integrity and assessment reliability. This issue is further compounded by the broader shortcomings in the design and effectiveness of AI-detection tools, as discussed next. 

\par The following studies examine whether LLMs can demonstrate domain knowledge proficiency by producing texts that match the linguistic and stylistic qualities of human-authored academic writing. Durak et al. \cite{durak2025} conducted a linguistic analysis of educational texts produced by humans and various LLMs, including ChatGPT-3.5, Gemini AI, and BingAI. The study implements a systematic preprocessing pipeline, which involves cleaning the text data and vectorizing them using term frequency–inverse document frequency (TF-IDF) on the top 1,000 terms. To classify the origin of each text, the authors apply a range of machine learning algorithms, including ensemble methods such as Random Forest, Gradient Boosting, AdaBoost, Bagging, and Extra Trees, as well as a transformer-based model (BERTurk). The results indicate distinct stylistic patterns, with ChatGPT-generated content tending to exhibit more repetitive vocabulary and longer but less syntactically varied sentences. In contrast, human-written texts are marked by greater lexical diversity and more complex sentence structures. Among the LLMs evaluated, ChatGPT produces text most closely resembling human writing, making it more difficult to distinguish from actual human-authored content. Building on this question of indistinguishability at the stylistic level, further studies turned to a deeper challenge: whether LLMs can convincingly replicate personal experiences and reflections \cite{kaliterna_2024}. Using 47 authentic student essays on challenges and ethical dilemmas to derive prompts, the researchers generated a matched set of essays with ChatGPT and Bard, then analysed both corpora using LIWC-22 (Linguistic Inquiry and Word Count), a tool that quantifies linguistic, cognitive, and psychological features. According to the results, LLM-generated essays contained more affective and analytical language, yet were nonetheless persuasive in mimicking authentic student reflections. 

\par Taken together, these findings raise critical questions about whether educators and evaluators themselves can reliably discern between student and AI-generated work. Fleckenstein et al. \cite{fleckenstein2024} tasked in-service and pre-service teachers with distinguishing between texts produced by ChatGPT and by students. The findings indicate that pre-service teachers consistently struggle to identify the source of the texts, regardless of their quality. In-service teachers perform somewhat better with high-quality texts, but frequently misattribute low-quality GPT-generated texts as student-written. This pattern suggests a limited awareness among experienced educators that LLMs can produce low-proficiency writing which mimics typical student errors. Both groups were overconfident in their assessments, particularly when they believe texts were written by students, indicating a persistent uncertainty about authorship. Although GPT-generated texts are not systematically rated more favourably, experienced teachers were more likely to assign higher scores to high-quality LLM outputs. 

\subsection{Teaching Capability} Effective teaching extends beyond content delivery, encompassing the ability to foster conceptual understanding, encourage critical thinking, adapt instruction to diverse learners, and build trust through empathy and ethical responsibility. Concerns have been raised regarding the pedagogical competence of LLMs, not only in terms of their factual knowledge, but also with respect to whether they embody the instructional values, pedagogical skills, and dispositions typically associated with effective teaching. Gong \cite{Gong} carried out a quantitative evaluation of six OpenAI language models using a custom framework (TeacherComp), based on the teacher competency model, encompassing four key dimensions (knowledge, skills, values, and traits) and 14 established psychological and educational scales. The responses of the models on prompts with question-and-answer format were compared against normative human data to identify both similarities and differences in pedagogical competence. The LLMs examined display heightened levels of deceptive behaviours and traits associated with the dark triad (narcissism, psychopathy, and Machiavellianism), which are misaligned with core pedagogical values. Poucke et al. \cite{Poucke} further present a comparative study examining the evaluative language in ChatGPT and human-generated responses to questions about Australian cultural practices. Their findings indicate that ChatGPT relies on impersonal 'it is' constructions, reflecting pre-programmed Anglocentric ideological biases embedded in its training data. The model's authoritative tone and high perceived credibility raise concern that young or vulnerable learners may accept its statements without question. In contrast, human responses are more likely to express personal perspectives using subjective language. The presence of such traits in LLMs is particularly concerning in educational contexts, where values such as openness, integrity and emotional stability are paramount.

\subsection{Risk Summary \& Mitigation Strategies}

\begin{table}[t]
  \centering
  \footnotesize
  \setlength{\tabcolsep}{6pt}\renewcommand{\arraystretch}{0.95}
  \caption{Risk frequency in operational effectiveness of LLMs.}
  \begin{tabular}{>{\raggedright\arraybackslash}p{0.58\columnwidth} r l}
    \toprule
    Risk & Count & \multicolumn{1}{c}{} \\
    \midrule
    Superficial Understanding       & 8 & \sparkbar{8}{8} \\
    Limited Robustness              & 5 & \sparkbar{5}{8} \\
    Anthropomorphic Characteristics & 5 & \sparkbar{5}{8} \\
    Hallucinations                  & 4 & \sparkbar{4}{8} \\
    Bias                            & 2 & \sparkbar{2}{8} \\
    Privacy                         & 1 & \sparkbar{1}{8} \\
    Knowledge Limitations           & 1 & \sparkbar{1}{8} \\
    \bottomrule
  \end{tabular}
\end{table}

Table 2 illustrates the most frequently identified risks, based on the number of articles in which they are mentioned. The leading concern is superficial understanding, referring to LLMs' limited capacity to interpret subjective or creative student responses. This limitation highlights the lack of nuanced comprehension, particularly in tasks requiring deeper semantic reasoning or pedagogical insight. Limited robustness encompasses several issues, including inaccuracies, inconsistencies, system fatigue, and unpredictable behaviour, all of which can undermine the reliability of LLM-based educational tools. The risk of the anthropomorphic characteristics is also frequently noted. When paired with the issue of hallucinations, this can negatively affect students' learning experiences, as they may over-trust LLM responses due to their human-like fluency. In contrast, privacy and bias are among the least frequently cited risks related to operational effectiveness. Privacy concerns in this context primarily relate to the handling of sensitive student data, particularly when LLMs are integrated into classroom settings. Lastly, bias arises when LLMs exhibit Anglocentric tendencies, potentially marginalizing diverse perspectives and reinforcing cultural or linguistic inequalities in learning content.

To address the diverse risks associated with the use of LLMs in education, the literature recommends a broad set of mitigation strategies involving multiple stakeholders. While many of these strategies are discussed in an interconnected manner, this sections presents them by stakeholder group- Educators \& Developers- to clarify distinct roles and responsibilities. Mitigation strategies specific to students are not identified, as this subset of reviewed studies primarily focus on educators and developers, rather than on learners themselves. The goal is to highlight how each group can contribute to minimizing risk within their scope of influence.

\subsubsection{Educational Personnel}

Including teachers, academic staff, and administrators, play a central role in shaping the pedagogical context in which LLMs are used. Key strategies include: 
\begin{itemize}
    \item Human Monitoring: LLM tools should not be solely relied upon for grading or evaluation. Instead, they should be used as supplementary aids, utilizing professional judgment to maintain academic standards \cite{Lim,lin2024}.
    \item Teacher Training: Increase educators’ technical knowledge and awareness of ethical issues related to LLM use in educational contexts \cite{Gong}.
    \item Data Privacy: Educators should avoid inputting personal or sensitive student information into systems that lack rigorous safeguard procedures.
    \item Student Guidance: Instructors can guide students on how to use LLMs responsibly, including the use of directive prompts (e.g. "please summarize"), rather than outsourcing entire tasks \cite{Poucke}.
    \item Transparency: Clear guidelines about the permissible use of LLM tools in coursework and research should be established and communicated to students.

\end{itemize}

\subsubsection{Developers}
Developers are responsible for ensuring that the underlying systems are trustworthy, equitable and adaptable to educational contexts. Recommended strategies include:
\begin{itemize}
    \item Robust Training and Evaluation: Models should be trained on diverse, high-quality datasets, with continuous calibration and audits to ensure fair performance across various populations \cite{trikoili2025}.
    \item Improved Detection Tools: Given the challenges in distinguishing LLM-generated content from human writing, developers should continue to fine-tune detection systems to increase accuracy (i.e. reduce false positives and negatives), and enhance academic integrity \cite{borges2024,malinka2023}.
\end{itemize}

\section{Personalized Applications of LLMs}

We examine whether LLM behaviour shifts based on characteristics such as gender, ethnicity, or linguistic background \cite{weissburgg,hu2020,Warr}, raising critical concerns about fairness and equity in learning environments. This effect manifests in grading inconsistencies \cite{Warr}, storytelling biases \cite{price2024}, stereotyped image generation \cite{kaufenberg, currie2024}, and culturally insensitive dialogues \cite{deroock2024}.
While discussions of algorithmic bias have gained prominence in academic discourse, it is vital to identify where and how such disparities arise. This is particularly important given that LLMs may encode and reproduce raciolinguistic ideologies, where race and language are not merely independent features, but co-constructed systems of discrimination \cite{Flores2015}. Without rigorous empirical analysis, these hidden biases may go unchallenged, thereby reinforcing structural inequalities and disproportionately disadvantaging marginalized learners.

\subsection{Grading \& Progress} Warr \cite{Warr} investigated whether LLMs exhibit implicit bias by analysing how ChatGPT assigns scores to identical student writing samples when paired with different demographic profiles. The study employs prompts describing hypothetical 8th-grade students who differ by race, socioeconomic status, and school type (e.g. Black vs. White, lower-class vs. upper-class, public vs. elite private school), while keeping the same writing sample. Grades assigned by ChatGPT were analysed using non-parametric statistical tests (Wilcoxon, Mann-Whitney U, Kruskal-Wallis, and Dunn’s) enabling the discovery of potential biases in the model’s scoring behaviour without requiring access to its internal mechanisms. The findings reveal that ChatGPT 3.5 exhibits subtle forms of  bias in its evaluation of student writing, modifying scores based on demographic cues related to race, socioeconomic status, and educational background. Indicators such as references to low-income backgrounds or under-performing public schools are associated with lower scoring. The model’s scoring behaviour is also influenced by the sequence of prompts within a conversation, suggesting that it contextualizes and compares demographic information as it generates responses. Even with developer-imposed safeguards aimed at curbing explicit discrimination, these biases persist, mirroring structural inequalities deeply rooted in the training data. 

\par Complementing this focus on grading bias, Idowu et al. \cite{Idowu2024} examined fairness in AI algorithms used to monitor student progress over time.  Their study analyses three key data sources: institutional data (e.g. student demographics), virtual learning environment data (e.g. students' online engagement patterns, such as clicks and logins), and assessment data (e.g. scores from quizzes \& exams). The findings indicate that institutional data is the primary contributor to algorithmic bias, particularly concerning disability, age and gender. These results underscore the importance of assessing not only how AI systems grade, but also how they monitor student trajectories, since both forms of algorithmic evaluation can reproduce existing inequalities.

\subsection{Content Generation} Content generation capabilities of LLMs span a range of applications, including the creation of narrative texts \cite{deroock2024}, classroom instructional materials \cite{Hamdam2024}, and visual content such as images \cite{currie2024}. While the generated content is  largely shaped by the prompts provided by users, concerns have emerged regarding the consistency, appropriateness, and potential biases embedded in these outputs. Price et al. \cite{price2024} explored how ChatGPT’s feedback mechanisms may reinforce prevailing educational narratives, specifically focusing on how the model reinforces assumptions about which students are perceived as mathematically competent. In this exploratory analysis, the researchers prompt ChatGPT to respond to math-related writing samples attributed to students with diverse academic and demographic profiles—including racial/ethnic identities (e.g. Black, Asian, White) and educational classifications (e.g. gifted, special education, English language learners). The results suggest that GPT outputs vary subtly according to these labels, often reflecting societal biases—for instance, portraying Black students as exceptional compared to White and Asian, which were classified as neutral. Expanding on the scope of this issue, Weissburg \cite{weissburgg} examined how LLMs tailor educational content based on students’ demographic characteristics, including race, ethnicity, income level, and disability status. The study assessed whether the nature of the material changes in response to these variables, and how. Statistical analysis was employed to identify systematic differences in content delivery across demographic groups, highlighting potential disparities in how LLM-generated instruction is personalized. The analysis reveals that even state-of-the-art LLMs exhibit measurable biases across multiple characteristics, with income and disability status showing the highest levels of disparity and gender and race the lowest. Similarly, studies testing ChatGPT’s generation of images of a chemist \cite{kaufenberg} and a medical student \cite{currie2024} reveal implicit biases embedded within these systems. The visual outputs reflect stereotypical representations, with both roles predominantly portrayed as male.

\par Further research has examined ChatGPT's  ability to create morally oriented narratives for Chinese students in English as a Foreign Language (EFL) classes, aligning its output with the Confucian ethical framework \cite{Danson2024}. Prompt development was guided by the CLEAR framework \cite{lo2023}, which emphasizes Conciseness, Logic, Explicitness, Adaptability, and Reflectiveness. To enhance cultural and contextual relevance, the researchers introduced a two-step “Navigation and Generation” strategy: first, the model was prompted to establish a coherent understanding of Confucian moral constructs (“navigation”), followed by the generation of instructional narratives grounded in those principles ("generation"). Prompts were issued in both English and Mandarin, and the outputs were iteratively refined based on sociocultural congruence. Martin and White’s appraisal framework \cite{hashemi2023} was employed as a qualitative linguistic tool to examine the evaluative stance and cultural framing within the LLM-generated texts. This interdisciplinary methodology—integrating prompt engineering, sociolinguistic analysis, and cultural sensitivity—enables both the creation and critical evaluation of LLM-generated pedagogical content tailored to specific moral and educational contexts. The analysis reveals that ChatGPT frequently incorporates WEIRD (Western, Educated, Industrialized, Rich, Democratic) cultural assumptions into its outputs, leading to sociocultural misalignments that may distort Confucian moral constructs. This is concerning, as students benefit from stories and examples that reflect their own culture, which support more meaningful engagement with moral issues. Moreover, the narratives often display problematic authority representations, gender bias, and culturally incongruent naming practices.

\par De Rook et al. \cite{deroock2024} employed a hybrid approach, referred to as a "blended, cyborg methodology", which combines cognitive ethnography with digital ethnography to explore the interactional landscape surrounding the use of ChatGPT. The focus is on the relationship  between the user and the LLM tool, with particular attention to the socio-linguistic norms shaping these interactions and their influence on the development of writing tasks. The study highlights two major concerns: the frequent hallucination of sources, where ChatGPT fabricates citations, and the presence of linguistic bias, where the model’s default language style is overly pompous and aligns with dominant white linguistic norms. Further, through modifying prompts, ChatGPT can be manipulated into generating malicious macro code, despite its inbuilt ethical safeguards. This vulnerability stems from the model's difficulty to maintain contextual consistency across sequential prompts, allowing users to circumvent restrictions simply by rephrasing their requests.

\subsection{Predictions and Recommendations} Recently, dropout prediction models have gained prominence as a response to the high attrition rates observed in Massive Open Online Courses (MOOCs), allowing educators to intervene proactively \cite{Xing}. In parallel, increasing attention has been given to the fairness of these predictive systems, prompting audits that examine potential biases across demographic groups. For instance, Rzepka \cite{rzepka_fairness_2022} analysed bias in three types of models—decision tree (DTE), k-nearest neighbours (KNN), and multilayer perceptron (MLP)—trained on clickstream data from over 52,000 learners. These models continuously update their predictions throughout a student's online session, recalculating dropout risk with each new user interaction. To evaluate fairness, the models use four standard metrics: predictive parity (PP), equal opportunity (EO), predictive equality (PE), and the area between ROC curves (ABROCA). These fairness checks apply across demographic variables such as gender, parental education level, first language, and home literacy environment. The results show that all three models exhibit notable bias across these measures, indicating that AI-driven risk prediction produces unequal outcomes for students from different backgrounds. In addition, the research demonstrates that common mitigation strategies, such as oversampling minority groups or building separate models not only fail to reduce bias, but may even exacerbate it for certain groups. This counter-intuitive effect occurs because oversampling can make the model rely too heavily on a small set of repeated examples, leading to exaggerated patterns for that group. Likewise, separate models can suffer from limited data for each group, producing uneven quality and reinforcing differences instead of removing them. 

\par Similar concerns emerged in Zheng's research \cite{zheng2023}, evaluating biases in GPT-3.5 Turbo’s major recommendations for users with varying demographic profiles. Consistent with Rzepka’s findings \cite{rzepka_fairness_2022}, the study reveals that marginalized groups—such as LGBTQ and Hispanic users—are less likely to receive recommendations for STEM fields compared to their heterosexual and White counterparts. Asian students are three times more likely to receive STEM recommendations than African-American students. Additionally, students from higher socio-economic backgrounds are 30\% more likely to be directed toward STEM domains than those from lower socio-economic backgrounds \cite{zheng2023}. Another pressing concern identified in the literature is the inability of models such as Bard, GPT-3.5, and PaLM to reliably detect bias. When these models are tasked with analysing texts from education courses, they successfully identified only about half of the biased content, questioning their effectiveness in supporting equitable and critical learning environments \cite{Albuquerque16032025}. 

\par Song \cite{song2025} investigated students' perceived fairness in AI-driven decision making systems, emphasizing the importance of considering learner's perspectives as primary stakeholders in educational environments. In this study, U.S college students interacted with a simulated AI system that predicted whether they would pass or fail a maths course. The system was experimentally manipulated to vary along three dimensions: algorithmic bias, transparency, and decision-making style (i.e. algorithm only vs human-AI collaboration). Although the prediction outcome (pass or fail) was randomly assigned, participants were led to believe it was AI-generated. A mixed-methods approach was employed, combining quantitative data (including maths knowledge tests and webcam-based eye-tracking) with qualitative interviews. The results reveal that students highly value fairness, especially in high-stakes academic settings. Transparency significantly influences perceived fairness; when students understand how they system generated its predictions, they are more likely to consider the outcomes as just. However, concerns were raised regarding the system's objectivity and explainability, especially when AI predictions conflict with students' self-perceptions. These findings highlight a critical risk: students may reject otherwise technically robust AI systems if they lack transparency. This underscores the need for explainable, human-centred AI in educational contexts to foster trust and meaningful engagement.

\subsection{Risk Summary \& Mitigation Strategies}
Table 3 demonstrates the cited risks in the personalized applications of LLMs. The most pronounced risk is the bias of the model, which ranges from the evaluation of student writing, content generation and predictions and recommendations, ultimately producing unequal outcomes for students from different backgrounds. Mitigation strategies involve educators—who are the main users of LLM personalization features—and developers, who are responsible for addressing bias through improvements in model design, training data, and fine-tuning.

\begin{table}[t]
  \centering
  \footnotesize
  \setlength{\tabcolsep}{6pt}
  \renewcommand{\arraystretch}{0.95}
  \caption{Risk frequency in personalized applications of LLMs.}
  \label{tab:pa-risks-spark}
  \begin{tabular}{@{}>{\raggedright\arraybackslash}p{0.52\columnwidth} r r c@{}}
    \toprule
    Risk & Count & \multicolumn{1}{c}{} \\
    \midrule
    Bias           & 8 & \sparkbar{8}{8} \\
    Hallucinations & 1 & \sparkbar{1}{8} \\
    \bottomrule
  \end{tabular}
\end{table}

\subsubsection{Educational Personnel}
\begin{itemize}
    \item Bias Awareness in Grading: Exercise caution when using LLMs for grading or personalization, as linguistic patterns may encode biases even in the absence of explicit demographic identifiers \cite{Warr, price2024, Danson2024, weissburgg}.
    \item Active Intervention: Engage in deliberate prompt engineering and critical evaluation of LLM-generated content to support ongoing Diversity, Equity, Inclusion, and Respect (DEIR) initiatives in educational settings \cite{kaufenberg}.
    \item Ethical Deployment: Review and verify LLM-generated content before incorporating it into instructional materials or classroom activities \cite{kaufenberg,Hamdam2024,currie2024}.
\end{itemize}

\subsubsection{Developers}
\begin{itemize}
\item Diagnose bias early \& continuously: Apply bias-reduction at data, training, and inference stages; verify downstream effects on educational tasks \cite{Warr,cheng,weissburgg,zheng2023,Idowu2024}.
\item Design for inclusion: Use participatory methods and diverse stakeholder input to avoid reproducing systemic inequalities \cite{cheng}. 
\item Make models legible: Provide explanations, uncertainty cues, and documentation; keep reviewable logs to support contestability and trust \cite{song2025}.
\item Human-in-the-loop governance: Educator oversight for high-stakes uses, clear escalation paths, and periodic re-audits with context-specific mitigations \cite{song2025,zheng2023}.
\end{itemize}

\section{Interactive Learning Tools}

Empirical research has increasingly explored the pedagogical dimensions of LLM integration in education, with studies examining how LLMs influence student engagement \cite{alimardani2024,Achiam2023,strozyna2024,Radtke2024,zhan2025,Radtke2024,Alimardani2025,Woerner,weidinger,AlGhamdi,Guner2025,Haindl2024}, cognitive development \cite{Kosmyna_2025,essien2024,Zhang2025} and learning and academic performance \cite{zhu_impact_2025,keith_harnessing_2025,Jost2024,shi2025,Maher2023,Korpimies_2024,Caccavale2024,Chun2025,liu2024,shi2025}. In the following, we examine how students interact with LLM systems and the resulting impact across the identified dimensions.

\subsection{Student Engagement}

\par Student engagement refers to the degree of attention, interest and active participation that students exhibit during the learning process. The real-word consequences of passively accepting information from LLM tools as fact, are increasingly evident. A notable example involves a lawyer who relied on LLMs to prepare for a court hearing, inadvertently submitting a list of fabricated citations \cite{taylor_melbourne_2024}. This begs the question: how can we effectively educate individuals about the risks associated with LLMs, and how can we ensure that outputs are critically evaluated before use? Addressing this concern, Alimardani et al. \cite{Alimardani2025} conducted a study examining how students engage with ChatGPT after completing an undergraduate elective course on 'Law and Emerging Technologies'. Students were introduced to topics in law, ethics, AI and neuroscience, with a distinct emphasis on the responsible use of AI tools in academic work. The end-of-semester assignment required students to write a policy brief advising political representatives on the legal and ethical implications of autonomous vehicles. Despite being made aware of LLMs’ limitations during the course, the results reveal a phenomenon termed "verification drift" - a tendency for students to reduce critical evaluation because of the anthropomorphic characteristics, authoritative tone, and seemingly accurate and detailed responses of LLMs \cite{alimardani2024,Achiam2023}. Moreover, the "anchoring effect" is exhibited, whereby individuals place undue weight on the first piece of information they encounter \cite{Cho2017}. As a result of this, students tend to adopt a more narrow way of thinking. Building on these insights, Stróżyna et al. \cite{strozyna2024} conducted an experiment in which students were instructed to generate and critically evaluate texts produced by LLMs. The primary aim was to identify key limitations of LLMs from students' perspectives. These findings reveal persistent concerns: LLMs frequently hallucinate, are vulnerable to manipulation and their outputs vary significantly across platforms (e.g. ChatGPT vs. Bard). Although the tendency of these models to fabricate plausible-sounding but inaccurate content is widely recognized, this study highlights how students must still actively engage in identifying and correcting such errors. 

\par Radtke \cite{Radtke2024} investigated how authorship labels influence revision behaviour by presenting learners with two texts; one labelled as peer-written and the other as AI-generated. The findings reveal a statistically significant difference in revision time: participants spend considerably less time revising texts identified as AI-generated compared to those labelled as peer-written. This suggests that learners may perceive AI-generated content as less worthy of critical engagement, potentially overlooking important issues such as factual inaccuracies or hallucinated information. Furthermore, the study explores whether trust in AI systems influences revision practices. Learners who express high trust in AI-based text generators, having frequent exposure to AI in the media, or hold more favourable attitudes toward AI spend significantly less time revising the AI-labelled texts. These findings raise important concerns about students’ critical engagement and epistemic responsibility when interacting with AI-generated content. The introduction of LLMs is also associated with a shift in classroom dynamics. Once students begin interacting with an LLM, they tend to deviate from the instructions given, preferring to operate autonomously, exhibiting difficulty in reorientating their attention back to the teacher \cite{zhu_impact_2025}. Similar patterns are observed in \cite{Guner2025}, where despite an initial critical engagement with LLMs, students gradually adopt a reliance on direct answers in later sessions. 

\par Qualitative analyses across studies highlight student concerns regarding the use of LLMs in education. Key issues include the use of outdated training data, privacy risks and fears that over-reliance on these tools may hinder the development of essential skills (e.g. programming) \cite{Guner2025,Haindl2024}. The integration of LLMs into English composition courses that emphasizes the development of academic writing skills has been the subject of recent investigation \cite{Woerner}. In this study, students were permitted to engage with LLMs freely, and their written submissions were subsequently analysed. Writing quality was assessed based on several criteria, including structure, coherence, argumentation, grammatical accuracy and critical thinking. The findings indicate that students who did not use ChatGPT demonstrate greater engagement, stronger critical thinking, and more advanced research skills. However, their writing exhibits lower grammatical accuracy and less conformity to academic style. In contrast, LLM-assisted texts show higher grammatical correctness and stylistic consistency, but are notably weaker in terms of engagement, analytical depth, and research quality. Furthermore, students who rely exclusively on LLM tools exhibit lower attendance rates and reduced interest in the course content.

\par Students' engagement with feedback generated from LLMs in writing tasks is examined \cite{zhan2025}. Undergraduate student's interactions, including the prompts and the subsequent revisions made to their texts, were recorded and transcribed for analysis. While students actively engage with the feedback at a cognitive level by extracting and incorporating relevant information, their metacognitive engagement is limited. That is, students demonstrate minimal reflection on how to plan, regulate, or evaluate the use feedback provided. This lack of metacognitive activity poses a risk to the development of higher-order thinking skills and may lead to what has been coined "metacognitive laziness"- a growing dependence on technology rather than on one's own cognitive abilities \cite{fan2025}. The impact of ChatGPT-generated feedback on students’ technical writing skills is examined in a qualitative blind study \cite{AlGhamdi}. While students generally valued the clarity and consistency of the GPT-generated feedback, many express concern about its generic tone, occasional contradictions, and lack of personalization. Although such responses tend to encourage surface-level revisions, they do not consistently prompt deeper reflection, further reinforcing concerns about metacognitive stagnation. These findings echo broader critiques of LLM feedback systems, suggesting that without human oversight, such tools risk fostering dependency rather than independent engagement.

\subsection{Cognitive Impact} 
Kosmyna \cite{Kosmyna_2025} examined how different types of tools influence neural activity during writing tasks. Fifty-four students completed SAT-style essays across four sessions while high-density EEG recorded information flow among 32 brain regions. Participants were divided into three groups: a brain-only group with no external assistance, a search engine group using conventional web search tools, and an LLM-assisted group. The findings reveal a systematic decrease in brain connectivity corresponding to the level of external support. The brain-only group exhibited the most extensive and robust neural activation, also displaying the strongest localized activity beyond the visual cortex, particularly in the left parietal, right temporal, and anterior frontal regions, which are associated with semantic integration, creative ideation, and executive self-monitoring. The search engine group showed intermediate engagement. In the LLM condition, however, diminished strength in the dorsolateral prefrontal cortex and other frontal and temporal brain regions suggests that learners bypass deeper memory encoding processes, relying instead on externally provided text without integrating it meaningfully into memory networks. Overall, the results support the view that tools not only alter task performance, but also reshape the underlying cognitive architecture. When students rely on LLMs as substitutes for their own reasoning or critical engagement, the learning process is compromised. 

\par The learning process entails the development of key cognitive skills, such as analysing, synthesizing and evaluating information, which emerge progressively through observation, experience, reflection and reasoning. This process is fundamental in fostering learners who can think critically and act independently and should not be bypassed. Optimal cognitive development occurs within the learner’s Zone of Proximal Development (ZPD), where the difficulty of a task is appropriately matched to the level of available support and guidance \cite{vygotsky1978}. However, when tools such as ChatGPT provide support that significantly exceeds the learner’s current needs, students may become under-stimulated and operate within a cognitive “comfort zone”, thereby impeding deeper learning and intellectual growth. This is closely tied to the principle of experience-based \textit{neuro-plasticity}, the brain’s capacity to reorganize itself by forming new neural pathways in response to experiences, which is at full force during the schooling years \cite{goldberg2022}. Neuroscientific research indicates that activating specific neural circuits through active learning enhances retention and understanding, particularly when learning is salient, intense, and repeated \cite{carey2012}. When students become passive recipients of LLM-generated output, they may miss the opportunity to strengthen those neural pathways essential for deep learning and remain stagnant in the comfort zone with limited learning and neuroplasticity. Beyond the structural implications for neural development, the reliance on LLM tools interferes with how the brain’s reward system responds to learning. Human cognition evolved under conditions that reward effort and delay gratification. In contrast, the immediate feedback and constant availability of LLMs provide quick rewards, which can stimulate dopamine release in the brain. This may reinforce habitual use of the tool, while reducing the satisfaction and motivation that usually come from solving problems through effort \cite{berridge2009,berridge2018}. This neuro-chemical imbalance risks reducing students’ intrinsic motivation, weakening their cognitive resilience, and undermining their sense of agency in the learning process. 

\par The presence of embedded biases in LLMs emerges as a consistent concern across all the studies reviewed in our paper. Zhang \cite{Zhang2025} specifically investigated how these biases influence users’ cognitive processing, particularly in shaping their understanding of religious doctrines and perceptions of cultural diversity. The study begins with a baseline assessment of participants’ existing beliefs and attitudes toward various religions. Participants were then divided into two groups: an LLM-interference group (including positive, negative, and neutral viewpoints on different religions) and a non-LLM-interference group (conventional, non-LLM generated religious descriptions). Following this intervention, all participants completed a post-test identical to the baseline survey, measuring any changes in perceptions. Initial survey responses reveal relatively neutral attitudes toward all religions. However, after exposure to LLM-generated content, participants in the experimental group show statistically significant shifts. Evaluations of Christianity and Judaism became more favourable, while perceptions of Islam declined. Multivariate regression analysis, which accounts for possible confounding variables, revealed that the influence of LLM-generated content on religious cognition is independent of other variables. The findings underscore the potential of LLMs to subtly shape users’ religious beliefs, particularly when such content is consumed without critical scrutiny. This has important implications for religious education, where continued exposure to biased AI-generated narratives could skew students’ understanding of world religions. 

\subsection{Learning \& Academic Performance}

Academic performance relates to how well students have understood the material being taught. Several studies adopt quasi-experimental or mixed-methods to simulate authentic learning scenarios, often comparing AI-assisted cohorts with control groups, and analyse results using non-parametric statistical tests \cite{Chun2025,keith_harnessing_2025,Jost2024,Caccavale2024}. Zhu \cite{zhu_impact_2025} evaluated the impact of LLMs on secondary school students across multiple dimensions, including assignment performance, learning outcomes and overall experience. In an eleven-week classroom intervention, an LLM was implemented and used to facilitate a series of academic tasks. Students in the LLM-assisted group achieved higher submission rates, produced longer written outputs, and earned higher scores on assignment performance metrics. However, they under-performed on creativity and the administered post-knowledge tests. These results align with prior findings in undergraduate programming courses \cite{Jost2024}, which highlight the trade-off between LLM-facilitated efficiency and the development of problem-solving capabilities. While LLMs are increasingly deployed into programming workflows, upcoming developers must cultivate the foundational software engineering competencies necessary, such as identifying errors and design principles, which cannot be solely delegated to LLMs \cite{Pudari2023}. \cite{Korpimies_2024} extended this line of inquiry by examining students’ experiences with LLMs in a software project course, where learners were allowed to use the tools freely except for unit test generation. Surveys conducted assessed the extent of LLM usage, methods of application, and perceived impact on learning. Results reveal diverse usage patterns: many students reported efficiency and problem-solving benefits, though concerns were raised about over-reliance and poor-quality outputs. Moreover, students who predominantly relied on LLMs for code generation or documentation performed below the overall course median.

\par On the other hand, targeted use of LLMs has been associated with improved learning outcomes in specific contexts. Liu et al. \cite{liu_application_2024} found that students benefit from LLM-assisted instruction when learning MATLAB programming. By breaking down complex programming tasks into more manageable sub-problems, LLMs can provide step-by-step guidance. Nevertheless, these benefits are counterbalanced by notable limitations: when tasked with highly specific or advanced problems, particularly those involving modelling and simulation, LLMs often struggle. Among students using LLMs for programming, 30.99\% reported encountering incorrect outputs, while 80.28\% stated that the responses failed to meet their expectations. Maher et al. \cite{Maher2023} underscore the importance of individual differences in shaping students’ interactions with ChatGPT. For instance, students who express high confidence in ChatGPT’s capabilities tend to skip verification of generated code, whereas more cautious learners are more likely to review and assess the output. 

\par Shi \& Arvandoust \cite{shi2025} compared the effects of ChatGPT and traditional Automated Writing Evaluation (AWE) tools (e.g. Grammarly and Pigai) on students’ writing performance and shifts in Ideal L2 Self (ILS). ILS is a motivational construct, representing a learner’s aspirational identity as a proficient writer in a second language \cite{dornyei2014,papi2021}. Quantitative data were analysed using paired-sample t-tests to compare within and between group effects. The writing tasks were assessed using an analytic scoring rubric and motivational change was measured via the ILS writing self-scale. Despite the ChatGPT group demonstrating an increase in post-test writing performance, this group also scored significantly lower on aspects of  ILS associated with reduced motivation, lower goal orientation and weaker long-term commitment to language learning. In effect, students may become less invested in developing their own writing abilities when overly-reliant on LLM generated content. This notion is further corroborated by students’ own concerns about relying on LLM tools, despite their awareness that the generated output may be incomplete, misleading, or outdated \cite{essien2024}. These outcomes align with concerns around “AI enfeeblement” \cite{carlson2024}, a phenomenon in which habitual dependence on LLM systems undermines users’ cognitive engagement, sense of ownership, and capacity for autonomous skill development. 

\subsection{Risk Summary \& Mitigation Strategies}

\begin{table}[t]
  \centering
  \footnotesize
  \setlength{\tabcolsep}{6pt}
  \renewcommand{\arraystretch}{0.95}
  \caption{Risk frequency in interactive learning tool usage of LLMs.}
  \label{tab:ilt-risks-spark}
  \begin{tabular}{@{}>{\raggedright\arraybackslash}p{0.58\columnwidth} r r c@{}}
    \toprule
    Risk & Count  & \multicolumn{1}{c}{} \\
    \midrule
    Cognitive \& Metacognitive            & 9 & \sparkbar{9}{9} \\
    Behavioural                           & 7 & \sparkbar{7}{9} \\
    Anthropomorphic Characteristics       & 2 & \sparkbar{2}{9} \\
    Limited Robustness                    & 4 & \sparkbar{4}{9} \\
    Privacy                               & 4 &  \sparkbar{4}{9} \\
    Hallucinations                        & 1 &  \sparkbar{1}{9} \\
    \bottomrule
  \end{tabular}
\end{table}

Table 4 shows the frequency of risks in the interactive usage of LLMs. Cognitive and metacognitive risks are the most frequently cited. Although distinct in definition, cognitive and metacognitive risks are grouped together in the bar chart because they often co-occur in practice and are closely intertwined in how LLMs affect learning. For example, when a student accepts an LLM answer without scrutiny, there is both a cognitive consequence (superficial knowledge acquisition, reduced memory encoding) and a metacognitive consequence (failure to monitor accuracy or regulate strategy use). Behavioural risks are also prominent, encompassing over-reliance on LLMs, decreased revision behaviour, weakened independent learning skills, and reduced student agency. Additional risks include limited robustness and privacy concerns, as models may exhibit unreliable or inconsistent behaviour and expose sensitive information. By contrast, the least frequently cited risks relate to the anthropomorphic characteristics of LLMs and the generation of hallucinations. In addition to educational personnel and developers, the following mitigation strategies involve students and governments: students, as the primary users interacting with these tools, and governments, as regulatory bodies uniquely positioned to establish safeguards and standards.

\subsubsection{Educational Personnel}
\begin{itemize}
    \item Revise Learning Objectives \& Assessment Standards: Shift the focus toward higher-order cognitive skills such as critical analysis and knowledge transfer, rather than emphasizing simple content production \cite{zhu_impact_2025,alimardani2024}.
    \item LLM Literacy Training: Educators should guide students in collaborating with LLM tools, fostering independent learning and critical appraisal of LLM-generated content \cite{zhu_impact_2025,keith_harnessing_2025,weidinger,shi2025}.
\end{itemize}

\subsubsection{Developers}
\begin{itemize}
    \item Learner-Centred Design: Prioritize interfaces and features that foster inquiry, creativity, and critical thinking, while minimizing risks such as over-reliance and hallucinations \cite{zhu_impact_2025}.
    \item Explainability \& Transparency: Incorporate explainable features and interactive communication channels that allow students to question, understand, and receive constructive feedback on AI decisions \cite{song2025}.
    \item Human–in-the-loop Collaboration: Involve educators in the decision-making process \cite{song2025}.
\end{itemize}

\subsubsection{Students}
\begin{itemize}
    \item Verification: Cross-check LLM outputs with reliable sources and critically question all LLM outputs \cite{Alimardani2025}.
    \item Responsible Use: Treat LLMs as support tools rather than substitutes and remain transparent about their use in coursework \cite{Guner2025,strozyna2024}.  
    \item AI Literacy: Actively engage with the limitations of LLM models and reflect on their cognitive and ethical implications \cite{liu_application_2024,Zhang2025}. 
\end{itemize}

\subsubsection{Governments}
\begin{itemize}
    \item Data Governance: Develop robust policies regulating data privacy, usage and integration of LLMs to safeguard student welfare \cite{zhu_impact_2025}.
    \item Stakeholder Collaboration: Collaborative efforts among educators, policymakers, and technologists are essential to harness LLMs educational potential responsibly and equitably \cite{Chun2025}.
\end{itemize}

\section{Discussion}

This section synthesizes cross-study patterns and introduces the LLM–Risk Adapted Learning Model, a conceptual blueprint grounded in Self-Regulated Learning (SRL) \cite{Zimmerman2002}. We adapt SRL because the theory synthesizes perspectives from both Cognitive Learning Theory \cite{piaget_1971} and the Zone of Proximal Development \cite{vygotsky1978}, which have been central in explaining the educational risks of LLMs throughout this paper. Rather than presenting a finalized or empirically validated framework, our model synthesizes findings into a structured lens for understanding how risks emerge and evolve when LLMs are incorporated into educational practice, as shown in Figure 6.

\par The model reconceptualizes the learning cycle into three stages—Interaction, Monitoring, and Outcome (IMO)—to capture how risks identified in the literature accumulate and shape the learning process. While SRL offers a holistic view of how students regulate learning, it does not account for the unique dynamics introduced by LLM tools, as it assumes regulation is generated internally. The IMO adaptation addresses this gap by situating regulation within learner–LLM interactions and tracing how risks manifest across stages. The cycle underscores how interactions with LLMs shape learning processes, drawing attention to the cognitive and behavioural consequences of their use and the risks that accumulate across stages of the learning cycle.

\begin{figure}
    \centering
    \includegraphics[width=\linewidth]{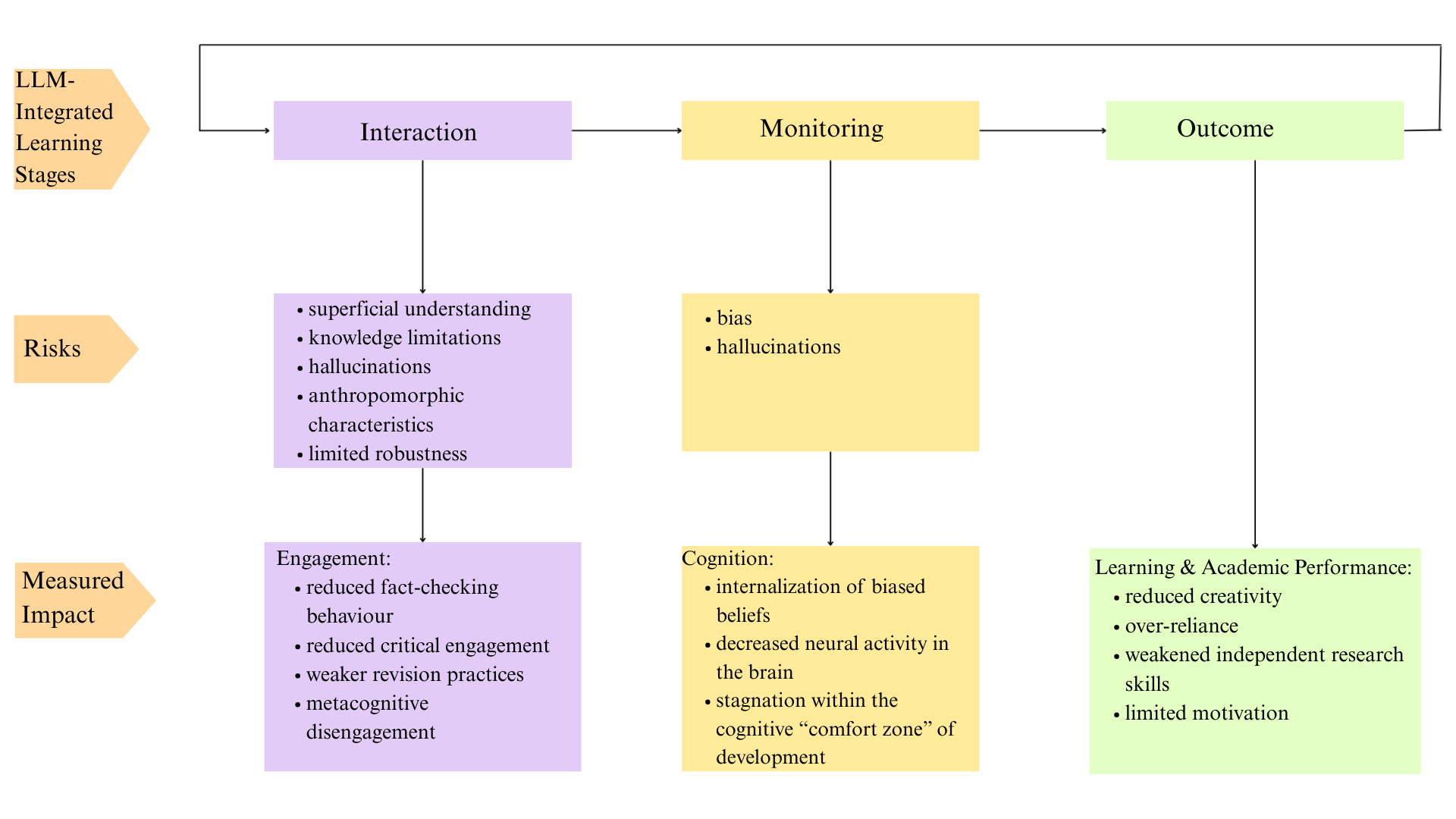}
    \Description{Diagram of the Interaction, Monitoring, \& Outcome (IMO) model.}
    \caption{LLM-Risk Adapted Learning Model (IMO), adapted from SRL theory, illustrating how risks emerge and compound across Interaction, Monitoring, and Outcome stages.}
    \label{fig:placeholder2}
    \vspace{-4mm}
\end{figure}

\par \textbf{Interaction} marks the stage where students engage with the LLM, encompassing their attention, interest, and the manner in which they work with the tool. Unlike models of learning that position planning or forethought as the initial stage—such as SRL \cite{Zimmerman2002} and metacognitive models \cite{Flavell}—our model suggests that when LLMs are integrated into the learning cycle, the opening phase shifts from reflective planning to immediate interaction. In this context, entering a prompt does not constitute genuine planning that requires a deliberate cognitive process, where learners actively consider goals, strategies, and approaches to the task. The impact of LLMs at this stage can be assessed through indicators of student engagement, including attention, interest, and active participation. Key risks include the models’ superficial understanding, knowledge limitations, hallucinations, anthropomorphic characteristics, and limited robustness, i.e., risks tied to the model’s operational effectiveness (see Section 4). The anthropomorphic characteristics of LLMs, as discussed throughout this paper, can foster a perceived sense of authority that encourages students to passively accept outputs rather than critically interrogate them. This tendency reduces fact-checking and weakens critical engagement, contributing to metacognitive disengagement.

\par \textbf{Monitoring}, according to SRL, refers to the stage where learners process and evaluate the outputs they receive, shaping not only their immediate understanding but also the neural activity and memory patterns that underlie learning. At this point, students do not simply receive information, but actively \emph{interpret} it—deciding what to trust, how to connect it to prior knowledge, and whether to revise or extend their understanding. These cognitive processes shape immediate comprehension and influence long-term memory formation. The primary risks here arise from model bias and hallucinations: because students focus directly on the generated output—often personalized in nature (see Section 5)—their comprehension can become subtly distorted. When learners interpret information passively, they fail to integrate it meaningfully into memory networks, which weakens long-term retention and may reinforce shallow neural activation patterns that hinder deeper learning. This tendency is further compounded by a cognitive comfort zone, where students rely on surface-level engagement with LLM output, rather than pushing themselves toward effortful, higher-order processing.

\par \textbf{Outcome} denotes the learner’s subsequent performance and development, encompassing self-reflection, evaluation of progress, and the transfer of knowledge and skills. Although most risks emerge during earlier stages of interaction and monitoring, their consequences become most evident here. At this phase, students may display reduced creativity, over-reliance on LLMs, weakened independent learning skills, or limited motivation.

\par As the IMO model illustrates, the stages of the LLM-Risk Adapted Learning Model should not be understood as strictly linear. Instead, risks introduced at each stage accumulate and compound over time, feeding forward into subsequent stages and, ultimately, looping back to shape the next cycle of interaction. For example, risks encountered during Interaction —such as anthropomorphic characteristics, hallucinations, or superficial engagement—directly weaken how learners interpret outputs, increasing susceptibility to bias and shallow processing during Monitoring. These interpretive distortions, in turn, surface at the Outcome stage as reduced creativity, over-reliance, or limited motivation. The feedback loop in the model highlights how weakened outcomes in turn influences the next round of Interaction, reinforcing passive engagement and further bypassing reflective planning. This cyclical pattern underscores that risks in LLM-mediated learning are not isolated events, but temporal dynamics that unfold across the learning process, with each stage amplifying vulnerabilities in the next.

\par The IMO model highlights opportunities for intervention at each stage of the cycle:
\begin{itemize}
    \item \textbf{Interaction}—design prompts, planning templates, reflection nudges, and expose model limitations (e.g., anthropomorphic characteristics).
    \item \textbf{Monitoring}—provide source transparency, contrastive explanations, and lightweight verification prompts to elicit active critique.
    \item \textbf{Outcome}— support autonomy with self-reflection and optional usage dashboards that make reliance patterns visible and encourage gradual tapering.
\end{itemize}
\section{Conclusion \& Future Directions}

\par This systematic review examined the empirical literature on LLMs in education and their affiliated risks. Guided by four research questions, we identified which educational applications of LLMs have been most frequently investigated (RQ1), how their impacts have been measured (RQ2), what risks are most prominently reported (RQ3), and which mitigation strategies have been proposed or evaluated (RQ4). By consolidating findings from computer science, education, and cognitive psychology, we offer the first comprehensive survey of empirically documented LLM-related risks in educational contexts. A further contribution lies in the synthesis of the diverse methodologies and evaluation metrics across diverse disciplines, which provides a practical reference point for researchers. It reveals how measures such as engagement, cognitive performance, and learning outcomes have been operationalized to date, and where important gaps remain. Future work can build on this synthesis in two ways: first, by refining and extending existing measures to capture subtler dimensions of metacognition, collaboration, or trust; and second, by stress-testing these metrics with newer and more powerful models to assess their robustness across different LLM generations. In this way, the review not only consolidates prior knowledge, but also equips the community with methodological tools to advance rigorous, comparable, and cumulative research on LLMs in education.

\par The review also highlights broader implications for the design of LLM-based educational tools. Technical optimization alone is insufficient: systems must also support learners’ autonomy, competence, and relatedness. Key design challenges include ensuring transparency without cognitive overload, scaffolding student agency without fostering over-reliance, and balancing personalization with equity and accessibility. By framing these design priorities alongside our mapping of applications, methods, and metrics, we advance an evidence-based, human-centred agenda for integrating LLMs in education. Responsible design and deployment requires acknowledging the interplay between technical constraints, cognitive processes, and societal values. In doing so, we aim to guide future research toward educational technologies that are not only effective and innovative but also equitable, transparent, and supportive of learners’ wellbeing.

\par However, our review is not without limitations. The time-frame and search strategy may have excluded the most recent developments, particularly as LLM capabilities evolve rapidly, and our focus on English-language publications may under-represent findings reported elsewhere. Moreover, most evidence stems from short-term interventions, leaving open questions about sustained cognitive and social impacts in classrooms. These constraints point to important directions for future research: longitudinal studies of sustained effects, comparative work across cultural and institutional contexts, and design-oriented investigations that embed mitigation strategies directly into everyday interactions with LLM-powered tools. Our findings highlight a deeper concern: even with models such as GPT-3.5 or GPT-4, we observe emerging patterns of metacognitive and behavioural changes. If such tendencies already manifest with current technologies, what will they mean in the context of systems approaching superintelligence? Addressing this question underscores the urgency of developing educational frameworks and sociotechnical safeguards that preserve human agency, critical reasoning, and meaningful learning in the face of ever more powerful generative models.

\appendix

\bibliographystyle{ACM-Reference-Format}
\bibliography{References}


\begin{thebibliography}{139}


\ifx \showCODEN    \undefined \def \showCODEN     #1{\unskip}     \fi
\ifx \showISBNx    \undefined \def \showISBNx     #1{\unskip}     \fi
\ifx \showISBNxiii \undefined \def \showISBNxiii  #1{\unskip}     \fi
\ifx \showISSN     \undefined \def \showISSN      #1{\unskip}     \fi
\ifx \showLCCN     \undefined \def \showLCCN      #1{\unskip}     \fi
\ifx \shownote     \undefined \def \shownote      #1{#1}          \fi
\ifx \showarticletitle \undefined \def \showarticletitle #1{#1}   \fi
\ifx \showURL      \undefined \def \showURL       {\relax}        \fi
\providecommand\bibfield[2]{#2}
\providecommand\bibinfo[2]{#2}
\providecommand\natexlab[1]{#1}
\providecommand\showeprint[2][]{arXiv:#2}

\bibitem[Abadi et~al\mbox{.}(2016)]%
        {Abadi}
\bibfield{author}{\bibinfo{person}{Martin Abadi}, \bibinfo{person}{Andy Chu}, \bibinfo{person}{Ian Goodfellow}, \bibinfo{person}{H.~Brendan McMahan}, \bibinfo{person}{Ilya Mironov}, \bibinfo{person}{Kunal Talwar}, {and} \bibinfo{person}{Li Zhang}.} \bibinfo{year}{2016}\natexlab{}.
\newblock \showarticletitle{Deep Learning with Differential Privacy}. In \bibinfo{booktitle}{\emph{Proceedings of the 2016 ACM SIGSAC Conference on Computer and Communications Security}} (Vienna, Austria) \emph{(\bibinfo{series}{CCS '16})}. \bibinfo{publisher}{Association for Computing Machinery}, \bibinfo{address}{New York, NY, USA}, \bibinfo{pages}{308–318}.
\newblock
\showISBNx{9781450341394}
\href{https://doi.org/10.1145/2976749.2978318}{doi:\nolinkurl{10.1145/2976749.2978318}}


\bibitem[Abdul~Razak et~al\mbox{.}(2024)]%
        {Abdul}
\bibfield{author}{\bibinfo{person}{Aina~Nabila Abdul~Razak}, \bibinfo{person}{May Lim}, \bibinfo{person}{Inmaculada Tomeo-Reyes}, {and} \bibinfo{person}{Darson~Dezheng Li}.} \bibinfo{year}{2024}\natexlab{}.
\newblock \showarticletitle{Exploring the Capabilities and Limitations of Generative AI in Providing Feedback on Engineering Drawings: A Case Study}. In \bibinfo{booktitle}{\emph{2024 World Engineering Education Forum - Global Engineering Deans Council (WEEF-GEDC)}}. \bibinfo{publisher}{IEEE}, \bibinfo{address}{location}, \bibinfo{pages}{1--9}.
\newblock
\href{https://doi.org/10.1109/WEEF-GEDC63419.2024.10854935}{doi:\nolinkurl{10.1109/WEEF-GEDC63419.2024.10854935}}


\bibitem[Akgun and Toker(2024)]%
        {akgun2024}
\bibfield{author}{\bibinfo{person}{Mahir Akgun} {and} \bibinfo{person}{Sacip Toker}.} \bibinfo{year}{2024}\natexlab{}.
\newblock \showarticletitle{Evaluating the Effect of Pretesting with Conversational AI on Retention of Needed Information}.
\newblock \bibinfo{journal}{\emph{Journal Name}} \bibinfo{volume}{1}, \bibinfo{number}{1} (\bibinfo{date}{12} \bibinfo{year}{2024}).
\newblock
\href{https://doi.org/10.48550/arXiv.2412.13487}{doi:\nolinkurl{10.48550/arXiv.2412.13487}}


\bibitem[Akgun and Greenhow(2022)]%
        {akgun}
\bibfield{author}{\bibinfo{person}{Selin Akgun} {and} \bibinfo{person}{Christine Greenhow}.} \bibinfo{year}{2022}\natexlab{}.
\newblock \showarticletitle{Artificial intelligence in education: {Addressing} ethical challenges in {K}-12 settings}.
\newblock \bibinfo{journal}{\emph{AI and Ethics}} \bibinfo{volume}{2}, \bibinfo{number}{3} (\bibinfo{date}{Aug.} \bibinfo{year}{2022}), \bibinfo{pages}{431--440}.
\newblock
\showISSN{2730-5953, 2730-5961}
\href{https://doi.org/10.1007/s43681-021-00096-7}{doi:\nolinkurl{10.1007/s43681-021-00096-7}}


\bibitem[Albuquerque et~al\mbox{.}(2025)]%
        {Albuquerque16032025}
\bibfield{author}{\bibinfo{person}{Josmario Albuquerque}, \bibinfo{person}{Bart Rienties}, \bibinfo{person}{Wayne Holmes}, {and} \bibinfo{person}{Martin Hlosta}.} \bibinfo{year}{2025}\natexlab{}.
\newblock \showarticletitle{From hype to evidence: exploring large language models for inter-group bias classification in higher education}.
\newblock \bibinfo{journal}{\emph{Interactive Learning Environments}} \bibinfo{volume}{33}, \bibinfo{number}{3} (\bibinfo{year}{2025}), \bibinfo{pages}{2332--2354}.
\newblock
\href{https://doi.org/10.1080/10494820.2024.2408554}{doi:\nolinkurl{10.1080/10494820.2024.2408554}}
\showeprint{https://doi.org/10.1080/10494820.2024.2408554}


\bibitem[AlGhamdi(2024)]%
        {AlGhamdi}
\bibfield{author}{\bibinfo{person}{Rayed AlGhamdi}.} \bibinfo{year}{2024}\natexlab{}.
\newblock \showarticletitle{Exploring the impact of ChatGPT-generated feedback on technical writing skills of computing students: A blinded study}.
\newblock \bibinfo{journal}{\emph{Education and Information Technologies}}  \bibinfo{volume}{29} (\bibinfo{date}{03} \bibinfo{year}{2024}), \bibinfo{pages}{18901--18926}.
\newblock
\href{https://doi.org/10.1007/s10639-024-12594-2}{doi:\nolinkurl{10.1007/s10639-024-12594-2}}


\bibitem[Alimardani(2024)]%
        {alimardani2024}
\bibfield{author}{\bibinfo{person}{Armin Alimardani}.} \bibinfo{year}{2024}\natexlab{}.
\newblock \showarticletitle{Generative artificial intelligence vs. law students: an empirical study on criminal law exam performance}.
\newblock \bibinfo{journal}{\emph{Law, Innovation and Technology}} \bibinfo{volume}{16}, \bibinfo{number}{2} (\bibinfo{year}{2024}), \bibinfo{pages}{777--819}.
\newblock
\href{https://doi.org/10.1080/17579961.2024.2392932}{doi:\nolinkurl{10.1080/17579961.2024.2392932}}
\newblock
\shownote{Publisher: Routledge \_eprint: https://doi.org/10.1080/17579961.2024.2392932}.


\bibitem[Alimardani(2025)]%
        {Alimardani2025}
\bibfield{author}{\bibinfo{person}{Armin Alimardani}.} \bibinfo{year}{2025}\natexlab{}.
\newblock \showarticletitle{Borderline Disaster: An Empirical Study on Student Usage of GenAI in a Law Assignment}.
\newblock \bibinfo{journal}{\emph{IEEE Transactions on Technology and Society}} \bibinfo{volume}{6}, \bibinfo{number}{2} (\bibinfo{year}{2025}), \bibinfo{pages}{210--219}.
\newblock
\href{https://doi.org/10.1109/TTS.2025.3540978}{doi:\nolinkurl{10.1109/TTS.2025.3540978}}


\bibitem[Alzoubi and Mishra(2024)]%
        {alzoubi}
\bibfield{author}{\bibinfo{person}{Yehia~Ibrahim Alzoubi} {and} \bibinfo{person}{Alok Mishra}.} \bibinfo{year}{2024}\natexlab{}.
\newblock \showarticletitle{Green artificial intelligence initiatives: {Potentials} and challenges}.
\newblock \bibinfo{journal}{\emph{Journal of Cleaner Production}}  \bibinfo{volume}{468} (\bibinfo{year}{2024}), \bibinfo{pages}{143090}.
\newblock
\showISSN{0959-6526}
\href{https://doi.org/10.1016/j.jclepro.2024.143090}{doi:\nolinkurl{10.1016/j.jclepro.2024.143090}}


\bibitem[Awidi(2024)]%
        {awidi2024}
\bibfield{author}{\bibinfo{person}{Isaiah~T. Awidi}.} \bibinfo{year}{2024}\natexlab{}.
\newblock \showarticletitle{Comparing expert tutor evaluation of reflective essays with marking by generative artificial intelligence ({AI}) tool}.
\newblock \bibinfo{journal}{\emph{Computers and Education: Artificial Intelligence}}  \bibinfo{volume}{6} (\bibinfo{year}{2024}), \bibinfo{pages}{100226}.
\newblock
\showISSN{2666-920X}
\href{https://doi.org/10.1016/j.caeai.2024.100226}{doi:\nolinkurl{10.1016/j.caeai.2024.100226}}


\bibitem[Bai et~al\mbox{.}(2025)]%
        {bai2025}
\bibfield{author}{\bibinfo{person}{Zechen Bai}, \bibinfo{person}{Pichao Wang}, \bibinfo{person}{Tianjun Xiao}, \bibinfo{person}{Tong He}, \bibinfo{person}{Zongbo Han}, \bibinfo{person}{Zheng Zhang}, {and} \bibinfo{person}{Mike~Zheng Shou}.} \bibinfo{year}{2025}\natexlab{}.
\newblock \bibinfo{title}{Hallucination of Multimodal Large Language Models: A Survey}.
\newblock
\showeprint[arxiv]{2404.18930}~[cs.CV]
\urldef\tempurl%
\url{https://arxiv.org/abs/2404.18930}
\showURL{%
\tempurl}


\bibitem[Banihashem et~al\mbox{.}(2024)]%
        {banihashem2024}
\bibfield{author}{\bibinfo{person}{Seyyed~Kazem Banihashem}, \bibinfo{person}{Nafiseh~Taghizadeh Kerman}, \bibinfo{person}{Omid Noroozi}, \bibinfo{person}{Jewoong Moon}, {and} \bibinfo{person}{Hendrik Drachsler}.} \bibinfo{year}{2024}\natexlab{}.
\newblock \showarticletitle{Feedback sources in essay writing: peer-generated or {AI}-generated feedback?}
\newblock \bibinfo{journal}{\emph{International Journal of Educational Technology in Higher Education}} \bibinfo{volume}{21}, \bibinfo{number}{1} (\bibinfo{date}{April} \bibinfo{year}{2024}), \bibinfo{pages}{23}.
\newblock
\showISSN{2365-9440}
\href{https://doi.org/10.1186/s41239-024-00455-4}{doi:\nolinkurl{10.1186/s41239-024-00455-4}}


\bibitem[Bender et~al\mbox{.}(2021)]%
        {Bender}
\bibfield{author}{\bibinfo{person}{Emily~M. Bender}, \bibinfo{person}{Timnit Gebru}, \bibinfo{person}{Angelina McMillan-Major}, {and} \bibinfo{person}{Shmargaret Shmitchell}.} \bibinfo{year}{2021}\natexlab{}.
\newblock \showarticletitle{On the Dangers of Stochastic Parrots: Can Language Models Be Too Big?}. In \bibinfo{booktitle}{\emph{Proceedings of the 2021 ACM Conference on Fairness, Accountability, and Transparency}} (Virtual Event, Canada) \emph{(\bibinfo{series}{FAccT '21})}. \bibinfo{publisher}{Association for Computing Machinery}, \bibinfo{address}{New York, NY, USA}, \bibinfo{pages}{610–623}.
\newblock
\showISBNx{9781450383097}
\href{https://doi.org/10.1145/3442188.3445922}{doi:\nolinkurl{10.1145/3442188.3445922}}


\bibitem[Berridge(2018)]%
        {berridge2018}
\bibfield{author}{\bibinfo{person}{Kent~C Berridge}.} \bibinfo{year}{2018}\natexlab{}.
\newblock \showarticletitle{Evolving concepts of emotion and motivation}.
\newblock \bibinfo{journal}{\emph{Frontiers in psychology}}  \bibinfo{volume}{9} (\bibinfo{year}{2018}), \bibinfo{pages}{317391}.
\newblock


\bibitem[Berridge et~al\mbox{.}(2009)]%
        {berridge2009}
\bibfield{author}{\bibinfo{person}{Kent~C Berridge}, \bibinfo{person}{Terry~E Robinson}, {and} \bibinfo{person}{J~Wayne Aldridge}.} \bibinfo{year}{2009}\natexlab{}.
\newblock \showarticletitle{Dissecting components of reward:‘liking’,‘wanting’, and learning}.
\newblock \bibinfo{journal}{\emph{Current opinion in pharmacology}} \bibinfo{volume}{9}, \bibinfo{number}{1} (\bibinfo{year}{2009}), \bibinfo{pages}{65--73}.
\newblock


\bibitem[Bommasani et~al\mbox{.}(2022)]%
        {bommasani2022}
\bibfield{author}{\bibinfo{person}{Rishi Bommasani}, \bibinfo{person}{Drew~A. Hudson}, {and} \bibinfo{person}{Ehsan Adeli}.} \bibinfo{year}{2022}\natexlab{}.
\newblock \bibinfo{title}{On the Opportunities and Risks of Foundation Models}.
\newblock
\showeprint[arxiv]{2108.07258}~[cs.LG]
\urldef\tempurl%
\url{https://arxiv.org/abs/2108.07258}
\showURL{%
\tempurl}


\bibitem[Borges et~al\mbox{.}(2024)]%
        {borges2024}
\bibfield{author}{\bibinfo{person}{Beatriz Borges}, \bibinfo{person}{Negar Foroutan}, \bibinfo{person}{Deniz Bayazit}, \bibinfo{person}{Anna Sotnikova}, \bibinfo{person}{Syrielle Montariol}, \bibinfo{person}{Tanya Nazaretsky}, \bibinfo{person}{Mohammadreza Banaei}, \bibinfo{person}{Alireza Sakhaeirad}, \bibinfo{person}{Philippe Servant}, \bibinfo{person}{Seyed~Parsa Neshaei}, \bibinfo{person}{Jibril Frej}, \bibinfo{person}{Angelika Romanou}, \bibinfo{person}{Gail Weiss}, \bibinfo{person}{Sepideh Mamooler}, \bibinfo{person}{Zeming Chen}, \bibinfo{person}{Simin Fan}, \bibinfo{person}{Silin Gao}, \bibinfo{person}{Mete Ismayilzada}, \bibinfo{person}{Debjit Paul}, \bibinfo{person}{Philippe Schwaller}, \bibinfo{person}{Sacha Friedli}, \bibinfo{person}{Patrick Jermann}, \bibinfo{person}{Tanja Käser}, \bibinfo{person}{Antoine Bosselut}, \bibinfo{person}{{EPFL Grader Consortium}}, {and} \bibinfo{person}{{EPFL Data Consortium}}.} \bibinfo{year}{2024}\natexlab{}.
\newblock \showarticletitle{Could {ChatGPT} get an engineering degree? {Evaluating} higher education vulnerability to {AI} assistants}.
\newblock \bibinfo{journal}{\emph{Proceedings of the National Academy of Sciences}} \bibinfo{volume}{121}, \bibinfo{number}{49} (\bibinfo{date}{Dec.} \bibinfo{year}{2024}), \bibinfo{pages}{e2414955121}.
\newblock
\showISSN{0027-8424, 1091-6490}
\href{https://doi.org/10.1073/pnas.2414955121}{doi:\nolinkurl{10.1073/pnas.2414955121}}


\bibitem[Borji(2023)]%
        {borji2023s}
\bibfield{author}{\bibinfo{person}{Ali Borji}.} \bibinfo{year}{2023}\natexlab{}.
\newblock \bibinfo{title}{A Categorical Archive of ChatGPT Failures}.
\newblock
\showeprint[arxiv]{2302.03494}~[cs.CL]
\urldef\tempurl%
\url{https://arxiv.org/abs/2302.03494}
\showURL{%
\tempurl}


\bibitem[Bouchard(2025)]%
        {bouchard}
\bibfield{author}{\bibinfo{person}{Jeremie Bouchard}.} \bibinfo{year}{2025}\natexlab{}.
\newblock \showarticletitle{{ChatGPT} and the separation between knowledge and knower}.
\newblock \bibinfo{journal}{\emph{Education and Information Technologies}} \bibinfo{volume}{30}, \bibinfo{number}{8} (\bibinfo{date}{June} \bibinfo{year}{2025}), \bibinfo{pages}{10091--10110}.
\newblock
\showISSN{1360-2357, 1573-7608}
\href{https://doi.org/10.1007/s10639-024-13249-y}{doi:\nolinkurl{10.1007/s10639-024-13249-y}}


\bibitem[Caccavale et~al\mbox{.}(2024)]%
        {Caccavale2024}
\bibfield{author}{\bibinfo{person}{Fiammetta Caccavale}, \bibinfo{person}{Carina L.~Gargalo}, \bibinfo{person}{Krist Gernaey}, {and} \bibinfo{person}{Ulrich Krühne}.} \bibinfo{year}{2024}\natexlab{}.
\newblock \showarticletitle{Towards Education 4.0: The Role of Large Language Models as Virtual Tutors in Chemical Engineering}.
\newblock \bibinfo{journal}{\emph{Education for Chemical Engineers}}  \bibinfo{volume}{49} (\bibinfo{date}{07} \bibinfo{year}{2024}).
\newblock
\href{https://doi.org/10.1016/j.ece.2024.07.002}{doi:\nolinkurl{10.1016/j.ece.2024.07.002}}


\bibitem[Carey(2012)]%
        {carey2012}
\bibfield{author}{\bibinfo{person}{Leeanne~M Carey}.} \bibinfo{year}{2012}\natexlab{}.
\newblock \bibinfo{booktitle}{\emph{Stroke rehabilitation: insights from neuroscience and imaging}}.
\newblock \bibinfo{publisher}{Oxford university press}, \bibinfo{address}{Oxford}.
\newblock


\bibitem[Carlson et~al\mbox{.}(2024)]%
        {carlson2024}
\bibfield{author}{\bibinfo{person}{Makenna Carlson}, \bibinfo{person}{Austin Pack}, {and} \bibinfo{person}{Juan Escalante}.} \bibinfo{year}{2024}\natexlab{}.
\newblock \showarticletitle{Utilizing {OpenAI}'s {GPT}-4 for written feedback}.
\newblock \bibinfo{journal}{\emph{TESOL Journal}} \bibinfo{volume}{15}, \bibinfo{number}{2} (\bibinfo{date}{June} \bibinfo{year}{2024}), \bibinfo{pages}{Advance online publication}.
\newblock
\showISSN{1056-7941, 1949-3533}
\href{https://doi.org/10.1002/tesj.759}{doi:\nolinkurl{10.1002/tesj.759}}
\newblock
\shownote{Publisher: Wiley}.


\bibitem[Carter et~al\mbox{.}(2020)]%
        {Carter}
\bibfield{author}{\bibinfo{person}{Lemuria Carter}, \bibinfo{person}{Dapeng Liu}, {and} \bibinfo{person}{Caley Cantrell}.} \bibinfo{year}{2020}\natexlab{}.
\newblock \showarticletitle{Exploring the Intersection of the Digital Divide and Artificial Intelligence: A Hermeneutic Literature Review}.
\newblock \bibinfo{journal}{\emph{AIS Transactions on Human-Computer Interaction}}  \bibinfo{volume}{12} (\bibinfo{date}{12} \bibinfo{year}{2020}), \bibinfo{pages}{253--275}.
\newblock
\href{https://doi.org/10.17705/1thci.00138}{doi:\nolinkurl{10.17705/1thci.00138}}


\bibitem[Chai et~al\mbox{.}(2024)]%
        {chai}
\bibfield{author}{\bibinfo{person}{Fangyuan Chai}, \bibinfo{person}{Jiajia Ma}, \bibinfo{person}{Yi Wang}, \bibinfo{person}{Jun Zhu}, {and} \bibinfo{person}{Tingting Han}.} \bibinfo{year}{2024}\natexlab{}.
\newblock \showarticletitle{Grading by {AI} makes me feel fairer? {How} different evaluators affect college students’ perception of fairness}.
\newblock \bibinfo{journal}{\emph{Frontiers in Psychology}}  \bibinfo{volume}{Volume 15 - 2024} (\bibinfo{year}{2024}), \bibinfo{pages}{1221177}.
\newblock
\showISSN{1664-1078}
\href{https://doi.org/10.3389/fpsyg.2024.1221177}{doi:\nolinkurl{10.3389/fpsyg.2024.1221177}}


\bibitem[Cheng et~al\mbox{.}(2023)]%
        {cheng}
\bibfield{author}{\bibinfo{person}{Myra Cheng}, \bibinfo{person}{Esin Durmus}, {and} \bibinfo{person}{Dan Jurafsky}.} \bibinfo{year}{2023}\natexlab{}.
\newblock \showarticletitle{Marked Personas: Using Natural Language Prompts to Measure Stereotypes in Language Models}. In \bibinfo{booktitle}{\emph{Proceedings of the 61st Annual Meeting of the Association for Computational Linguistics (Volume 1: Long Papers)}}, \bibfield{editor}{\bibinfo{person}{Anna Rogers}, \bibinfo{person}{Jordan Boyd-Graber}, {and} \bibinfo{person}{Naoaki Okazaki}} (Eds.). \bibinfo{publisher}{Association for Computational Linguistics}, \bibinfo{address}{Toronto, Canada}, \bibinfo{pages}{1504--1532}.
\newblock
\href{https://doi.org/10.18653/v1/2023.acl-long.84}{doi:\nolinkurl{10.18653/v1/2023.acl-long.84}}


\bibitem[Cho et~al\mbox{.}(2017)]%
        {Cho2017}
\bibfield{author}{\bibinfo{person}{Isaac Cho}, \bibinfo{person}{Ryan Wesslen}, \bibinfo{person}{Alireza Karduni}, \bibinfo{person}{Sashank Santhanam}, \bibinfo{person}{Samira Shaikh}, {and} \bibinfo{person}{Wenwen Dou}.} \bibinfo{year}{2017}\natexlab{}.
\newblock \showarticletitle{The Anchoring Effect in Decision-Making with Visual Analytics}. In \bibinfo{booktitle}{\emph{2017 IEEE Conference on Visual Analytics Science and Technology (VAST)}}. \bibinfo{publisher}{IEEE}, \bibinfo{address}{Piscataway, NJ}, \bibinfo{pages}{116--126}.
\newblock
\href{https://doi.org/10.1109/VAST.2017.8585665}{doi:\nolinkurl{10.1109/VAST.2017.8585665}}


\bibitem[Chun et~al\mbox{.}(2025)]%
        {Chun2025}
\bibfield{author}{\bibinfo{person}{Jiyong Chun}, \bibinfo{person}{Jeongsoo Kim}, \bibinfo{person}{Hyejin Kim}, \bibinfo{person}{Geumgu Lee}, \bibinfo{person}{Sanggoo Cho}, \bibinfo{person}{Changshik Kim}, \bibinfo{person}{Yeesook Chung}, {and} \bibinfo{person}{Seoyoon Heo}.} \bibinfo{year}{2025}\natexlab{}.
\newblock \showarticletitle{A Comparative Analysis of On-Device AI-Driven, Self-Regulated Learning and Traditional Pedagogy in University Health Sciences Education}.
\newblock \bibinfo{journal}{\emph{Applied Sciences}}  \bibinfo{volume}{15} (\bibinfo{date}{02} \bibinfo{year}{2025}), \bibinfo{pages}{1815}.
\newblock
\href{https://doi.org/10.3390/app15041815}{doi:\nolinkurl{10.3390/app15041815}}


\bibitem[Cooper(2023)]%
        {cooper2023}
\bibfield{author}{\bibinfo{person}{Grant Cooper}.} \bibinfo{year}{2023}\natexlab{}.
\newblock \showarticletitle{Examining {Science} {Education} in {ChatGPT}: {An} {Exploratory} {Study} of {Generative} {Artificial} {Intelligence}}.
\newblock \bibinfo{journal}{\emph{Journal of Science Education and Technology}} \bibinfo{volume}{32}, \bibinfo{number}{3} (\bibinfo{date}{June} \bibinfo{year}{2023}), \bibinfo{pages}{444--452}.
\newblock
\showISSN{1059-0145, 1573-1839}
\href{https://doi.org/10.1007/s10956-023-10039-y}{doi:\nolinkurl{10.1007/s10956-023-10039-y}}


\bibitem[Couldry and Mejias(2019)]%
        {couldry2019}
\bibfield{author}{\bibinfo{person}{Nick Couldry} {and} \bibinfo{person}{Ulises~A. Mejias}.} \bibinfo{year}{2019}\natexlab{}.
\newblock \showarticletitle{Data {Colonialism}: {Rethinking} {Big} {Data}’s {Relation} to the {Contemporary} {Subject}}.
\newblock \bibinfo{journal}{\emph{Television \& New Media}} \bibinfo{volume}{20}, \bibinfo{number}{4} (\bibinfo{date}{May} \bibinfo{year}{2019}), \bibinfo{pages}{336--349}.
\newblock
\showISSN{1527-4764, 1552-8316}
\href{https://doi.org/10.1177/1527476418796632}{doi:\nolinkurl{10.1177/1527476418796632}}
\newblock
\shownote{Publisher: SAGE Publications}.


\bibitem[Currie et~al\mbox{.}(2024)]%
        {currie2024}
\bibfield{author}{\bibinfo{person}{Geoffrey Currie}, \bibinfo{person}{Josie Currie}, \bibinfo{person}{Sam Anderson}, {and} \bibinfo{person}{Johnathan Hewis}.} \bibinfo{year}{2024}\natexlab{}.
\newblock \showarticletitle{Gender bias in generative artificial intelligence text-to-image depiction of medical students}.
\newblock \bibinfo{journal}{\emph{Health Education Journal}} \bibinfo{volume}{83}, \bibinfo{number}{7} (\bibinfo{date}{Nov.} \bibinfo{year}{2024}), \bibinfo{pages}{732--746}.
\newblock
\showISSN{0017-8969, 1748-8176}
\href{https://doi.org/10.1177/00178969241274621}{doi:\nolinkurl{10.1177/00178969241274621}}


\bibitem[Danesh et~al\mbox{.}(2024)]%
        {danesh2024}
\bibfield{author}{\bibinfo{person}{Arman Danesh}, \bibinfo{person}{Hirad Pazouki}, \bibinfo{person}{Farzad Danesh}, \bibinfo{person}{Arsalan Danesh}, {and} \bibinfo{person}{Saynur Vardar‐Sengul}.} \bibinfo{year}{2024}\natexlab{}.
\newblock \showarticletitle{Artificial intelligence in dental education: {ChatGPT}'s performance on the periodontic in‐service examination}.
\newblock \bibinfo{journal}{\emph{Journal of Periodontology}} \bibinfo{volume}{95}, \bibinfo{number}{7} (\bibinfo{date}{July} \bibinfo{year}{2024}), \bibinfo{pages}{682--687}.
\newblock
\showISSN{0022-3492, 1943-3670}
\href{https://doi.org/10.1002/JPER.23-0514}{doi:\nolinkurl{10.1002/JPER.23-0514}}


\bibitem[Davies et~al\mbox{.}(2024)]%
        {davies2024}
\bibfield{author}{\bibinfo{person}{Nathan~P Davies}, \bibinfo{person}{Robert Wilson}, \bibinfo{person}{Madeleine~S Winder}, \bibinfo{person}{Simon~J Tunster}, \bibinfo{person}{Kathryn McVicar}, \bibinfo{person}{Shivan Thakrar}, \bibinfo{person}{Joe Williams}, {and} \bibinfo{person}{Allan Reid}.} \bibinfo{year}{2024}\natexlab{}.
\newblock \showarticletitle{{ChatGPT} sits the {DFPH} exam: large language model performance and potential to support public health learning}.
\newblock \bibinfo{journal}{\emph{BMC Medical Education}} \bibinfo{volume}{24}, \bibinfo{number}{1} (\bibinfo{date}{Jan.} \bibinfo{year}{2024}), \bibinfo{pages}{57}.
\newblock
\showISSN{1472-6920}
\href{https://doi.org/10.1186/s12909-024-05042-9}{doi:\nolinkurl{10.1186/s12909-024-05042-9}}


\bibitem[De~Roock(2024)]%
        {deroock2024}
\bibfield{author}{\bibinfo{person}{Roberto~Santiago De~Roock}.} \bibinfo{year}{2024}\natexlab{}.
\newblock \showarticletitle{To {Become} an {Object} {Among} {Objects}: {Generative} {Artificial} “{Intelligence},” {Writing}, and {Linguistic} {White} {Supremacy}}.
\newblock \bibinfo{journal}{\emph{Reading Research Quarterly}} \bibinfo{volume}{59}, \bibinfo{number}{4} (\bibinfo{date}{Oct.} \bibinfo{year}{2024}), \bibinfo{pages}{590--608}.
\newblock
\showISSN{0034-0553, 1936-2722}
\href{https://doi.org/10.1002/rrq.569}{doi:\nolinkurl{10.1002/rrq.569}}


\bibitem[Deshpande et~al\mbox{.}(2023)]%
        {deshpande}
\bibfield{author}{\bibinfo{person}{Ameet Deshpande}, \bibinfo{person}{Vishvak Murahari}, \bibinfo{person}{Tanmay Rajpurohit}, \bibinfo{person}{Ashwin Kalyan}, {and} \bibinfo{person}{Karthik Narasimhan}.} \bibinfo{year}{2023}\natexlab{}.
\newblock \showarticletitle{Toxicity in chatgpt: Analyzing persona-assigned language models}. In \bibinfo{booktitle}{\emph{Findings of the Association for Computational Linguistics: EMNLP 2023}}, \bibfield{editor}{\bibinfo{person}{Houda Bouamor}, \bibinfo{person}{Juan Pino}, {and} \bibinfo{person}{Kalika Bali}} (Eds.). \bibinfo{publisher}{Association for Computational Linguistics}, \bibinfo{address}{Singapore}, \bibinfo{pages}{1236--1270}.
\newblock
\href{https://doi.org/10.18653/v1/2023.findings-emnlp.88}{doi:\nolinkurl{10.18653/v1/2023.findings-emnlp.88}}


\bibitem[Dong et~al\mbox{.}(2024)]%
        {Dong_2024}
\bibfield{author}{\bibinfo{person}{Bingyu Dong}, \bibinfo{person}{Jie Bai}, \bibinfo{person}{Tao Xu}, {and} \bibinfo{person}{Yun Zhou}.} \bibinfo{year}{2024}\natexlab{}.
\newblock \showarticletitle{Large Language Models in Education: A Systematic Review}. In \bibinfo{booktitle}{\emph{2024 6th International Conference on Computer Science and Technologies in Education (CSTE)}}. \bibinfo{publisher}{IEEE}, \bibinfo{address}{New York, NY, USA}, \bibinfo{pages}{131--134}.
\newblock
\href{https://doi.org/10.1109/CSTE62025.2024.00031}{doi:\nolinkurl{10.1109/CSTE62025.2024.00031}}


\bibitem[Dörnyei(2014)]%
        {dornyei2014}
\bibfield{author}{\bibinfo{person}{Zoltán Dörnyei}.} \bibinfo{year}{2014}\natexlab{}.
\newblock \bibinfo{booktitle}{\emph{The {Psychology} of the {Language} {Learner}} (\bibinfo{edition}{0} ed.)}.
\newblock \bibinfo{publisher}{Routledge}, \bibinfo{address}{London}.
\newblock
\showISBNx{978-1-135-70478-0}
\href{https://doi.org/10.4324/9781410613349}{doi:\nolinkurl{10.4324/9781410613349}}


\bibitem[Essien et~al\mbox{.}(2024)]%
        {essien2024}
\bibfield{author}{\bibinfo{person}{Aniekan Essien}, \bibinfo{person}{Oyegoke~Teslim Bukoye}, \bibinfo{person}{Xianghan O’Dea}, {and} \bibinfo{person}{Marios Kremantzis}.} \bibinfo{year}{2024}\natexlab{}.
\newblock \showarticletitle{The influence of {AI} text generators on critical thinking skills in {UK} business schools}.
\newblock \bibinfo{journal}{\emph{Studies in Higher Education}} \bibinfo{volume}{49}, \bibinfo{number}{5} (\bibinfo{date}{May} \bibinfo{year}{2024}), \bibinfo{pages}{865--882}.
\newblock
\showISSN{0307-5079, 1470-174X}
\href{https://doi.org/10.1080/03075079.2024.2316881}{doi:\nolinkurl{10.1080/03075079.2024.2316881}}
\newblock
\shownote{Publisher: Informa UK Limited}.


\bibitem[Fan et~al\mbox{.}(2025)]%
        {fan2025}
\bibfield{author}{\bibinfo{person}{Yizhou Fan}, \bibinfo{person}{Luzhen Tang}, \bibinfo{person}{Huixiao Le}, \bibinfo{person}{Kejie Shen}, \bibinfo{person}{Shufang Tan}, \bibinfo{person}{Yueying Zhao}, \bibinfo{person}{Yuan Shen}, \bibinfo{person}{Xinyu Li}, {and} \bibinfo{person}{Dragan Gašević}.} \bibinfo{year}{2025}\natexlab{}.
\newblock \showarticletitle{Beware of metacognitive laziness: {Effects} of generative artificial intelligence on learning motivation, processes, and performance}.
\newblock \bibinfo{journal}{\emph{British Journal of Educational Technology}} \bibinfo{volume}{56}, \bibinfo{number}{2} (\bibinfo{date}{March} \bibinfo{year}{2025}), \bibinfo{pages}{489--530}.
\newblock
\showISSN{0007-1013, 1467-8535}
\href{https://doi.org/10.1111/bjet.13544}{doi:\nolinkurl{10.1111/bjet.13544}}
\newblock
\shownote{Publisher: Wiley}.


\bibitem[Flavell(1979)]%
        {Flavell}
\bibfield{author}{\bibinfo{person}{John Flavell}.} \bibinfo{year}{1979}\natexlab{}.
\newblock \showarticletitle{Metacognition and Cognitive Monitoring: A New Area of Cognitive-Developmental Inquiry}.
\newblock \bibinfo{journal}{\emph{American Psychologist}}  \bibinfo{volume}{34} (\bibinfo{date}{10} \bibinfo{year}{1979}), \bibinfo{pages}{906--911}.
\newblock
\href{https://doi.org/10.1037/0003-066X.34.10.906}{doi:\nolinkurl{10.1037/0003-066X.34.10.906}}


\bibitem[Fleckenstein et~al\mbox{.}(2024)]%
        {fleckenstein2024}
\bibfield{author}{\bibinfo{person}{Johanna Fleckenstein}, \bibinfo{person}{Jennifer Meyer}, \bibinfo{person}{Thorben Jansen}, \bibinfo{person}{Stefan~D. Keller}, \bibinfo{person}{Olaf Köller}, {and} \bibinfo{person}{Jens Möller}.} \bibinfo{year}{2024}\natexlab{}.
\newblock \showarticletitle{Do teachers spot {AI}? {Evaluating} the detectability of {AI}-generated texts among student essays}.
\newblock \bibinfo{journal}{\emph{Computers and Education: Artificial Intelligence}}  \bibinfo{volume}{6} (\bibinfo{year}{2024}), \bibinfo{pages}{100209}.
\newblock
\showISSN{2666-920X}
\href{https://doi.org/10.1016/j.caeai.2024.100209}{doi:\nolinkurl{10.1016/j.caeai.2024.100209}}


\bibitem[Flores and Rosa(2015)]%
        {Flores2015}
\bibfield{author}{\bibinfo{person}{Nelson Flores} {and} \bibinfo{person}{Jonathan Rosa}.} \bibinfo{year}{2015}\natexlab{}.
\newblock \showarticletitle{Undoing Appropriateness: Raciolinguistic Ideologies and Language Diversity in Education}.
\newblock \bibinfo{journal}{\emph{Harvard Educational Review}}  \bibinfo{volume}{85} (\bibinfo{date}{06} \bibinfo{year}{2015}), \bibinfo{pages}{149--171}.
\newblock
\href{https://doi.org/10.17763/0017-8055.85.2.149}{doi:\nolinkurl{10.17763/0017-8055.85.2.149}}


\bibitem[Funa and Gabay(2025)]%
        {funa}
\bibfield{author}{\bibinfo{person}{Aaron~A. Funa} {and} \bibinfo{person}{Renz Alvin~E. Gabay}.} \bibinfo{year}{2025}\natexlab{}.
\newblock \showarticletitle{Policy guidelines and recommendations on {AI} use in teaching and learning: {A} meta-synthesis study}.
\newblock \bibinfo{journal}{\emph{Social Sciences \& Humanities Open}}  \bibinfo{volume}{11} (\bibinfo{year}{2025}), \bibinfo{pages}{101221}.
\newblock
\showISSN{2590-2911}
\href{https://doi.org/10.1016/j.ssaho.2024.101221}{doi:\nolinkurl{10.1016/j.ssaho.2024.101221}}


\bibitem[Ge et~al\mbox{.}(2025)]%
        {yubin2025}
\bibfield{author}{\bibinfo{person}{Yubin Ge}, \bibinfo{person}{Neeraja Kirtane}, \bibinfo{person}{Hao Peng}, {and} \bibinfo{person}{Dilek Hakkani-Tür}.} \bibinfo{year}{2025}\natexlab{}.
\newblock \bibinfo{title}{LLMs are Vulnerable to Malicious Prompts Disguised as Scientific Language}.
\newblock
\showeprint[arxiv]{2501.14073}~[cs.CL]
\urldef\tempurl%
\url{https://arxiv.org/abs/2501.14073}
\showURL{%
\tempurl}


\bibitem[Gogus(2012)]%
        {bloom}
\bibfield{author}{\bibinfo{person}{Aytac Gogus}.} \bibinfo{year}{2012}\natexlab{}.
\newblock \showarticletitle{Bloom's {Taxonomy} of {Learning} {Objectives}}.
\newblock In \bibinfo{booktitle}{\emph{Encyclopedia of the {Sciences} of {Learning}}}, \bibfield{editor}{\bibinfo{person}{Norbert~M. Seel}} (Ed.). \bibinfo{publisher}{Springer US}, \bibinfo{address}{Boston, MA}, \bibinfo{pages}{469--473}.
\newblock
\showISBNx{978-1-4419-1428-6}
\href{https://doi.org/10.1007/978-1-4419-1428-6_141}{doi:\nolinkurl{10.1007/978-1-4419-1428-6_141}}


\bibitem[Goldberg(2022)]%
        {goldberg2022}
\bibfield{author}{\bibinfo{person}{Hagar Goldberg}.} \bibinfo{year}{2022}\natexlab{}.
\newblock \showarticletitle{Growing {Brains}, {Nurturing} {Minds}—{Neuroscience} as an {Educational} {Tool} to {Support} {Students}’ {Development} as {Life}-{Long} {Learners}}.
\newblock \bibinfo{journal}{\emph{Brain Sciences}} \bibinfo{volume}{12}, \bibinfo{number}{12} (\bibinfo{date}{Nov.} \bibinfo{year}{2022}), \bibinfo{pages}{1622}.
\newblock
\showISSN{2076-3425}
\href{https://doi.org/10.3390/brainsci12121622}{doi:\nolinkurl{10.3390/brainsci12121622}}
\newblock
\shownote{Publisher: MDPI AG}.


\bibitem[Gong et~al\mbox{.}(2025)]%
        {Gong}
\bibfield{author}{\bibinfo{person}{Liuying Gong}, \bibinfo{person}{Jingyuan Chen}, {and} \bibinfo{person}{Fei Wu}.} \bibinfo{year}{2025}\natexlab{}.
\newblock \showarticletitle{Is ChatGPT a Competent Teacher? Systematic Evaluation of Large Language Models on the Competency Model}.
\newblock \bibinfo{journal}{\emph{IEEE Trans. Learn. Technol.}}  \bibinfo{volume}{18} (\bibinfo{date}{April} \bibinfo{year}{2025}), \bibinfo{pages}{530–541}.
\newblock
\showISSN{1939-1382}
\href{https://doi.org/10.1109/TLT.2025.3564177}{doi:\nolinkurl{10.1109/TLT.2025.3564177}}


\bibitem[Guizani et~al\mbox{.}(2025)]%
        {guizani_systematic_2025}
\bibfield{author}{\bibinfo{person}{Sghaier Guizani}, \bibinfo{person}{Tehseen Mazhar}, \bibinfo{person}{Tariq Shahzad}, \bibinfo{person}{Wasim Ahmad}, \bibinfo{person}{Afsha Bibi}, {and} \bibinfo{person}{Habib Hamam}.} \bibinfo{year}{2025}\natexlab{}.
\newblock \showarticletitle{A systematic literature review to implement large language model in higher education: issues and solutions}.
\newblock \bibinfo{journal}{\emph{Discover Education}} \bibinfo{volume}{4}, \bibinfo{number}{1} (\bibinfo{date}{Feb.} \bibinfo{year}{2025}), \bibinfo{pages}{35}.
\newblock
\showISSN{2731-5525}
\href{https://doi.org/10.1007/s44217-025-00424-7}{doi:\nolinkurl{10.1007/s44217-025-00424-7}}


\bibitem[Güner and Er(2025)]%
        {Guner2025}
\bibfield{author}{\bibinfo{person}{Hacer Güner} {and} \bibinfo{person}{Erkan Er}.} \bibinfo{year}{2025}\natexlab{}.
\newblock \showarticletitle{AI in the classroom: Exploring students’ interaction with ChatGPT in programming learning}.
\newblock \bibinfo{journal}{\emph{Education and Information Technologies}}  \bibinfo{volume}{30} (\bibinfo{date}{01} \bibinfo{year}{2025}), \bibinfo{pages}{12681--12707}.
\newblock
\href{https://doi.org/10.1007/s10639-025-13337-7}{doi:\nolinkurl{10.1007/s10639-025-13337-7}}


\bibitem[Haindl and Weinberger(2024)]%
        {Haindl2024}
\bibfield{author}{\bibinfo{person}{Philipp Haindl} {and} \bibinfo{person}{Gerald Weinberger}.} \bibinfo{year}{2024}\natexlab{}.
\newblock \showarticletitle{Students’ Experiences of Using ChatGPT in an Undergraduate Programming Course}.
\newblock \bibinfo{journal}{\emph{IEEE Access}}  \bibinfo{volume}{PP} (\bibinfo{date}{01} \bibinfo{year}{2024}), \bibinfo{pages}{1--1}.
\newblock
\href{https://doi.org/10.1109/ACCESS.2024.3380909}{doi:\nolinkurl{10.1109/ACCESS.2024.3380909}}


\bibitem[Hamdan(2024)]%
        {Hamdam2024}
\bibfield{author}{\bibinfo{person}{Basil Hamdan}.} \bibinfo{year}{2024}\natexlab{}.
\newblock \showarticletitle{Integrating ChatGPT in Cybersecurity Education: Use Cases and Implications}.
\newblock \bibinfo{journal}{\emph{J. Comput. Sci. Coll.}} \bibinfo{volume}{40}, \bibinfo{number}{2} (\bibinfo{date}{Oct.} \bibinfo{year}{2024}), \bibinfo{pages}{105–114}.
\newblock
\showISSN{1937-4771}


\bibitem[Hartmann et~al\mbox{.}(2023)]%
        {hartmann2023}
\bibfield{author}{\bibinfo{person}{Jochen Hartmann}, \bibinfo{person}{Mark Heitmann}, \bibinfo{person}{Christian Siebert}, {and} \bibinfo{person}{Christina Schamp}.} \bibinfo{year}{2023}\natexlab{}.
\newblock \showarticletitle{More than a {Feeling}: {Accuracy} and {Application} of {Sentiment} {Analysis}}.
\newblock \bibinfo{journal}{\emph{International Journal of Research in Marketing}} \bibinfo{volume}{40}, \bibinfo{number}{1} (\bibinfo{year}{2023}), \bibinfo{pages}{75--87}.
\newblock
\showISSN{0167-8116}
\href{https://doi.org/10.1016/j.ijresmar.2022.05.005}{doi:\nolinkurl{10.1016/j.ijresmar.2022.05.005}}


\bibitem[Harvey et~al\mbox{.}(2025)]%
        {harvey}
\bibfield{author}{\bibinfo{person}{Emma Harvey}, \bibinfo{person}{Allison Koenecke}, {and} \bibinfo{person}{Rene~F. Kizilcec}.} \bibinfo{year}{2025}\natexlab{}.
\newblock \showarticletitle{"Don't Forget the Teachers": Towards an Educator-Centered Understanding of Harms from Large Language Models in Education}. In \bibinfo{booktitle}{\emph{Proceedings of the 2025 CHI Conference on Human Factors in Computing Systems}} \emph{(\bibinfo{series}{CHI '25})}. \bibinfo{publisher}{Association for Computing Machinery}, \bibinfo{address}{New York, NY, USA}, Article \bibinfo{articleno}{1064}, \bibinfo{numpages}{19}~pages.
\newblock
\showISBNx{9798400713941}
\href{https://doi.org/10.1145/3706598.3713210}{doi:\nolinkurl{10.1145/3706598.3713210}}


\bibitem[Hashemi and Mahdavirad(2023)]%
        {hashemi2023}
\bibfield{author}{\bibinfo{person}{Ali Hashemi} {and} \bibinfo{person}{Fatemeh Mahdavirad}.} \bibinfo{year}{2023}\natexlab{}.
\newblock \showarticletitle{A cross-cultural, cross-disciplinary, and cross-gender study on {Appraisal} resources in {PhD} dissertation abstracts: {Martin} \& {White}'s (2005) {Appraisal} {Theory} in focus}.
\newblock \bibinfo{journal}{\emph{Heliyon}} \bibinfo{volume}{9}, \bibinfo{number}{11} (\bibinfo{date}{Nov.} \bibinfo{year}{2023}), \bibinfo{pages}{e22074}.
\newblock
\showISSN{2405-8440}
\href{https://doi.org/10.1016/j.heliyon.2023.e22074}{doi:\nolinkurl{10.1016/j.heliyon.2023.e22074}}


\bibitem[Hernandez et~al\mbox{.}(2022)]%
        {hernandez2022}
\bibfield{author}{\bibinfo{person}{Evan Hernandez}, \bibinfo{person}{Sarah Schwettmann}, \bibinfo{person}{David Bau}, \bibinfo{person}{Teona Bagashvili}, \bibinfo{person}{Antonio Torralba}, {and} \bibinfo{person}{Jacob Andreas}.} \bibinfo{year}{2022}\natexlab{}.
\newblock \bibinfo{title}{Natural Language Descriptions of Deep Visual Features}.
\newblock
\showeprint[arxiv]{2201.11114}~[cs.CV]
\urldef\tempurl%
\url{https://arxiv.org/abs/2201.11114}
\showURL{%
\tempurl}


\bibitem[Hosseini et~al\mbox{.}(2025)]%
        {hosseini}
\bibfield{author}{\bibinfo{person}{Mohammad Hosseini}, \bibinfo{person}{Peng Gao}, {and} \bibinfo{person}{Carolina Vivas-Valencia}.} \bibinfo{year}{2025}\natexlab{}.
\newblock \showarticletitle{A social-environmental impact perspective of generative artificial intelligence}.
\newblock \bibinfo{journal}{\emph{Environmental Science and Ecotechnology}}  \bibinfo{volume}{23} (\bibinfo{year}{2025}), \bibinfo{pages}{100520}.
\newblock
\showISSN{2666-4984}
\href{https://doi.org/10.1016/j.ese.2024.100520}{doi:\nolinkurl{10.1016/j.ese.2024.100520}}


\bibitem[Hu and Rangwala(2020)]%
        {hu2020}
\bibfield{author}{\bibinfo{person}{Qian Hu} {and} \bibinfo{person}{Huzefa Rangwala}.} \bibinfo{year}{2020}\natexlab{}.
\newblock \showarticletitle{Towards Fair Educational Data Mining: A Case Study on Detecting At-Risk Students.}
\newblock \bibinfo{journal}{\emph{International Educational Data Mining Society}} \bibinfo{volume}{1}, \bibinfo{number}{1} (\bibinfo{year}{2020}), \bibinfo{pages}{1--7}.
\newblock


\bibitem[Huang(2023)]%
        {huang23}
\bibfield{author}{\bibinfo{person}{Huiman Huang}.} \bibinfo{year}{2023}\natexlab{}.
\newblock \showarticletitle{Performance of {ChatGPT} on {Registered} {Nurse} {License} {Exam} in {Taiwan}: {A} {Descriptive} {Study}}.
\newblock \bibinfo{journal}{\emph{Healthcare}} \bibinfo{volume}{11}, \bibinfo{number}{21} (\bibinfo{date}{Oct.} \bibinfo{year}{2023}), \bibinfo{pages}{2855}.
\newblock
\showISSN{2227-9032}
\href{https://doi.org/10.3390/healthcare11212855}{doi:\nolinkurl{10.3390/healthcare11212855}}


\bibitem[Idowu et~al\mbox{.}(2024)]%
        {Idowu2024}
\bibfield{author}{\bibinfo{person}{Jamiu Idowu}, \bibinfo{person}{Adriano Koshiyama}, {and} \bibinfo{person}{Philip Treleaven}.} \bibinfo{year}{2024}\natexlab{}.
\newblock \showarticletitle{Investigating Algorithmic Bias in Student Progress Monitoring}.
\newblock \bibinfo{journal}{\emph{Computers and Education: Artificial Intelligence}}  \bibinfo{volume}{7} (\bibinfo{date}{07} \bibinfo{year}{2024}), \bibinfo{pages}{100267}.
\newblock
\href{https://doi.org/10.1016/j.caeai.2024.100267}{doi:\nolinkurl{10.1016/j.caeai.2024.100267}}


\bibitem[Jiang et~al\mbox{.}(2024)]%
        {jiang2024}
\bibfield{author}{\bibinfo{person}{Tristan Jiang}, \bibinfo{person}{Elina Liu}, \bibinfo{person}{Tasawar Baig}, {and} \bibinfo{person}{Qingrong Li}.} \bibinfo{year}{2024}\natexlab{}.
\newblock \showarticletitle{Enhancing decision‐making in higher education: {Exploring} the integration of {ChatGPT} and data visualization tools in data analysis}.
\newblock \bibinfo{journal}{\emph{New Directions for Higher Education}} \bibinfo{volume}{2024}, \bibinfo{number}{207} (\bibinfo{date}{Sept.} \bibinfo{year}{2024}), \bibinfo{pages}{15--29}.
\newblock
\showISSN{0271-0560, 1536-0741}
\href{https://doi.org/10.1002/he.20510}{doi:\nolinkurl{10.1002/he.20510}}


\bibitem[Jose et~al\mbox{.}(2025)]%
        {jose_cognitive_2025}
\bibfield{author}{\bibinfo{person}{Binny Jose}, \bibinfo{person}{Jaya Cherian}, \bibinfo{person}{Alie~Molly Verghis}, \bibinfo{person}{Sony~Mary Varghise}, \bibinfo{person}{Mumthas S}, {and} \bibinfo{person}{Sibichan Joseph}.} \bibinfo{year}{2025}\natexlab{}.
\newblock \showarticletitle{The cognitive paradox of {AI} in education: between enhancement and erosion}.
\newblock \bibinfo{journal}{\emph{Frontiers in Psychology}}  \bibinfo{volume}{Volume 16 - 2025} (\bibinfo{year}{2025}), \bibinfo{pages}{xx--yy}.
\newblock
\showISSN{1664-1078}
\href{https://doi.org/10.3389/fpsyg.2025.1550621}{doi:\nolinkurl{10.3389/fpsyg.2025.1550621}}


\bibitem[Jošt et~al\mbox{.}(2024)]%
        {Jost2024}
\bibfield{author}{\bibinfo{person}{Gregor Jošt}, \bibinfo{person}{Viktor Taneski}, {and} \bibinfo{person}{Sašo Karakatič}.} \bibinfo{year}{2024}\natexlab{}.
\newblock \showarticletitle{The Impact of Large Language Models on Programming Education and Student Learning Outcomes}.
\newblock \bibinfo{journal}{\emph{Applied Sciences}}  \bibinfo{volume}{14} (\bibinfo{date}{05} \bibinfo{year}{2024}), \bibinfo{pages}{4115}.
\newblock
\href{https://doi.org/10.3390/app14104115}{doi:\nolinkurl{10.3390/app14104115}}


\bibitem[Kaliterna et~al\mbox{.}(2024)]%
        {kaliterna_2024}
\bibfield{author}{\bibinfo{person}{Mariano Kaliterna}, \bibinfo{person}{Marija Žuljević}, \bibinfo{person}{Luka Ursić}, \bibinfo{person}{Jakov Krka}, {and} \bibinfo{person}{Darko Duplancic}.} \bibinfo{year}{2024}\natexlab{}.
\newblock \showarticletitle{Testing the capacity of {Bard} and {ChatGPT} for writing essays on ethical dilemmas: {A} cross-sectional study}.
\newblock \bibinfo{journal}{\emph{Scientific Reports}}  \bibinfo{volume}{14} (\bibinfo{date}{Oct.} \bibinfo{year}{2024}), \bibinfo{pages}{to appear}.
\newblock
\href{https://doi.org/10.1038/s41598-024-77576-3}{doi:\nolinkurl{10.1038/s41598-024-77576-3}}


\bibitem[Kasneci et~al\mbox{.}(2023)]%
        {kasneci2023}
\bibfield{author}{\bibinfo{person}{Enkelejda Kasneci}, \bibinfo{person}{Kathrin Sessler}, \bibinfo{person}{Stefan Küchemann}, \bibinfo{person}{Maria Bannert}, \bibinfo{person}{Daryna Dementieva}, \bibinfo{person}{Frank Fischer}, \bibinfo{person}{Urs Gasser}, \bibinfo{person}{Georg Groh}, \bibinfo{person}{Stephan Günnemann}, \bibinfo{person}{Eyke Hüllermeier}, \bibinfo{person}{Stephan Krusche}, \bibinfo{person}{Gitta Kutyniok}, \bibinfo{person}{Tilman Michaeli}, \bibinfo{person}{Claudia Nerdel}, \bibinfo{person}{Jürgen Pfeffer}, \bibinfo{person}{Oleksandra Poquet}, \bibinfo{person}{Michael Sailer}, \bibinfo{person}{Albrecht Schmidt}, \bibinfo{person}{Tina Seidel}, \bibinfo{person}{Matthias Stadler}, \bibinfo{person}{Jochen Weller}, \bibinfo{person}{Jochen Kuhn}, {and} \bibinfo{person}{Gjergji Kasneci}.} \bibinfo{year}{2023}\natexlab{}.
\newblock \showarticletitle{{ChatGPT} for good? {On} opportunities and challenges of large language models for education}.
\newblock \bibinfo{journal}{\emph{Learning and Individual Differences}}  \bibinfo{volume}{103} (\bibinfo{year}{2023}), \bibinfo{pages}{102274}.
\newblock
\showISSN{1041-6080}
\href{https://doi.org/10.1016/j.lindif.2023.102274}{doi:\nolinkurl{10.1016/j.lindif.2023.102274}}


\bibitem[Kaufenberg-Lashua et~al\mbox{.}(2024)]%
        {kaufenberg}
\bibfield{author}{\bibinfo{person}{Meagan~M. Kaufenberg-Lashua}, \bibinfo{person}{Joseph~K. West}, \bibinfo{person}{Jaime~J. Kelly}, {and} \bibinfo{person}{Valeria~A. Stepanova}.} \bibinfo{year}{2024}\natexlab{}.
\newblock \showarticletitle{What {Does} {AI} {Think} a {Chemist} {Looks} {Like}? {An} {Analysis} of {Diversity} in {Generative} {AI}}.
\newblock \bibinfo{journal}{\emph{Journal of Chemical Education}} \bibinfo{volume}{101}, \bibinfo{number}{11} (\bibinfo{year}{2024}), \bibinfo{pages}{4704--4713}.
\newblock
\href{https://doi.org/10.1021/acs.jchemed.4c00249}{doi:\nolinkurl{10.1021/acs.jchemed.4c00249}}
\newblock
\shownote{\_eprint: https://doi.org/10.1021/acs.jchemed.4c00249}.


\bibitem[Keith et~al\mbox{.}(2025)]%
        {keith_harnessing_2025}
\bibfield{author}{\bibinfo{person}{Matthew Keith}, \bibinfo{person}{Eleanor Keiller}, \bibinfo{person}{Christopher Windows-Yule}, \bibinfo{person}{Iain Kings}, {and} \bibinfo{person}{Phillip Robbins}.} \bibinfo{year}{2025}\natexlab{}.
\newblock \showarticletitle{Harnessing generative {AI} in chemical engineering education: {Implementation} and evaluation of the large language model {ChatGPT} v3.5}.
\newblock \bibinfo{journal}{\emph{Education for Chemical Engineers}}  \bibinfo{volume}{51} (\bibinfo{date}{April} \bibinfo{year}{2025}), \bibinfo{pages}{20--33}.
\newblock
\showISSN{17497728}
\href{https://doi.org/10.1016/j.ece.2025.01.002}{doi:\nolinkurl{10.1016/j.ece.2025.01.002}}


\bibitem[Klayklung et~al\mbox{.}(2023)]%
        {Klayklung}
\bibfield{author}{\bibinfo{person}{Prapasiri Klayklung}, \bibinfo{person}{Piyawatjana Chocksathaporn}, \bibinfo{person}{Pongsakorn Limna}, \bibinfo{person}{Tanpat Kraiwanit}, {and} \bibinfo{person}{Kris Jangjarat}.} \bibinfo{year}{2023}\natexlab{}.
\newblock \showarticletitle{Revolutionizing Education with ChatGPT: Enhancing Learning Through Conversational AI}.
\newblock \bibinfo{journal}{\emph{Journal of Language Teaching and Research}}  \bibinfo{volume}{2} (\bibinfo{date}{09} \bibinfo{year}{2023}), \bibinfo{pages}{217--225}.
\newblock


\bibitem[Kooli and Yusuf(2025)]%
        {kooli}
\bibfield{author}{\bibinfo{person}{Chokri Kooli} {and} \bibinfo{person}{Nadia Yusuf}.} \bibinfo{year}{2025}\natexlab{}.
\newblock \showarticletitle{Transforming {Educational} {Assessment}: {Insights} {Into} the {Use} of {ChatGPT} and {Large} {Language} {Models} in {Grading}}.
\newblock \bibinfo{journal}{\emph{International Journal of Human–Computer Interaction}} \bibinfo{volume}{41}, \bibinfo{number}{5} (\bibinfo{year}{2025}), \bibinfo{pages}{3388--3399}.
\newblock
\href{https://doi.org/10.1080/10447318.2024.2338330}{doi:\nolinkurl{10.1080/10447318.2024.2338330}}
\newblock
\shownote{Publisher: Taylor \& Francis \_eprint: https://doi.org/10.1080/10447318.2024.2338330}.


\bibitem[Korpimies et~al\mbox{.}(2024)]%
        {Korpimies_2024}
\bibfield{author}{\bibinfo{person}{Kai Korpimies}, \bibinfo{person}{Antti Laaksonen}, {and} \bibinfo{person}{Matti Luukkainen}.} \bibinfo{year}{2024}\natexlab{}.
\newblock \showarticletitle{Unrestricted Use of LLMs in a Software Project Course: Student Perceptions on Learning and Impact on Course Performance}. In \bibinfo{booktitle}{\emph{Proceedings of the 24th Koli Calling International Conference on Computing Education Research}} \emph{(\bibinfo{series}{Koli Calling '24})}. \bibinfo{publisher}{Association for Computing Machinery}, \bibinfo{address}{New York, NY, USA}, Article \bibinfo{articleno}{23}, \bibinfo{numpages}{7}~pages.
\newblock
\showISBNx{9798400710384}
\href{https://doi.org/10.1145/3699538.3699541}{doi:\nolinkurl{10.1145/3699538.3699541}}


\bibitem[Kosmyna et~al\mbox{.}(2025)]%
        {Kosmyna_2025}
\bibfield{author}{\bibinfo{person}{Nataliya Kosmyna}, \bibinfo{person}{Eugene Hauptmann}, \bibinfo{person}{Ye Yuan}, \bibinfo{person}{Jessica Situ}, \bibinfo{person}{Xian-Hao Liao}, \bibinfo{person}{Ashly Beresnitzky}, \bibinfo{person}{Iris Braunstein}, {and} \bibinfo{person}{Pattie Maes}.} \bibinfo{year}{2025}\natexlab{}.
\newblock \showarticletitle{Your Brain on ChatGPT: Accumulation of Cognitive Debt when Using an AI Assistant for Essay Writing Task}.
\newblock \bibinfo{journal}{\emph{arXiv preprint}} \bibinfo{volume}{1}, \bibinfo{number}{1} (\bibinfo{date}{06} \bibinfo{year}{2025}).
\newblock
\href{https://doi.org/10.48550/arXiv.2506.08872}{doi:\nolinkurl{10.48550/arXiv.2506.08872}}


\bibitem[Kwak and Pardos(2024)]%
        {kwak2024}
\bibfield{author}{\bibinfo{person}{Yerin Kwak} {and} \bibinfo{person}{Zachary~A. Pardos}.} \bibinfo{year}{2024}\natexlab{}.
\newblock \showarticletitle{Bridging large language model disparities: {Skill} tagging of multilingual educational content}.
\newblock \bibinfo{journal}{\emph{British Journal of Educational Technology}} \bibinfo{volume}{55}, \bibinfo{number}{5} (\bibinfo{date}{Sept.} \bibinfo{year}{2024}), \bibinfo{pages}{2039--2057}.
\newblock
\showISSN{0007-1013, 1467-8535}
\href{https://doi.org/10.1111/bjet.13465}{doi:\nolinkurl{10.1111/bjet.13465}}


\bibitem[Kwet(2019)]%
        {kwet}
\bibfield{author}{\bibinfo{person}{Michael Kwet}.} \bibinfo{year}{2019}\natexlab{}.
\newblock \bibinfo{title}{Digital colonialism is threatening the {Global} {South}}.
\newblock
\urldef\tempurl%
\url{https://www.aljazeera.com/opinions/2019/3/13/digital-colonialism-is-threatening-the-global-south}
\showURL{%
\tempurl}


\bibitem[Lee et~al\mbox{.}(2025)]%
        {lee}
\bibfield{author}{\bibinfo{person}{Hao-Ping~(Hank) Lee}, \bibinfo{person}{Advait Sarkar}, \bibinfo{person}{Lev Tankelevitch}, \bibinfo{person}{Ian Drosos}, \bibinfo{person}{Sean Rintel}, \bibinfo{person}{Richard Banks}, {and} \bibinfo{person}{Nicholas Wilson}.} \bibinfo{year}{2025}\natexlab{}.
\newblock \showarticletitle{The {Impact} of {Generative} {AI} on {Critical} {Thinking}: {Self}-{Reported} {Reductions} in {Cognitive} {Effort} and {Confidence} {Effects} {From} a {Survey} of {Knowledge} {Workers}}. In \bibinfo{booktitle}{\emph{{CHI} 2025}}. \bibinfo{publisher}{ACM}, \bibinfo{address}{Honolulu, HI, USA}, \bibinfo{pages}{1--22}.
\newblock
\urldef\tempurl%
\url{https://www.microsoft.com/en-us/research/publication/the-impact-of-generative-ai-on-critical-thinking-self-reported-reductions-in-cognitive-effort-and-confidence-effects-from-a-survey-of-knowledge-workers/}
\showURL{%
\tempurl}


\bibitem[Li et~al\mbox{.}(2024)]%
        {Li}
\bibfield{author}{\bibinfo{person}{Kangkang Li}, \bibinfo{person}{Qian Yang}, {and} \bibinfo{person}{Xianmin Yang}.} \bibinfo{year}{2024}\natexlab{}.
\newblock \showarticletitle{Can Autograding of Student-Generated Questions Quality by ChatGPT Match Human Experts?}
\newblock \bibinfo{journal}{\emph{IEEE Transactions on Learning Technologies}}  \bibinfo{volume}{17} (\bibinfo{year}{2024}), \bibinfo{pages}{1574--1584}.
\newblock
\href{https://doi.org/10.1109/TLT.2024.3394807}{doi:\nolinkurl{10.1109/TLT.2024.3394807}}


\bibitem[Lim(2024)]%
        {Lim}
\bibfield{author}{\bibinfo{person}{May Lim}.} \bibinfo{year}{2024}\natexlab{}.
\newblock \showarticletitle{WIP: Just-in-Time AI Assisted Formative Feedback for Written, Oral, Team-Based Assessment Tasks: What Worked, What Didn't and Why}. In \bibinfo{booktitle}{\emph{2024 IEEE Frontiers in Education Conference (FIE)}}. \bibinfo{publisher}{IEEE}, \bibinfo{address}{USA}, \bibinfo{pages}{1--4}.
\newblock
\href{https://doi.org/10.1109/FIE61694.2024.10893330}{doi:\nolinkurl{10.1109/FIE61694.2024.10893330}}


\bibitem[Lin and Crosthwaite(2024)]%
        {lin2024}
\bibfield{author}{\bibinfo{person}{Shiming Lin} {and} \bibinfo{person}{Peter Crosthwaite}.} \bibinfo{year}{2024}\natexlab{}.
\newblock \showarticletitle{The grass is not always greener: {Teacher} vs. {GPT}-assisted written corrective feedback}.
\newblock \bibinfo{journal}{\emph{System}}  \bibinfo{volume}{127} (\bibinfo{year}{2024}), \bibinfo{pages}{103529}.
\newblock
\showISSN{0346-251X}
\href{https://doi.org/10.1016/j.system.2024.103529}{doi:\nolinkurl{10.1016/j.system.2024.103529}}


\bibitem[Liu and Yang(2024)]%
        {liu_application_2024}
\bibfield{author}{\bibinfo{person}{Chao Liu} {and} \bibinfo{person}{Shengyi Yang}.} \bibinfo{year}{2024}\natexlab{}.
\newblock \showarticletitle{Application of large language models in engineering education: {A} case study of system modeling and simulation courses}.
\newblock \bibinfo{journal}{\emph{International Journal of Mechanical Engineering Education}}  \bibinfo{volume}{Advance online publication} (\bibinfo{date}{Aug.} \bibinfo{year}{2024}), \bibinfo{pages}{03064190241272728}.
\newblock
\showISSN{0306-4190, 2050-4586}
\href{https://doi.org/10.1177/03064190241272728}{doi:\nolinkurl{10.1177/03064190241272728}}


\bibitem[Liu et~al\mbox{.}(2024)]%
        {liu2024}
\bibfield{author}{\bibinfo{person}{Hanchao Liu}, \bibinfo{person}{Wenyuan Xue}, \bibinfo{person}{Yifei Chen}, \bibinfo{person}{Dapeng Chen}, \bibinfo{person}{Xiutian Zhao}, \bibinfo{person}{Ke Wang}, \bibinfo{person}{Liping Hou}, \bibinfo{person}{Rongjun Li}, {and} \bibinfo{person}{Wei Peng}.} \bibinfo{year}{2024}\natexlab{}.
\newblock \bibinfo{title}{A Survey on Hallucination in Large Vision-Language Models}.
\newblock
\showeprint[arxiv]{2402.00253}~[cs.CV]
\urldef\tempurl%
\url{https://arxiv.org/abs/2402.00253}
\showURL{%
\tempurl}


\bibitem[Lo(2023)]%
        {lo2023}
\bibfield{author}{\bibinfo{person}{Leo~S. Lo}.} \bibinfo{year}{2023}\natexlab{}.
\newblock \showarticletitle{The {CLEAR} path: {A} framework for enhancing information literacy through prompt engineering}.
\newblock \bibinfo{journal}{\emph{The Journal of Academic Librarianship}} \bibinfo{volume}{49}, \bibinfo{number}{4} (\bibinfo{year}{2023}), \bibinfo{pages}{102720}.
\newblock
\showISSN{0099-1333}
\href{https://doi.org/10.1016/j.acalib.2023.102720}{doi:\nolinkurl{10.1016/j.acalib.2023.102720}}


\bibitem[Maher et~al\mbox{.}(2023)]%
        {Maher2023}
\bibfield{author}{\bibinfo{person}{Mary~Lou Maher}, \bibinfo{person}{Sri~Yash Tadimalla}, {and} \bibinfo{person}{Dhruv Dhamani}.} \bibinfo{year}{2023}\natexlab{}.
\newblock \showarticletitle{An Exploratory Study on the Impact of AI tools on the Student Experience in Programming Courses: an Intersectional Analysis Approach}. In \bibinfo{booktitle}{\emph{2023 IEEE Frontiers in Education Conference (FIE)}}. \bibinfo{publisher}{IEEE}, \bibinfo{address}{Piscataway, NJ, USA}, \bibinfo{pages}{1--5}.
\newblock
\href{https://doi.org/10.1109/FIE58773.2023.10343037}{doi:\nolinkurl{10.1109/FIE58773.2023.10343037}}


\bibitem[Malinka et~al\mbox{.}(2023)]%
        {malinka2023}
\bibfield{author}{\bibinfo{person}{Kamil Malinka}, \bibinfo{person}{Martin Peresíni}, \bibinfo{person}{Anton Firc}, \bibinfo{person}{Ondrej Hujnák}, {and} \bibinfo{person}{Filip Janus}.} \bibinfo{year}{2023}\natexlab{}.
\newblock \showarticletitle{On the {Educational} {Impact} of {ChatGPT}: {Is} {Artificial} {Intelligence} {Ready} to {Obtain} a {University} {Degree}?}. In \bibinfo{booktitle}{\emph{Proceedings of the 2023 {Conference} on {Innovation} and {Technology} in {Computer} {Science} {Education} {V}. 1}}. \bibinfo{publisher}{ACM}, \bibinfo{address}{Turku Finland}, \bibinfo{pages}{47--53}.
\newblock
\showISBNx{979-8-4007-0138-2}
\href{https://doi.org/10.1145/3587102.3588827}{doi:\nolinkurl{10.1145/3587102.3588827}}


\bibitem[McGee(2023)]%
        {mcgee2023}
\bibfield{author}{\bibinfo{person}{Robert~W. McGee}.} \bibinfo{year}{2023}\natexlab{}.
\newblock \showarticletitle{Is {Chat} {Gpt} {Biased} {Against} {Conservatives}? {An} {Empirical} {Study}}.
\newblock \bibinfo{journal}{\emph{SSRN Electronic Journal}} \bibinfo{volume}{na}, \bibinfo{number}{na} (\bibinfo{year}{2023}), \bibinfo{pages}{1--19}.
\newblock
\showISSN{1556-5068}
\href{https://doi.org/10.2139/ssrn.4359405}{doi:\nolinkurl{10.2139/ssrn.4359405}}


\bibitem[Mormul et~al\mbox{.}(2024)]%
        {Mormul2024}
\bibfield{author}{\bibinfo{person}{Yevhenii Mormul}, \bibinfo{person}{Jan Przybyszewski}, \bibinfo{person}{Andrew Nakoud}, {and} \bibinfo{person}{Paul Cuffe}.} \bibinfo{year}{2024}\natexlab{}.
\newblock \showarticletitle{Reliance on Artificial Intelligence Tools May Displace Research Skills Acquisition Within Engineering Doctoral Programmes: Examples and Implications}. In \bibinfo{booktitle}{\emph{2024 21st International Conference on Information Technology Based Higher Education and Training (ITHET)}}. \bibinfo{publisher}{IEEE}, \bibinfo{address}{Location not specified}, \bibinfo{pages}{1--10}.
\newblock
\href{https://doi.org/10.1109/ITHET61869.2024.10837618}{doi:\nolinkurl{10.1109/ITHET61869.2024.10837618}}


\bibitem[Nguyen(2023)]%
        {Nguyen}
\bibfield{author}{\bibinfo{person}{Nathan Nguyen}.} \bibinfo{year}{2023}\natexlab{}.
\newblock \showarticletitle{Exploring the role of AI in education}.
\newblock \bibinfo{journal}{\emph{London Journal of Social Sciences}} \bibinfo{volume}{1}, \bibinfo{number}{1} (\bibinfo{date}{09} \bibinfo{year}{2023}), \bibinfo{pages}{84--95}.
\newblock
\href{https://doi.org/10.31039/ljss.2023.6.108}{doi:\nolinkurl{10.31039/ljss.2023.6.108}}


\bibitem[OpenAI et~al\mbox{.}(2023)]%
        {Achiam2023}
\bibfield{author}{\bibinfo{person}{OpenAI}, \bibinfo{person}{Josh Achiam}, \bibinfo{person}{Steven Adler}, \bibinfo{person}{Sandhini Agarwal}, \bibinfo{person}{Lama Ahmad}, \bibinfo{person}{Ilge Akkaya}, \bibinfo{person}{Florencia Aleman}, \bibinfo{person}{Diogo Almeida}, \bibinfo{person}{Janko Altenschmidt}, \bibinfo{person}{Sam Altman}, \bibinfo{person}{Shyamal Anadkat}, \bibinfo{person}{Red Avila}, \bibinfo{person}{Igor Babuschkin}, \bibinfo{person}{Suchir Balaji}, \bibinfo{person}{Valerie Balcom}, \bibinfo{person}{Paul Baltescu}, \bibinfo{person}{Haiming Bao}, \bibinfo{person}{Mohammad Bavarian}, \bibinfo{person}{Jeff Belgum}, {and} \bibinfo{person}{Barret Zoph}.} \bibinfo{year}{2023}\natexlab{}.
\newblock \showarticletitle{GPT-4 Technical Report}.
\newblock \bibinfo{journal}{\emph{arXiv preprint}}  \bibinfo{volume}{N/A} (\bibinfo{date}{03} \bibinfo{year}{2023}).
\newblock
\href{https://doi.org/10.48550/arXiv.2303.08774}{doi:\nolinkurl{10.48550/arXiv.2303.08774}}


\bibitem[Ouzzani et~al\mbox{.}(2016)]%
        {rayaan}
\bibfield{author}{\bibinfo{person}{Mourad Ouzzani}, \bibinfo{person}{Hossam Hammady}, \bibinfo{person}{Zbys Fedorowicz}, {and} \bibinfo{person}{Ahmed Elmagarmid}.} \bibinfo{year}{2016}\natexlab{}.
\newblock \showarticletitle{Rayyan—a web and mobile app for systematic reviews}.
\newblock \bibinfo{journal}{\emph{Systematic Reviews}} \bibinfo{volume}{5}, \bibinfo{number}{1} (\bibinfo{date}{Dec.} \bibinfo{year}{2016}), \bibinfo{pages}{210}.
\newblock
\showISSN{2046-4053}
\href{https://doi.org/10.1186/s13643-016-0384-4}{doi:\nolinkurl{10.1186/s13643-016-0384-4}}


\bibitem[Page et~al\mbox{.}(2021)]%
        {page_prisma_2021}
\bibfield{author}{\bibinfo{person}{Matthew~J Page}, \bibinfo{person}{Joanne~E McKenzie}, \bibinfo{person}{Patrick~M Bossuyt}, \bibinfo{person}{Isabelle Boutron}, \bibinfo{person}{Tammy~C Hoffmann}, \bibinfo{person}{Cynthia~D Mulrow}, \bibinfo{person}{Larissa Shamseer}, \bibinfo{person}{Jennifer~M Tetzlaff}, \bibinfo{person}{Elie~A Akl}, \bibinfo{person}{Sue~E Brennan}, \bibinfo{person}{Roger Chou}, \bibinfo{person}{Julie Glanville}, \bibinfo{person}{Jeremy~M Grimshaw}, \bibinfo{person}{Asbjørn Hróbjartsson}, \bibinfo{person}{Manoj~M Lalu}, \bibinfo{person}{Tianjing Li}, \bibinfo{person}{Elizabeth~W Loder}, \bibinfo{person}{Evan Mayo-Wilson}, \bibinfo{person}{Steve McDonald}, \bibinfo{person}{Luke~A McGuinness}, \bibinfo{person}{Lesley~A Stewart}, \bibinfo{person}{James Thomas}, \bibinfo{person}{Andrea~C Tricco}, \bibinfo{person}{Vivian~A Welch}, \bibinfo{person}{Penny Whiting}, {and} \bibinfo{person}{David Moher}.} \bibinfo{year}{2021}\natexlab{}.
\newblock \showarticletitle{The {PRISMA} 2020 statement: an updated guideline for reporting systematic reviews}.
\newblock \bibinfo{journal}{\emph{BMJ}} \bibinfo{volume}{372}, \bibinfo{number}{n71} (\bibinfo{date}{March} \bibinfo{year}{2021}), \bibinfo{pages}{n71}.
\newblock
\showISSN{1756-1833}
\href{https://doi.org/10.1136/bmj.n71}{doi:\nolinkurl{10.1136/bmj.n71}}


\bibitem[Papi and Khajavy(2021)]%
        {papi2021}
\bibfield{author}{\bibinfo{person}{Mostafa Papi} {and} \bibinfo{person}{Gholam~Hassan Khajavy}.} \bibinfo{year}{2021}\natexlab{}.
\newblock \showarticletitle{Motivational {Mechanisms} {Underlying} {Second} {Language} {Achievement}: {A} {Regulatory} {Focus} {Perspective}}.
\newblock \bibinfo{journal}{\emph{Language Learning}} \bibinfo{volume}{71}, \bibinfo{number}{2} (\bibinfo{date}{June} \bibinfo{year}{2021}), \bibinfo{pages}{537--572}.
\newblock
\showISSN{0023-8333, 1467-9922}
\href{https://doi.org/10.1111/lang.12443}{doi:\nolinkurl{10.1111/lang.12443}}
\newblock
\shownote{Publisher: Wiley}.


\bibitem[Peláez-Sánchez et~al\mbox{.}(2024)]%
        {pelaez_2024}
\bibfield{author}{\bibinfo{person}{Cristina Peláez-Sánchez}, \bibinfo{person}{Davis Velarde~Camaqui}, {and} \bibinfo{person}{Leonardo Glasserman-Morales}.} \bibinfo{year}{2024}\natexlab{}.
\newblock \showarticletitle{The impact of large language models on higher education: exploring the connection between {AI} and {Education} 4.0}.
\newblock \bibinfo{journal}{\emph{Frontiers in Education}}  \bibinfo{volume}{9} (\bibinfo{date}{June} \bibinfo{year}{2024}), \bibinfo{pages}{to appear}.
\newblock
\href{https://doi.org/10.3389/feduc.2024.1392091}{doi:\nolinkurl{10.3389/feduc.2024.1392091}}


\bibitem[Pereira and Ferreira~Mello(2025)]%
        {Pereira2025}
\bibfield{author}{\bibinfo{person}{Andre~Fabiano Pereira} {and} \bibinfo{person}{Rafael Ferreira~Mello}.} \bibinfo{year}{2025}\natexlab{}.
\newblock \showarticletitle{A Systematic Literature Review on Large Language Models Applications in Computer Programming Teaching Evaluation Process}.
\newblock \bibinfo{journal}{\emph{IEEE Access}}  \bibinfo{volume}{13} (\bibinfo{year}{2025}), \bibinfo{pages}{113449--113460}.
\newblock
\href{https://doi.org/10.1109/ACCESS.2025.3584060}{doi:\nolinkurl{10.1109/ACCESS.2025.3584060}}


\bibitem[Perrigo(2023)]%
        {perrigo2023}
\bibfield{author}{\bibinfo{person}{Billy Perrigo}.} \bibinfo{year}{2023}\natexlab{}.
\newblock \bibinfo{title}{Exclusive: {The} \$2 {Per} {Hour} {Workers} {Who} {Made} {ChatGPT} {Safer}}.
\newblock
\urldef\tempurl%
\url{https://time.com/6247678/openai-chatgpt-kenya-workers/}
\showURL{%
\tempurl}


\bibitem[Perry(1981)]%
        {perry}
\bibfield{author}{\bibinfo{person}{WGJ Perry}.} \bibinfo{year}{1981}\natexlab{}.
\newblock \showarticletitle{Cognitive and ethical growth: the making of meaning, the modern American college}.
\newblock \bibinfo{journal}{\emph{Ed. Chickering AW San Francisco: Jossey-Boss}} \bibinfo{volume}{1}, \bibinfo{number}{1} (\bibinfo{year}{1981}), \bibinfo{pages}{1--10}.
\newblock


\bibitem[Piaget(1971)]%
        {piaget_1971}
\bibfield{author}{\bibinfo{person}{Jean Piaget}.} \bibinfo{year}{1971}\natexlab{}.
\newblock \showarticletitle{The theory of stages in cognitive development}.
\newblock In \bibinfo{booktitle}{\emph{Measurement and {Piaget}}}. \bibinfo{publisher}{McGraw-Hill}, \bibinfo{address}{New York, NY, US}, \bibinfo{pages}{ix, 283--ix, 283}.
\newblock


\bibitem[Price et~al\mbox{.}(2024)]%
        {price2024}
\bibfield{author}{\bibinfo{person}{Madeline~Day Price}, \bibinfo{person}{Erin Smith}, {and} \bibinfo{person}{R.~Alex Smith}.} \bibinfo{year}{2024}\natexlab{}.
\newblock \showarticletitle{“{Exceptional} {Talent} and {Enthusiasm} for {Math}”: {An} {Examination} of {Storylines} {Circulated} by {ChatGPT} about {Mathematical} {Learners}}.
\newblock \bibinfo{journal}{\emph{International Journal of Education in Mathematics, Science and Technology}} \bibinfo{volume}{12}, \bibinfo{number}{6} (\bibinfo{date}{Sept.} \bibinfo{year}{2024}), \bibinfo{pages}{1620--1637}.
\newblock
\showISSN{2147-611X}
\href{https://doi.org/10.46328/ijemst.4461}{doi:\nolinkurl{10.46328/ijemst.4461}}


\bibitem[Pudari and Ernst(2023)]%
        {Pudari2023}
\bibfield{author}{\bibinfo{person}{Rohith Pudari} {and} \bibinfo{person}{Neil Ernst}.} \bibinfo{year}{2023}\natexlab{}.
\newblock \showarticletitle{From Copilot to Pilot: Towards AI Supported Software Development}.
\newblock \bibinfo{journal}{\emph{arXiv}}  \bibinfo{volume}{N/A} (\bibinfo{date}{03} \bibinfo{year}{2023}).
\newblock
\href{https://doi.org/10.48550/arXiv.2303.04142}{doi:\nolinkurl{10.48550/arXiv.2303.04142}}


\bibitem[Qin(2024)]%
        {qin2024}
\bibfield{author}{\bibinfo{person}{Ying Qin}.} \bibinfo{year}{2024}\natexlab{}.
\newblock \showarticletitle{Is {ChatGPT} {Reliable} in {Scoring} {Learner}'s {Translation} {Quality}?}. In \bibinfo{booktitle}{\emph{2024 6th {International} {Conference} on {Computer} {Science} and {Technologies} in {Education} ({CSTE})}}. \bibinfo{publisher}{IEEE}, \bibinfo{address}{Xi'an, China}, \bibinfo{pages}{146--149}.
\newblock
\showISBNx{979-8-3503-5180-4}
\href{https://doi.org/10.1109/CSTE62025.2024.00034}{doi:\nolinkurl{10.1109/CSTE62025.2024.00034}}


\bibitem[Radtke and Rummel(2024)]%
        {Radtke2024}
\bibfield{author}{\bibinfo{person}{Anna Radtke} {and} \bibinfo{person}{Nikol Rummel}.} \bibinfo{year}{2024}\natexlab{}.
\newblock \showarticletitle{Generative AI in academic writing: Does information on authorship impact learners’ revision behavior?}
\newblock \bibinfo{journal}{\emph{Computers and Education: Artificial Intelligence}}  \bibinfo{volume}{8} (\bibinfo{date}{12} \bibinfo{year}{2024}), \bibinfo{pages}{100350}.
\newblock
\href{https://doi.org/10.1016/j.caeai.2024.100350}{doi:\nolinkurl{10.1016/j.caeai.2024.100350}}


\bibitem[Raihan et~al\mbox{.}(2025)]%
        {Raihan2025}
\bibfield{author}{\bibinfo{person}{Nishat Raihan}, \bibinfo{person}{Mohammed~Latif Siddiq}, \bibinfo{person}{Joanna~C.S. Santos}, {and} \bibinfo{person}{Marcos Zampieri}.} \bibinfo{year}{2025}\natexlab{}.
\newblock \showarticletitle{Large Language Models in Computer Science Education: A Systematic Literature Review}. In \bibinfo{booktitle}{\emph{Proceedings of the 56th ACM Technical Symposium on Computer Science Education V. 1}} (Pittsburgh, PA, USA) \emph{(\bibinfo{series}{SIGCSETS 2025})}. \bibinfo{publisher}{Association for Computing Machinery}, \bibinfo{address}{New York, NY, USA}, \bibinfo{pages}{938–944}.
\newblock
\showISBNx{9798400705311}
\href{https://doi.org/10.1145/3641554.3701863}{doi:\nolinkurl{10.1145/3641554.3701863}}


\bibitem[Ramaswamy et~al\mbox{.}(2020)]%
        {Ramaswamy}
\bibfield{author}{\bibinfo{person}{Swaroop Ramaswamy}, \bibinfo{person}{Om Thakkar}, \bibinfo{person}{Rajiv Mathews}, \bibinfo{person}{Galen Andrew}, \bibinfo{person}{H.~Brendan McMahan}, {and} \bibinfo{person}{Françoise Beaufays}.} \bibinfo{year}{2020}\natexlab{}.
\newblock \bibinfo{title}{Training Production Language Models without Memorizing User Data}.
\newblock
\showeprint[arxiv]{2009.10031}~[cs.LG]
\urldef\tempurl%
\url{https://arxiv.org/abs/2009.10031}
\showURL{%
\tempurl}


\bibitem[Ryan and Deci(2000)]%
        {ryan}
\bibfield{author}{\bibinfo{person}{Richard~M. Ryan} {and} \bibinfo{person}{Edward~L. Deci}.} \bibinfo{year}{2000}\natexlab{}.
\newblock \showarticletitle{Self-determination theory and the facilitation of intrinsic motivation, social development, and well-being.}
\newblock \bibinfo{journal}{\emph{American Psychologist}} \bibinfo{volume}{55}, \bibinfo{number}{1} (\bibinfo{year}{2000}), \bibinfo{pages}{68--78}.
\newblock
\showISSN{1935-990X, 0003-066X}
\href{https://doi.org/10.1037/0003-066X.55.1.68}{doi:\nolinkurl{10.1037/0003-066X.55.1.68}}


\bibitem[Rzepka et~al\mbox{.}(2022)]%
        {rzepka_fairness_2022}
\bibfield{author}{\bibinfo{person}{Nathalie Rzepka}, \bibinfo{person}{Katharina Simbeck}, \bibinfo{person}{Hans-Georg Müller}, {and} \bibinfo{person}{Niels Pinkwart}.} \bibinfo{year}{2022}\natexlab{}.
\newblock \showarticletitle{Fairness of {In}-session {Dropout} {Prediction}:}. In \bibinfo{booktitle}{\emph{Proceedings of the 14th {International} {Conference} on {Computer} {Supported} {Education}}}. \bibinfo{publisher}{SCITEPRESS - Science and Technology Publications}, \bibinfo{address}{Online Streaming, --- Select a Country ---}, \bibinfo{pages}{316--326}.
\newblock
\showISBNx{978-989-758-562-3}
\href{https://doi.org/10.5220/0010962100003182}{doi:\nolinkurl{10.5220/0010962100003182}}


\bibitem[Sachs(2023)]%
        {sachs2023}
\bibfield{author}{\bibinfo{person}{Goldman Sachs}.} \bibinfo{year}{2023}\natexlab{}.
\newblock \showarticletitle{AI poised to drive 160\% increase in power demand}.
\newblock \bibinfo{journal}{\emph{Goldman Sachs Intelligence}} \bibinfo{volume}{X}, \bibinfo{number}{Y} (\bibinfo{year}{2023}), \bibinfo{pages}{1--2}.
\newblock


\bibitem[Salinas et~al\mbox{.}(2025)]%
        {salinas}
\bibfield{author}{\bibinfo{person}{Alejandro Salinas}, \bibinfo{person}{Amit Haim}, {and} \bibinfo{person}{Julian Nyarko}.} \bibinfo{year}{2025}\natexlab{}.
\newblock \bibinfo{title}{What's in a Name? Auditing Large Language Models for Race and Gender Bias}.
\newblock
\showeprint[arxiv]{2402.14875}~[cs.CL]
\urldef\tempurl%
\url{https://arxiv.org/abs/2402.14875}
\showURL{%
\tempurl}


\bibitem[Sathish et~al\mbox{.}(2024)]%
        {sathish2024}
\bibfield{author}{\bibinfo{person}{Vishwas Sathish}, \bibinfo{person}{Hannah Lin}, \bibinfo{person}{Aditya~K Kamath}, {and} \bibinfo{person}{Anish Nyayachavadi}.} \bibinfo{year}{2024}\natexlab{}.
\newblock \bibinfo{title}{LLeMpower: Understanding Disparities in the Control and Access of Large Language Models}.
\newblock
\showeprint[arxiv]{2404.09356}~[cs.CY]
\urldef\tempurl%
\url{https://arxiv.org/abs/2404.09356}
\showURL{%
\tempurl}


\bibitem[Shi et~al\mbox{.}(2025)]%
        {shi2025}
\bibfield{author}{\bibinfo{person}{Huawei Shi}, \bibinfo{person}{Ching~Sing Chai}, \bibinfo{person}{Sihan Zhou}, {and} \bibinfo{person}{Scott Aubrey}.} \bibinfo{year}{2025}\natexlab{}.
\newblock \showarticletitle{Comparing the effects of {ChatGPT} and automated writing evaluation on students’ writing and ideal {L2} writing self}.
\newblock \bibinfo{journal}{\emph{Computer Assisted Language Learning}} \bibinfo{volume}{XX}, \bibinfo{number}{YY} (\bibinfo{date}{Feb.} \bibinfo{year}{2025}), \bibinfo{pages}{1--28}.
\newblock
\showISSN{0958-8221, 1744-3210}
\href{https://doi.org/10.1080/09588221.2025.2454541}{doi:\nolinkurl{10.1080/09588221.2025.2454541}}
\newblock
\shownote{Publisher: Informa UK Limited}.


\bibitem[Song et~al\mbox{.}(2025)]%
        {song2025}
\bibfield{author}{\bibinfo{person}{Yukyeong Song}, \bibinfo{person}{Chenglu Li}, \bibinfo{person}{Wanli Xing}, \bibinfo{person}{Bailing Lyu}, {and} \bibinfo{person}{Wangda Zhu}.} \bibinfo{year}{2025}\natexlab{}.
\newblock \showarticletitle{Investigating perceived fairness of {AI} prediction system for math learning: {A} mixed-methods study with college students}.
\newblock \bibinfo{journal}{\emph{The Internet and Higher Education}}  \bibinfo{volume}{65} (\bibinfo{date}{April} \bibinfo{year}{2025}), \bibinfo{pages}{101000}.
\newblock
\showISSN{1096-7516}
\href{https://doi.org/10.1016/j.iheduc.2025.101000}{doi:\nolinkurl{10.1016/j.iheduc.2025.101000}}
\newblock
\shownote{Publisher: Elsevier BV}.


\bibitem[Srinivasa et~al\mbox{.}(2022)]%
        {srinivasa}
\bibfield{author}{\bibinfo{person}{K.~G. Srinivasa}, \bibinfo{person}{Muralidhar Kurni}, {and} \bibinfo{person}{Kuppala Saritha}.} \bibinfo{year}{2022}\natexlab{}.
\newblock \showarticletitle{Harnessing the {Power} of {AI} to {Education}}.
\newblock In \bibinfo{booktitle}{\emph{Learning, {Teaching}, and {Assessment} {Methods} for {Contemporary} {Learners}: {Pedagogy} for the {Digital} {Generation}}}. \bibinfo{publisher}{Springer Nature Singapore}, \bibinfo{address}{Singapore}, \bibinfo{pages}{311--342}.
\newblock
\showISBNx{978-981-19-6734-4}
\href{https://doi.org/10.1007/978-981-19-6734-4_13}{doi:\nolinkurl{10.1007/978-981-19-6734-4_13}}


\bibitem[Stadler et~al\mbox{.}(2024)]%
        {stadler2024}
\bibfield{author}{\bibinfo{person}{Matthias Stadler}, \bibinfo{person}{Maria Bannert}, {and} \bibinfo{person}{Michael Sailer}.} \bibinfo{year}{2024}\natexlab{}.
\newblock \showarticletitle{Cognitive ease at a cost: {LLMs} reduce mental effort but compromise depth in student scientific inquiry}.
\newblock \bibinfo{journal}{\emph{Computers in Human Behavior}}  \bibinfo{volume}{160} (\bibinfo{date}{Nov.} \bibinfo{year}{2024}), \bibinfo{pages}{108386}.
\newblock
\showISSN{07475632}
\href{https://doi.org/10.1016/j.chb.2024.108386}{doi:\nolinkurl{10.1016/j.chb.2024.108386}}


\bibitem[Stróżyna et~al\mbox{.}(2024)]%
        {strozyna2024}
\bibfield{author}{\bibinfo{person}{Milena Stróżyna}, \bibinfo{person}{Krzysztof Węcel}, \bibinfo{person}{Piotr Stolarski}, \bibinfo{person}{Ewelina Księżniak}, \bibinfo{person}{Marcin Sawiński}, \bibinfo{person}{Włodzimierz Lewoniewski}, {and} \bibinfo{person}{Witold Abramowicz}.} \bibinfo{year}{2024}\natexlab{}.
\newblock \showarticletitle{Exploring the {Challenges} and {Potential} of {Generative} {AI}: {Insights} from an {Empirical} {Study}}.
\newblock \bibinfo{journal}{\emph{Procedia Computer Science}}  \bibinfo{volume}{246} (\bibinfo{year}{2024}), \bibinfo{pages}{2042--2051}.
\newblock
\showISSN{1877-0509}
\href{https://doi.org/10.1016/j.procs.2024.09.658}{doi:\nolinkurl{10.1016/j.procs.2024.09.658}}
\newblock
\shownote{Publisher: Elsevier BV}.


\bibitem[Stuchlikova and Weis(2024)]%
        {Stuchlikova}
\bibfield{author}{\bibinfo{person}{Lubica Stuchlikova} {and} \bibinfo{person}{Martin Weis}.} \bibinfo{year}{2024}\natexlab{}.
\newblock \showarticletitle{From Information to Insight: Reimagining Critical Thinking Pedagogy in the Age of Artificial Intelligence}. In \bibinfo{booktitle}{\emph{2024 International Conference on Emerging eLearning Technologies and Applications (ICETA)}}. \bibinfo{publisher}{IEEE}, \bibinfo{address}{Slovakia}, \bibinfo{pages}{591--598}.
\newblock
\href{https://doi.org/10.1109/ICETA63795.2024.10850787}{doi:\nolinkurl{10.1109/ICETA63795.2024.10850787}}


\bibitem[Stureborg et~al\mbox{.}(2024)]%
        {stureborg}
\bibfield{author}{\bibinfo{person}{Rickard Stureborg}, \bibinfo{person}{Dimitris Alikaniotis}, {and} \bibinfo{person}{Yoshi Suhara}.} \bibinfo{year}{2024}\natexlab{}.
\newblock \bibinfo{title}{Large Language Models are Inconsistent and Biased Evaluators}.
\newblock
\showeprint[arxiv]{2405.01724}~[cs.CL]
\urldef\tempurl%
\url{https://arxiv.org/abs/2405.01724}
\showURL{%
\tempurl}


\bibitem[Suzgun et~al\mbox{.}(2024)]%
        {suzgun2024}
\bibfield{author}{\bibinfo{person}{Mirac Suzgun}, \bibinfo{person}{Tayfun Gur}, \bibinfo{person}{Federico Bianchi}, \bibinfo{person}{Daniel~E. Ho}, \bibinfo{person}{Thomas Icard}, \bibinfo{person}{Dan Jurafsky}, {and} \bibinfo{person}{James Zou}.} \bibinfo{year}{2024}\natexlab{}.
\newblock \bibinfo{title}{Belief in the Machine: Investigating Epistemological Blind Spots of Language Models}.
\newblock
\showeprint[arxiv]{2410.21195}~[cs.CL]
\urldef\tempurl%
\url{https://arxiv.org/abs/2410.21195}
\showURL{%
\tempurl}


\bibitem[Sweller(1988)]%
        {sweller1988}
\bibfield{author}{\bibinfo{person}{John Sweller}.} \bibinfo{year}{1988}\natexlab{}.
\newblock \showarticletitle{Cognitive {Load} {During} {Problem} {Solving}: {Effects} on {Learning}}.
\newblock \bibinfo{journal}{\emph{Cognitive Science}} \bibinfo{volume}{12}, \bibinfo{number}{2} (\bibinfo{date}{April} \bibinfo{year}{1988}), \bibinfo{pages}{257--285}.
\newblock
\showISSN{0364-0213, 1551-6709}
\href{https://doi.org/10.1207/s15516709cog1202_4}{doi:\nolinkurl{10.1207/s15516709cog1202_4}}


\bibitem[Tapalova et~al\mbox{.}(2022)]%
        {Tapalova}
\bibfield{author}{\bibinfo{person}{Olga Tapalova}, \bibinfo{person}{Nadezhda Zhiyenbayeva}, {and} \bibinfo{person}{Dmitry Gura}.} \bibinfo{year}{2022}\natexlab{}.
\newblock \showarticletitle{Artificial Intelligence in Education: AIEd for Personalised Learning Pathways}.
\newblock \bibinfo{journal}{\emph{Electronic Journal of e-Learning}}  \bibinfo{volume}{20} (\bibinfo{date}{12} \bibinfo{year}{2022}), \bibinfo{pages}{639--653}.
\newblock
\href{https://doi.org/10.34190/ejel.20.5.2597}{doi:\nolinkurl{10.34190/ejel.20.5.2597}}


\bibitem[Taylor(2024)]%
        {taylor_melbourne_2024}
\bibfield{author}{\bibinfo{person}{Josh Taylor}.} \bibinfo{year}{2024}\natexlab{}.
\newblock \showarticletitle{Melbourne lawyer referred to complaints body after {AI} generated made-up case citations in family court}.
\newblock \bibinfo{journal}{\emph{The Guardian}} \bibinfo{volume}{x}, \bibinfo{number}{y} (\bibinfo{date}{Oct.} \bibinfo{year}{2024}), \bibinfo{pages}{1--3}.
\newblock
\showISSN{0261-3077}
\urldef\tempurl%
\url{https://www.theguardian.com/law/2024/oct/10/melbourne-lawyer-referred-to-complaints-body-after-ai-generated-made-up-case-citations-in-family-court-ntwnfb}
\showURL{%
\tempurl}


\bibitem[Thorp(2023)]%
        {thorp2023}
\bibfield{author}{\bibinfo{person}{H.~Holden Thorp}.} \bibinfo{year}{2023}\natexlab{}.
\newblock \showarticletitle{{ChatGPT} is fun, but not an author}.
\newblock \bibinfo{journal}{\emph{Science}} \bibinfo{volume}{379}, \bibinfo{number}{6630} (\bibinfo{year}{2023}), \bibinfo{pages}{313--313}.
\newblock
\href{https://doi.org/10.1126/science.adg7879}{doi:\nolinkurl{10.1126/science.adg7879}}
\newblock
\shownote{\_eprint: https://www.science.org/doi/pdf/10.1126/science.adg7879}.


\bibitem[Trikoili et~al\mbox{.}(2025)]%
        {trikoili2025}
\bibfield{author}{\bibinfo{person}{Anna Trikoili}, \bibinfo{person}{Despoina Georgiou}, {and} \bibinfo{person}{Daniel~Pittich Pappa, Christina Ioanna~and}.} \bibinfo{year}{2025}\natexlab{}.
\newblock \showarticletitle{Critical {Thinking} {Assessment} in {Higher} {Education}: {A} {Mixed}-{Methods} {Comparative} {Analysis} of {AI} and {Human} {Evaluator}}.
\newblock \bibinfo{journal}{\emph{International Journal of Human–Computer Interaction}} \bibinfo{volume}{0}, \bibinfo{number}{0} (\bibinfo{year}{2025}), \bibinfo{pages}{1--14}.
\newblock
\href{https://doi.org/10.1080/10447318.2025.2499164}{doi:\nolinkurl{10.1080/10447318.2025.2499164}}
\newblock
\shownote{Publisher: Taylor \& Francis \_eprint: https://doi.org/10.1080/10447318.2025.2499164}.


\bibitem[Van~Poucke(2024)]%
        {Poucke}
\bibfield{author}{\bibinfo{person}{Margo Van~Poucke}.} \bibinfo{year}{2024}\natexlab{}.
\newblock \showarticletitle{ChatGPT, the perfect virtual teaching assistant? Ideological bias in learner-chatbot interactions}.
\newblock \bibinfo{journal}{\emph{Computers and Composition}}  \bibinfo{volume}{73} (\bibinfo{date}{06} \bibinfo{year}{2024}), \bibinfo{pages}{102871}.
\newblock
\href{https://doi.org/10.1016/j.compcom.2024.102871}{doi:\nolinkurl{10.1016/j.compcom.2024.102871}}


\bibitem[VanLehn(2011)]%
        {Kurt}
\bibfield{author}{\bibinfo{person}{Kurt VanLehn}.} \bibinfo{year}{2011}\natexlab{}.
\newblock \showarticletitle{The relative effectiveness of human tutoring, intelligent tutoring systems, and other tutoring systems}.
\newblock \bibinfo{journal}{\emph{Educational Psychologist}} \bibinfo{volume}{46}, \bibinfo{number}{4} (\bibinfo{year}{2011}), \bibinfo{pages}{197--221}.
\newblock
\showISSN{0046-1520}
\href{https://doi.org/10.1080/00461520.2011.611369}{doi:\nolinkurl{10.1080/00461520.2011.611369}}


\bibitem[Vygotsky and Cole(1978)]%
        {vygotsky1978}
\bibfield{author}{\bibinfo{person}{Lev~Semenovich Vygotsky} {and} \bibinfo{person}{Michael Cole}.} \bibinfo{year}{1978}\natexlab{}.
\newblock \bibinfo{booktitle}{\emph{Mind in society: Development of higher psychological processes}}.
\newblock \bibinfo{publisher}{Harvard university press}, \bibinfo{address}{Cambridge, MA}.
\newblock


\bibitem[Wang et~al\mbox{.}(2024)]%
        {wang_2024}
\bibfield{author}{\bibinfo{person}{Chenyue Wang}, \bibinfo{person}{Sophie~C Boerman}, \bibinfo{person}{Anne~C Kroon}, \bibinfo{person}{Judith Möller}, {and} \bibinfo{person}{Claes H~De~Vreese}.} \bibinfo{year}{2024}\natexlab{}.
\newblock \showarticletitle{The Artificial Itelligence Divide: {Who} is the Most Vulnerable?}
\newblock \bibinfo{journal}{\emph{New Media \& Society}} \bibinfo{volume}{27}, \bibinfo{number}{7} (\bibinfo{date}{Feb.} \bibinfo{year}{2024}), \bibinfo{pages}{3867--3889}.
\newblock
\showISSN{1461-4448, 1461-7315}
\href{https://doi.org/10.1177/14614448241232345}{doi:\nolinkurl{10.1177/14614448241232345}}


\bibitem[Warr et~al\mbox{.}(2024)]%
        {Warr}
\bibfield{author}{\bibinfo{person}{Melissa Warr}, \bibinfo{person}{Nicole Oster}, {and} \bibinfo{person}{Roger Isaac}.} \bibinfo{year}{2024}\natexlab{}.
\newblock \showarticletitle{Implicit bias in large language models: Experimental proof and implications for education}.
\newblock \bibinfo{journal}{\emph{Journal of Research on Technology in Education}} \bibinfo{volume}{1}, \bibinfo{number}{1} (\bibinfo{date}{08} \bibinfo{year}{2024}), \bibinfo{pages}{1--24}.
\newblock
\href{https://doi.org/10.1080/15391523.2024.2395295}{doi:\nolinkurl{10.1080/15391523.2024.2395295}}


\bibitem[Watts et~al\mbox{.}(2023)]%
        {watts3}
\bibfield{author}{\bibinfo{person}{Field~M. Watts}, \bibinfo{person}{Amber~J. Dood}, \bibinfo{person}{Ginger~V. Shultz}, {and} \bibinfo{person}{Jon-Marc~G. Rodriguez}.} \bibinfo{year}{2023}\natexlab{}.
\newblock \showarticletitle{Comparing {Student} and {Generative} {Artificial} {Intelligence} {Chatbot} {Responses} to {Organic} {Chemistry} {Writing}-to-{Learn} {Assignments}}.
\newblock \bibinfo{journal}{\emph{Journal of Chemical Education}} \bibinfo{volume}{100}, \bibinfo{number}{10} (\bibinfo{date}{Oct.} \bibinfo{year}{2023}), \bibinfo{pages}{3806--3817}.
\newblock
\showISSN{0021-9584}
\href{https://doi.org/10.1021/acs.jchemed.3c00664}{doi:\nolinkurl{10.1021/acs.jchemed.3c00664}}
\newblock
\shownote{Publisher: American Chemical Society}.


\bibitem[Weidinger et~al\mbox{.}(2021)]%
        {weidinger}
\bibfield{author}{\bibinfo{person}{Laura Weidinger}, \bibinfo{person}{John Mellor}, \bibinfo{person}{Maribeth Rauh}, \bibinfo{person}{Conor Griffin}, \bibinfo{person}{Jonathan Uesato}, \bibinfo{person}{Po-Sen Huang}, \bibinfo{person}{Myra Cheng}, \bibinfo{person}{Mia Glaese}, \bibinfo{person}{Borja Balle}, \bibinfo{person}{Atoosa Kasirzadeh}, \bibinfo{person}{Zac Kenton}, \bibinfo{person}{Sasha Brown}, \bibinfo{person}{Will Hawkins}, \bibinfo{person}{Tom Stepleton}, \bibinfo{person}{Courtney Biles}, \bibinfo{person}{Abeba Birhane}, \bibinfo{person}{Julia Haas}, \bibinfo{person}{Laura Rimell}, \bibinfo{person}{Lisa~Anne Hendricks}, \bibinfo{person}{William Isaac}, \bibinfo{person}{Sean Legassick}, \bibinfo{person}{Geoffrey Irving}, {and} \bibinfo{person}{Iason Gabriel}.} \bibinfo{year}{2021}\natexlab{}.
\newblock \bibinfo{title}{Ethical and social risks of harm from Language Models}.
\newblock
\showeprint[arxiv]{2112.04359}~[cs.CL]
\urldef\tempurl%
\url{https://arxiv.org/abs/2112.04359}
\showURL{%
\tempurl}


\bibitem[Weissburg et~al\mbox{.}(2025)]%
        {weissburgg}
\bibfield{author}{\bibinfo{person}{Iain Weissburg}, \bibinfo{person}{Sathvika Anand}, \bibinfo{person}{Sharon Levy}, {and} \bibinfo{person}{Haewon Jeong}.} \bibinfo{year}{2025}\natexlab{}.
\newblock \bibinfo{title}{LLMs are Biased Teachers: Evaluating LLM Bias in Personalized Education}.
\newblock
\showeprint[arxiv]{2410.14012}~[cs.CL]
\urldef\tempurl%
\url{https://arxiv.org/abs/2410.14012}
\showURL{%
\tempurl}


\bibitem[Woerner et~al\mbox{.}(2024)]%
        {Woerner}
\bibfield{author}{\bibinfo{person}{Jan~H.R. Woerner}, \bibinfo{person}{Aleksandra~P. Turtova}, {and} \bibinfo{person}{Andrew~S.I.D. Lang}.} \bibinfo{year}{2024}\natexlab{}.
\newblock \showarticletitle{Transformative Potentials and Ethical Considerations of AI Tools in Higher Education: Case Studies and Reflections}. In \bibinfo{booktitle}{\emph{SoutheastCon 2024}}. \bibinfo{publisher}{IEEE}, \bibinfo{address}{Location}, \bibinfo{pages}{510--515}.
\newblock
\href{https://doi.org/10.1109/SoutheastCon52093.2024.10500042}{doi:\nolinkurl{10.1109/SoutheastCon52093.2024.10500042}}


\bibitem[Wu et~al\mbox{.}(2023)]%
        {wu2023}
\bibfield{author}{\bibinfo{person}{Tianyu Wu}, \bibinfo{person}{Shizhu He}, \bibinfo{person}{Jingping Liu}, \bibinfo{person}{Siqi Sun}, \bibinfo{person}{Kang Liu}, \bibinfo{person}{Qing-Long Han}, {and} \bibinfo{person}{Yang Tang}.} \bibinfo{year}{2023}\natexlab{}.
\newblock \showarticletitle{A brief overview of ChatGPT: The history, status quo and potential future development}.
\newblock \bibinfo{journal}{\emph{IEEE/CAA Journal of Automatica Sinica}} \bibinfo{volume}{10}, \bibinfo{number}{5} (\bibinfo{year}{2023}), \bibinfo{pages}{1122--1136}.
\newblock


\bibitem[Xing and Du(2018)]%
        {Xing}
\bibfield{author}{\bibinfo{person}{Wanli Xing} {and} \bibinfo{person}{Dongping Du}.} \bibinfo{year}{2018}\natexlab{}.
\newblock \showarticletitle{Dropout Prediction in MOOCs: Using Deep Learning for Personalized Intervention}.
\newblock \bibinfo{journal}{\emph{Journal of Educational Computing Research}}  \bibinfo{volume}{57} (\bibinfo{date}{03} \bibinfo{year}{2018}), \bibinfo{pages}{073563311875701}.
\newblock
\href{https://doi.org/10.1177/0735633118757015}{doi:\nolinkurl{10.1177/0735633118757015}}


\bibitem[Xu et~al\mbox{.}(2024)]%
        {xu}
\bibfield{author}{\bibinfo{person}{Hanyi Xu}, \bibinfo{person}{Wensheng Gan}, \bibinfo{person}{Zhenlian Qi}, \bibinfo{person}{Jiayang Wu}, {and} \bibinfo{person}{Philip Yu}.} \bibinfo{year}{2024}\natexlab{}.
\newblock \bibinfo{title}{Large Language Models for Education: A Survey}.
\newblock
\href{https://doi.org/10.48550/arXiv.2405.13001}{doi:\nolinkurl{10.48550/arXiv.2405.13001}}


\bibitem[Yang et~al\mbox{.}(2024)]%
        {yang2024}
\bibfield{author}{\bibinfo{person}{Yeonsun Yang}, \bibinfo{person}{Ahyeon Shin}, \bibinfo{person}{Mincheol Kang}, \bibinfo{person}{Jiheon Kang}, {and} \bibinfo{person}{Jean~Young Song}.} \bibinfo{year}{2024}\natexlab{}.
\newblock \bibinfo{title}{Can {We} {Delegate} {Learning} to {Automation}?: {A} {Comparative} {Study} of {LLM} {Chatbots}, {Search} {Engines}, and {Books}}.
\newblock
\href{https://doi.org/10.48550/ARXIV.2410.01396}{doi:\nolinkurl{10.48550/ARXIV.2410.01396}}
\newblock
\shownote{Version Number: 1}.


\bibitem[Yildiz~Durak et~al\mbox{.}(2025)]%
        {durak2025}
\bibfield{author}{\bibinfo{person}{Hatice Yildiz~Durak}, \bibinfo{person}{Figen Eğin}, {and} \bibinfo{person}{Aytuğ Onan}.} \bibinfo{year}{2025}\natexlab{}.
\newblock \showarticletitle{A Comparison of Human‐Written Versus {AI}‐Generated Text in Discussions at Educational Settings: Investigating Features for {ChatGPT}, Gemini and {BingAI}}.
\newblock \bibinfo{journal}{\emph{European Journal of Education}} \bibinfo{volume}{60}, \bibinfo{number}{1} (\bibinfo{date}{March} \bibinfo{year}{2025}), \bibinfo{pages}{e70014}.
\newblock
\showISSN{0141-8211, 1465-3435}
\href{https://doi.org/10.1111/ejed.70014}{doi:\nolinkurl{10.1111/ejed.70014}}


\bibitem[Yu et~al\mbox{.}(2024)]%
        {yu2024}
\bibfield{author}{\bibinfo{person}{Lei Yu}, \bibinfo{person}{Meng Cao}, \bibinfo{person}{Jackie Chi~Kit Cheung}, {and} \bibinfo{person}{Yue Dong}.} \bibinfo{year}{2024}\natexlab{}.
\newblock \bibinfo{title}{Mechanistic Understanding and Mitigation of Language Model Non-Factual Hallucinations}.
\newblock
\showeprint[arxiv]{2403.18167}~[cs.CL]
\urldef\tempurl%
\url{https://arxiv.org/abs/2403.18167}
\showURL{%
\tempurl}


\bibitem[Zhai et~al\mbox{.}(2024)]%
        {zhai}
\bibfield{author}{\bibinfo{person}{Chunpeng Zhai}, \bibinfo{person}{Santoso Wibowo}, {and} \bibinfo{person}{Lily~D. Li}.} \bibinfo{year}{2024}\natexlab{}.
\newblock \showarticletitle{Evaluating the {AI} dialogue {System}'s intercultural, humorous, and empathetic dimensions in {English} language learning: {A} case study}.
\newblock \bibinfo{journal}{\emph{Computers and Education: Artificial Intelligence}}  \bibinfo{volume}{7} (\bibinfo{year}{2024}), \bibinfo{pages}{100262}.
\newblock
\showISSN{2666-920X}
\href{https://doi.org/10.1016/j.caeai.2024.100262}{doi:\nolinkurl{10.1016/j.caeai.2024.100262}}


\bibitem[Zhan(2025)]%
        {zhan2025}
\bibfield{author}{\bibinfo{person}{Zi~Yan Zhan, Ying~and}.} \bibinfo{year}{2025}\natexlab{}.
\newblock \showarticletitle{Students’ engagement with {ChatGPT} feedback: implications for student feedback literacy in the context of generative artificial intelligence}.
\newblock \bibinfo{journal}{\emph{Assessment \& Evaluation in Higher Education}} \bibinfo{volume}{0}, \bibinfo{number}{0} (\bibinfo{year}{2025}), \bibinfo{pages}{1--14}.
\newblock
\href{https://doi.org/10.1080/02602938.2025.2471821}{doi:\nolinkurl{10.1080/02602938.2025.2471821}}
\newblock
\shownote{Publisher: SRHE Website \_eprint: https://doi.org/10.1080/02602938.2025.2471821}.


\bibitem[Zhang et~al\mbox{.}(2025)]%
        {Zhang2025}
\bibfield{author}{\bibinfo{person}{Jing Zhang}, \bibinfo{person}{Wenlong Song}, {and} \bibinfo{person}{Yang Liu}.} \bibinfo{year}{2025}\natexlab{}.
\newblock \showarticletitle{Cognitive bias in generative AI influences religious education}.
\newblock \bibinfo{journal}{\emph{Scientific Reports}}  \bibinfo{volume}{15} (\bibinfo{date}{05} \bibinfo{year}{2025}).
\newblock
\href{https://doi.org/10.1038/s41598-025-99121-6}{doi:\nolinkurl{10.1038/s41598-025-99121-6}}


\bibitem[Zhang et~al\mbox{.}(2024)]%
        {zhang2024}
\bibfield{author}{\bibinfo{person}{Yuji Zhang}, \bibinfo{person}{Sha Li}, \bibinfo{person}{Jiateng Liu}, \bibinfo{person}{Pengfei Yu}, \bibinfo{person}{Yi~R. Fung}, \bibinfo{person}{Jing Li}, \bibinfo{person}{Manling Li}, {and} \bibinfo{person}{Heng Ji}.} \bibinfo{year}{2024}\natexlab{}.
\newblock \bibinfo{title}{Knowledge Overshadowing Causes Amalgamated Hallucination in Large Language Models}.
\newblock
\showeprint[arxiv]{2407.08039}~[cs.CL]
\urldef\tempurl%
\url{https://arxiv.org/abs/2407.08039}
\showURL{%
\tempurl}


\bibitem[Zheng(2023)]%
        {zheng2023}
\bibfield{author}{\bibinfo{person}{Alex Zheng}.} \bibinfo{year}{2023}\natexlab{}.
\newblock \bibinfo{title}{Dissecting Bias of ChatGPT in College Major Recommendations}.
\newblock
\showeprint[arxiv]{2401.11699}~[cs.CY]
\urldef\tempurl%
\url{https://arxiv.org/abs/2401.11699}
\showURL{%
\tempurl}


\bibitem[Zheng and Stewart(2024)]%
        {Danson2024}
\bibfield{author}{\bibinfo{person}{Y.~Danson Zheng} {and} \bibinfo{person}{Nicola Stewart}.} \bibinfo{year}{2024}\natexlab{}.
\newblock \showarticletitle{Improving EFL students’ cultural awareness: Reframing moral dilemmatic stories with ChatGPT}.
\newblock \bibinfo{journal}{\emph{Computers and Education Artificial Intelligence}}  \bibinfo{volume}{6} (\bibinfo{date}{04} \bibinfo{year}{2024}), \bibinfo{pages}{100223}.
\newblock
\href{https://doi.org/10.1016/j.caeai.2024.100223}{doi:\nolinkurl{10.1016/j.caeai.2024.100223}}


\bibitem[Zhu et~al\mbox{.}(2025)]%
        {zhu_impact_2025}
\bibfield{author}{\bibinfo{person}{Yumeng Zhu}, \bibinfo{person}{Caifeng Zhu}, \bibinfo{person}{Tao Wu}, \bibinfo{person}{Shulei Wang}, \bibinfo{person}{Yiyun Zhou}, \bibinfo{person}{Jingyuan Chen}, \bibinfo{person}{Fei Wu}, {and} \bibinfo{person}{Yan Li}.} \bibinfo{year}{2025}\natexlab{}.
\newblock \showarticletitle{Impact of assignment completion assisted by {Large} {Language} {Model}-based chatbot on middle school students’ learning}.
\newblock \bibinfo{journal}{\emph{Education and Information Technologies}} \bibinfo{volume}{30}, \bibinfo{number}{2} (\bibinfo{date}{Feb.} \bibinfo{year}{2025}), \bibinfo{pages}{2429--2461}.
\newblock
\showISSN{1360-2357, 1573-7608}
\href{https://doi.org/10.1007/s10639-024-12898-3}{doi:\nolinkurl{10.1007/s10639-024-12898-3}}
\newblock
\shownote{Publisher: Springer Science and Business Media LLC}.


\bibitem[Zimmerman(2002)]%
        {Zimmerman2002}
\bibfield{author}{\bibinfo{person}{Barry~J. Zimmerman}.} \bibinfo{year}{2002}\natexlab{}.
\newblock \showarticletitle{Becoming a Self-Regulated Learner: An Overview}.
\newblock \bibinfo{journal}{\emph{Theory Into Practice}} \bibinfo{volume}{41}, \bibinfo{number}{2} (\bibinfo{year}{2002}), \bibinfo{pages}{64--70}.
\newblock


\end{thebibliography}

\end{document}